\documentclass[aps,pra,reprint,longbibliography]{revtex4-2}
\usepackage[T1]{fontenc}
\usepackage[utf8]{inputenc}
\usepackage{amsmath,amssymb,amsthm,mathrsfs}
\usepackage{graphicx,booktabs,array,tabularx,placeins,float}
\usepackage[hidelinks]{hyperref}
\setcitestyle{numbers,square,sort&compress}

\newtheorem{theorem}{Theorem}
\newtheorem{proposition}[theorem]{Proposition}

\newtheorem{corollary}[theorem]{Corollary}
\newcommand{\Sref}[1]{\ref{#1}}
\newif\ifstandalonesupp

\begin{document}

\title{\textbf{The speed limit of gravitational entanglement\\ and the cost of noise}}
\author{Ekta Panwar}
\affiliation{Institute of Physics, Slovak Academy of Sciences, D\'ubravsk\'a cesta 9, 84511 Bratislava, Slovakia}
\author{Anubhav Chaturvedi}
\affiliation{Institute of Theoretical Physics and Astrophysics, Faculty of Mathematics, Physics and Informatics, University of Gda\'nsk, 80-308 Gda\'nsk, Poland}

\begin{abstract}
\noindent Is gravity quantum? A proposed test lets two small masses, each in a
superposition of two places, become entangled through their mutual gravity.
Gravity is weak, so the masses must stay coherent for seconds while random
forces disturb them. Here we ask how fast gravity can entangle two masses, and
what noise costs. Without noise we find the maximum rate exactly: it falls as
the inverse cube of the separation, and two positions per mass achieve it.
Above a noise level set by the masses and their separation, gravity's leading
pull can be imitated by classical signals, and every strategy is limited to a
rate falling at least as the inverse fourth power. Exact, continuous error
correction shields the masses from noise but hides them from each other: the
best rate falls as the inverse seventh power, and six positions on a hexagon come
within 10\% of it at large separation.
\end{abstract}
\maketitle

\section*{Introduction}

Nobody knows whether gravity obeys the rules of quantum mechanics. Every other
force of nature does, but gravity is so feeble between small objects that its
quantum side has never been seen. A way to look for it without a particle
accelerator was proposed a few years ago\cite{Bose2017,MarlettoVedral2017}. Take
two small masses and put each of them in a superposition of two places: in two
places at once, as quantum mechanics allows. If gravity can then entangle the
two masses, so that their joint state cannot be described by a separate pure state for
each mass, the gravitational
pull between them cannot be something that two separate laboratories could
imitate with ordinary measurements and classical messages. Many regard this as
a decisive tabletop test, and it has become the goal of a growing experimental
effort\cite{Carney2019,Krisnanda2020,Lami2024}. Related witnesses quantify the
nonclassicality of mediated interactions through the correlations they
generate\cite{Ganardi2024}.

\begin{figure*}[t]
\centering
\includegraphics[width=\textwidth]{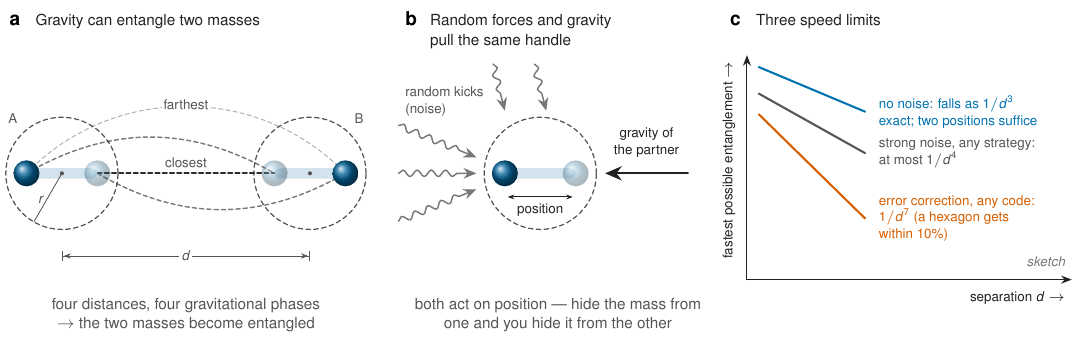}
\caption{\textbf{The experiment, the obstacle and three speed limits.}
\textbf{a}, Two masses, each in two places at once inside its own region of
radius $r$; the regions are a distance $d$ apart. The four combinations of
positions put the masses at four distances, gravity acts differently on each,
and the masses become entangled. \textbf{b}, Random kicks from the surroundings
and the pull of the partner mass both act on the position of a mass. A scheme
that hides the position from the kicks hides it from gravity too. \textbf{c},
The results of this paper, as a sketch. Without noise, the fastest possible
entanglement falls as $1/d^3$. Under strong random forces, no strategy can beat
a rate that falls as $1/d^4$. With exact error correction, the best rate falls
as $1/d^7$, and a six-position code nearly reaches it.}
\label{fig:story}
\end{figure*}

The mechanism is easy to picture (Fig.~\ref{fig:story}a). With each mass in two
places at once, there are four combinations of positions, and the two masses are
a different distance apart in each. Gravity is stronger when they are closer, so
each combination picks up a slightly different quantum phase, a kind of
internal clock reading. Wait long enough, and the four clocks drift apart in a
way that leaves the two masses entangled.

``Long enough'' is the problem. Gravity has been measured between gold spheres
two millimetres across\cite{Westphal2021}, and molecules of tens of thousands of
atomic mass units have been put in two places at once\cite{Fein2019}, but the
entanglement test needs both at the same time: objects heavy enough to
gravitate, in superpositions wide enough for gravity to tell the branches
apart. For the masses usually discussed, specks a couple of micrometres across
held a few hair-widths from each other, the gravitational effect takes many
seconds to build up. All the while, stray gas molecules, electric fields and
vibrations give the masses small random kicks, and a random kick is exactly what
destroys a superposition of two places.

There is a well-known cure for noise in quantum devices, called error
correction. One stores the quantum state not in two positions but in a cleverly
chosen pattern of several positions, a code. The information then sits in the
shape of the pattern, not in where the mass is, so that a random kick shifts
the mass without learning or spoiling what is stored, and its damage can be
undone. Codes of this kind are used in quantum
sensing\cite{ZhouEtAl2018,DemkowiczDobrzanski2017,LaydenCappellaro2018,LaydenEtAl2019}
and were recently proposed for gravitational entanglement by Wang, Pedernales
and Plenio\cite{WangPedernalesPlenio2026}. But here the cure has a side effect
(Fig.~\ref{fig:story}b). A random kick and the pull of the partner mass grip the
mass by the same handle: its position. A code that hides position from the noise
hides it from gravity as well. Wang and colleagues showed that exact correction
cancels the leading gravitational effect and found codes that keep a weaker,
shorter-ranged remnant\cite{WangPedernalesPlenio2026}. A related fact was known
for unprotected masses: if the random forces are strong enough, the leading
gravitational effect can be imitated by classical means and creates no
entanglement\cite{Kafri2014,LiMikiChen2026}.

These observations raise three plain questions, and this paper answers them.
How fast can gravity entangle two masses when nothing disturbs them? How much
of that speed does noise take away, whatever one does about it? And if one
shields the masses by error correction, how much speed can the best possible
shield keep? The answers are three speed limits (Fig.~\ref{fig:story}c). Each is
stated for the complete Newtonian pull between two masses that are free to move
inside two separate regions, and each holds for every strategy of the kind it
describes, including ones nobody has thought of.

\section*{Results}

\subsection*{A speed limit for gravity}

We begin with the ideal case of no noise at all. Two masses, $m_A$ and $m_B$, are
each free to be anywhere inside a spherical region of radius $r$; the centres of
the two regions are a distance $d>2r$ apart (Fig.~\ref{fig:story}a). The only
thing that connects the two laboratories is Newton's gravitational energy,
\begin{equation}
 V(\mathbf a,\mathbf b)=-\frac{Gm_Am_B}{|d\,\mathbf e_z+\mathbf b-\mathbf a|},
 \label{eq:potential}
\end{equation}
where $\mathbf a$ and $\mathbf b$ are the positions of the masses relative to
the centres of their regions and $\mathbf e_z$ points from one centre to the
other. Everything else is allowed and costs nothing. Each laboratory may prepare
any quantum state of its mass, use extra quantum systems as memories, measure,
and send classical messages to the other. The two laboratories start out
unentangled.

We measure entanglement in Bell pairs. A Bell pair is the basic unit of
entanglement, as the bit is the basic unit of information: two quantum bits,
one on each side, as entangled as two quantum bits can be. The \emph{entangling
capacity} is the largest number of Bell pairs per second that the laboratories
can produce, averaged over a long experiment in which many pairs are
made\cite{BennettHarrowLeungSmolin2003}. It is a speed limit: no protocol,
however elaborate, entangles faster.

Two quantities set the scale of everything that follows. One is a rate,
$g_0=Gm_Am_B/(\hbar d)$: the gravitational energy of the pair, turned into a
frequency by dividing by Planck's constant $\hbar$. For the masses we have in
mind it is a fraction of one per second. The other is the ratio $u=r/d$ of the
size of the regions to their separation, a number below one half. Our first
result is the exact capacity. Whenever the regions are not too close,
$d\ge(1+\sqrt5)\,r\approx3.24\,r$,
\begin{equation}
 C_G=\beta\,g_0\,\frac{2u^2}{1-4u^2},\qquad \beta=1.9123\ldots,
 \label{eq:capacity}
\end{equation}
where $\beta$ is a constant known from the theory of the simplest interaction
between two quantum
bits\cite{DuerVidalCiracLindenPopescu2001,ChildsLeungVerstraeteVidal2003,ChildsLeungVidal2004}.

The remarkable part is how little is needed. The best strategy is the simplest
one: put each mass in two places at once, as two narrow wavepackets
(well-localized lumps of the mass's quantum wave) at the two ends of the
diameter that points to the partner. Their four combinations sit at distances
$d-2r$, $d+2r$, $d$ and $d$, and the mismatch between these distances is all
the entangling power there is. Superpositions of many positions, motion
sideways, entanglement with local memories: none of it beats two points on a
line. That two points reach this rate was known; what is new is the proof that
nothing else does better. At large separation the speed limit falls as the cube
of the distance, $C_G\approx2\beta\,Gm_Am_Br^2/(\hbar d^3)$.

To make this concrete we use one example throughout: two masses of
$10^{-14}$\,kg in regions of radius $r=125\,\mu$m whose centres are
$d=450\,\mu$m apart. Then Eq.~\eqref{eq:capacity} gives $0.060$ Bell pairs per
second. At the very best, gravity delivers one Bell pair every 17 seconds.

Why can cleverness not help? In the proof (Methods), every possible position of
a mass is assigned a point in an abstract space, in such a way that the
gravitational energy of two positions is given by how their points line
up\cite{FrankLieb2010}. The entangling power available to any quantum state is
limited by how far apart these points can
be\cite{Bravyi2007,VanAcoleyen2013}. All of them lie inside a ball, and the two
ends of the axial diameter sit at opposite ends of it. Nothing can be farther
apart than that, which is why two positions are enough. Two caveats: the
capacity assumes ideal control with no limit on energy, and a fixed
kinetic-energy budget lowers it in the collinear equal-mass model; and when the two regions nearly touch, a third position helps
(Supplementary Sections~\Sref{sec:noiseless} and \Sref{sec:extensions}).

\subsection*{Enough noise closes gravity's main channel}

Now let the random kicks in. We model them in the simplest way: as a random
force that is the same everywhere in a region and has no memory from one moment
to the next. The forces on the two masses are independent, and nobody keeps a
record of them. Such noise is described by one number, its strength $\Lambda$,
and what it does is easy to state: a mass in two places a distance $L$ apart
loses its quantum coherence at the rate $\Lambda L^2$. Wider superpositions are
more fragile, by the square of their width. (A force with two-sided spectral density $S_F$ per Cartesian component
has $\Lambda=S_F/2\hbar^2$.)

Gravity's leading effect depends on size in the same way. When the masses are
far apart compared with the width of their superpositions, the gravitational
energy that distinguishes the four combinations of positions is proportional
to the product of the two widths: make each superposition $L$ wide, and the
entangling effect grows as $L^2$. Gravity and noise scale alike, the size of the
superposition cancels out, and the contest between them is decided by a
threshold that involves nothing but the two masses and their separation,
\begin{equation}
 \Lambda_c=\frac{Gm_Am_B}{\hbar d^3}.
 \label{eq:threshold}
\end{equation}

\begin{figure*}[t]
\centering
\includegraphics[width=\textwidth]{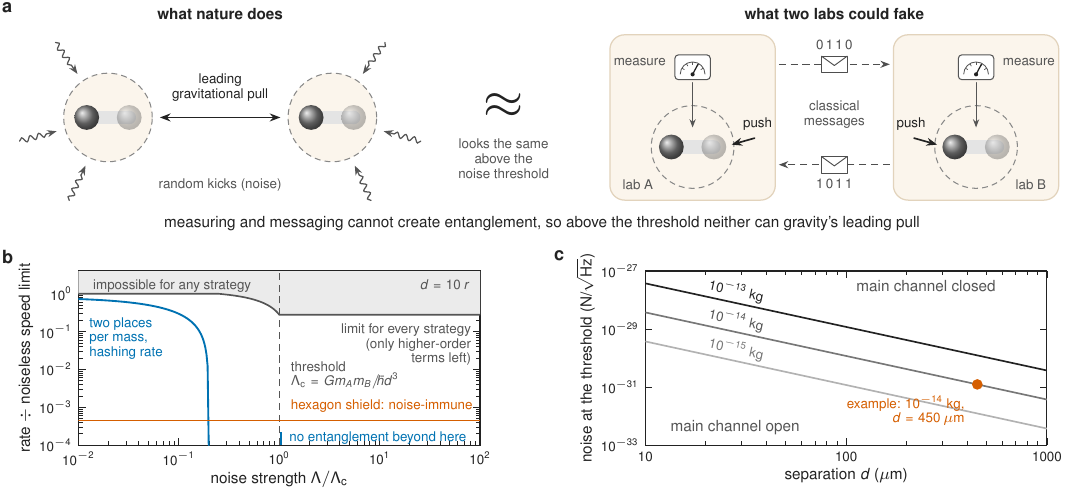}
\caption{\textbf{Enough noise closes gravity's main channel.} \textbf{a}, Above
a threshold of noise, gravity's leading pull together with the random kicks is
indistinguishable from two laboratories that measure the positions of their
masses, exchange the results as classical messages and push their masses
accordingly. Such a procedure cannot create entanglement. \textbf{b}, How fast
entanglement can be made (vertical axis, as a fraction of the noiseless speed
limit $C_G$) against the strength of the noise (horizontal axis, in units of
the threshold $\Lambda_c$ of Eq.~\eqref{eq:threshold}), for regions ten radii
apart. Grey: the limit for every strategy; the shaded area is impossible. The
limit drops as the noise approaches the threshold, because the main channel
closes and only the weaker effects of the potential are left
(Eq.~\eqref{eq:noisy} and Methods). Blue: a rate achievable with two unprotected
positions per mass; noise slows it, and beyond the marked point this scheme
produces no entanglement at all. Orange: the six-position error-correcting code
described below, which is immune to this noise but slow. Above the threshold,
the best possible rate lies between the orange and grey lines. \textbf{c}, The
noise at the threshold in laboratory units (newtons per root hertz; smaller is
quieter), against separation, for three masses. Above each line the main channel
is closed. The dot marks the running example.}
\label{fig:noise}
\end{figure*}

What happens at this threshold is striking (Fig.~\ref{fig:noise}a). Gravity's
leading effect, the one behind the $1/d^3$ law, couples the position of one
mass to the position of the other. When the noise on the two masses satisfies
$\sqrt{\Lambda_A\Lambda_B}\ge\Lambda_c$, this coupling together with the noise
can be imitated exactly by a mundane procedure: each laboratory keeps measuring
where its mass is, sends the result to the other as an ordinary, classical
message, and the other gives its mass a matching
push\cite{Kafri2014,LiMikiChen2026}. Laboratories that only measure and exchange
classical messages can never create entanglement; that is what makes
entanglement a quantum resource. So above the threshold gravity's main channel
is closed: it is still there, but nothing quantum can pass through it.

This much was known for the leading effect alone. The real potential has
further, weaker effects, and experimenters have further options: codes,
feedback, memories, keeping a record of which errors occurred. Our second
result covers all of it. Above the threshold, whatever the two laboratories do,
the capacity of the complete Newtonian interaction obeys
\begin{equation}
 \begin{aligned}
 C_{\rm noise}&\le\beta\,g_0\,B(u),\\
 B(u)&\approx\frac92\Bigl(\frac rd\Bigr)^{3}\qquad(u\to0),
 \end{aligned}
 \label{eq:noisy}
\end{equation}
with the complete expression given in Methods. The rate now falls at least as
fast as $1/d^4$: no ingenuity brings back the $1/d^3$ law. Compared with the
noiseless speed limit, the most that any strategy can reach is $54\%$ at
$d=6r$ and $28\%$ at $d=10r$, and the percentage keeps falling in proportion to
$r/d$ (Fig.~\ref{fig:noise}b). The idea of the proof is to split the true
evolution into the part that measuring and messaging can imitate and a
remainder, and to bound the entanglement that the remainder can produce while
the laboratories do whatever they like in between (Methods and Supplementary
Section~\Sref{sec:noise}).

Below the threshold the main channel is open, and two unprotected positions per
mass do become entangled, although the noise slows them down
(Fig.~\ref{fig:noise}b, blue). The threshold is not a knife edge. It marks
where gravity's leading effect stops working; the weaker effects let
unprotected superpositions entangle slightly beyond it (Methods).

In laboratory terms, the threshold for the running example is
$\Lambda_c=6.9\times10^{5}\,{\rm m^{-2}\,s^{-1}}$, which corresponds to a force
noise of $1.2\times10^{-31}\,{\rm N\,Hz^{-1/2}}$ (Fig.~\ref{fig:noise}c): in a
band of one hertz, fluctuations of less than a hundred-thousandth of the weight of a
single hydrogen atom. The threshold rises with the product of the masses and
falls with the cube of their separation, so heavier and closer masses can afford
more noise.

\subsection*{A shield that hides too much}

If the noise is too strong for unprotected masses, the natural response is to
shield them with error correction. To see what a shield does to gravity, look
at what a code must hide (Fig.~\ref{fig:shield}a).

A code stores one quantum bit in two states of a mass, the encoded states 0 and
1; each is a pattern of positions. Protection comes in two standard
grades\cite{KnillLaflamme1997}. A code \emph{detects} a random kick if one can
always tell that a kick has happened without disturbing what is stored. For
that, the two encoded states must have the same average position; otherwise the
kick, which acts on position, would treat them differently. A code
\emph{corrects} a kick if the damage can also be undone. For that the two
states must, in addition, have the same spread in every direction.

Gravity recognizes a mass by these same features. From a distance, a pattern of
positions pulls on its partner first through its average position (physicists
say its dipole), then through its spread (its quadrupole), then through its
three-lobed shape (its octupole), and so on, each feature reaching less far
than the one before. And a feature can help to entangle the two encoded bits
only if it differs between the encoded states; a feature they share tells the
partner nothing about which state is present.

Put the two together, and the consequence can be read off
Fig.~\ref{fig:shield}a. Without a shield, gravity uses the average positions,
and the encoded bits interact with a strength that falls as $1/d^3$. If kicks
can be detected, the average positions are hidden; the first usable feature is
the spread, and the interaction falls as $1/d^5$. If kicks can be corrected,
averages and spreads are both hidden; only the three-lobed shapes remain, and
the interaction falls as $1/d^7$. The shield and the price are one and the same
thing, and each further grade of protection costs two more powers of $r/d$.
(Detection by itself does not remove the noise. It is the stepping stone that
explains what correction costs.)

That the leading effect cancels under exact correction was shown by Wang and
colleagues, who also found the faster fall-off in one family of
codes\cite{WangPedernalesPlenio2026}. The question we now answer is whether a
better code could avoid the loss.

\subsection*{The best possible shield}

\begin{figure*}[t]
\centering
\includegraphics[width=\textwidth]{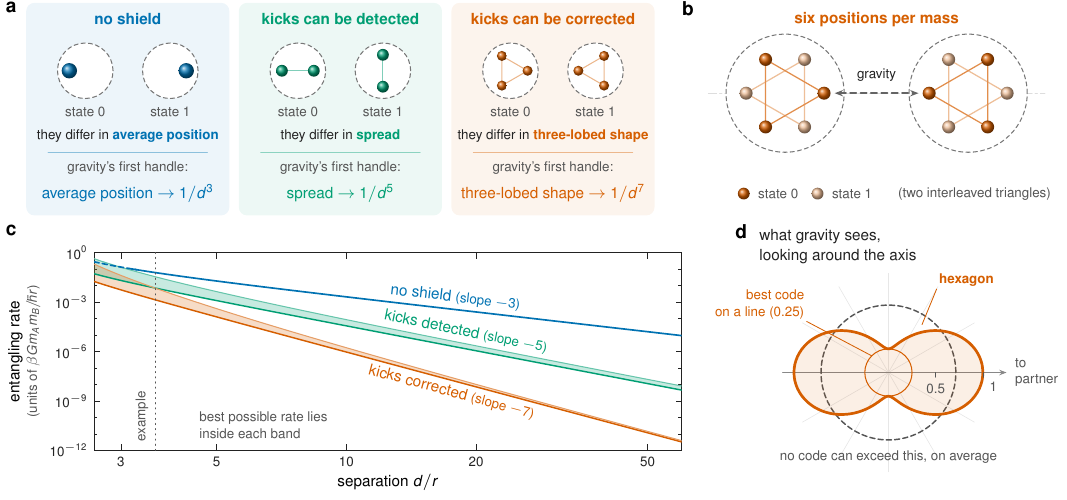}
\caption{\textbf{A shield against noise also hides a mass from gravity.}
\textbf{a}, The two encoded states of a code (0 and 1) are patterns of
positions. Without a shield they may differ in average position. To detect
random kicks they must share their average position and can first differ in
spread; to correct kicks they must share the spread as well and can first differ
in their three-lobed shape. Gravity can use only what differs, so the
interaction between the encoded states falls as $1/d^3$, $1/d^5$ and $1/d^7$.
\textbf{b}, The six-position code: the encoded states are the two triangles of a
hexagon, and the partner mass carries the mirror image. \textbf{c}, The fastest
possible entanglement against separation; the labels give the slopes, that is,
the powers of distance. Blue: the exact noiseless limit of
Eq.~\eqref{eq:capacity} (dashed where its proof does not apply). Green and
orange: codes that detect and codes that correct random kicks. Each band runs
from an explicit code (four positions on a square, six on a hexagon) up to the
limit that no code can exceed, so the best possible value lies inside the band.
For correcting codes this is a rate that can be reached under noise; for
detecting codes it is only the interaction between the encoded states.
\textbf{d}, Why a hexagon beats a line. Gravity looks at a code from every
direction $\phi$ around the line joining the masses; the distance from the
centre of the plot shows how strong a three-lobed signal it sees from each
direction. The best code on a line shows $1/4$ from all directions, the hexagon
$(1+3\cos^2\phi)/4$, and no code can show more than $0.713$ on
(root-mean-square) average (dashed).}
\label{fig:shield}
\end{figure*}

It cannot. Take any code that corrects random kicks exactly. It may use any
number of positions, store more than one quantum bit, and enlist extra quantum
memories in its laboratory. Let the correction be applied as often as one
likes, with the record of the errors discarded after each round, which is the
simplest way to use a code. Our third result is that, at large separation, the
best capacity $C_{\rm prot}$ of all such codes satisfies
\begin{equation}
 \frac{295}{32}\ \le\ \frac{C_{\rm prot}}{\beta\,g_0\,u^6}\ \le\ 10.177583\ldots+O(u).
 \label{eq:protected}
\end{equation}
So the best shielded rate falls as $1/d^7$, four powers of distance faster than
the unshielded limit, and the number in front is known to about 10\%. A
slightly weaker limit of the same form holds at every separation, not only far
away (Fig.~\ref{fig:shield}c), and the pattern continues: codes that correct
stronger classes of position errors lose four further powers of $r/d$ for each
class (Methods).

The lower value in Eq.~\eqref{eq:protected} comes from a code of appealing
simplicity (Fig.~\ref{fig:shield}b). Take six positions at the corners of a
regular hexagon, in a plane that contains the line between the masses, and let
the two encoded states be the two triangles formed by alternate corners. Each
triangle is the other one reflected through the centre. Such a reflection leaves
averages and spreads unchanged and reverses every three-lobed feature. The two
triangles therefore share their average position and their spread, as
correction demands, and have opposite three-lobed shapes, which is all that
gravity can still use. With the mirror-image hexagon at the partner mass, the
two encoded bits interact with a strength of at least
\begin{equation}
 g_{\rm hex}\ge\frac{295}{32}\,g_0\,u^{6},
 \label{eq:hex}
\end{equation}
and entangle at the rate $\beta g_{\rm hex}$. This is not only a property of an
idealized code: with smooth wavepackets and sufficiently frequent ideal
correction, the rate is approached under the uniform random forces of the previous
section, whatever their strength (Supplementary Section~\Sref{sec:correction}). Polygon
arrangements of this kind have been studied before under the name of quantum
spherical codes\cite{JainEtAl2024}.

How good is the hexagon? Among all codes whose positions lie in a plane, however
tilted and however much they store, $295/32$ is the best possible number in
Eq.~\eqref{eq:protected}. Over all codes in three dimensions the number cannot
exceed $10.18$, so at large separation the hexagon delivers at least 90\% of
what any code could. Codes confined to the line between the masses, by
contrast, cannot do better than $5/4$ in place of $295/32$, a value reached by a
code from quantum sensing\cite{LaydenEtAl2019}: the hexagon beats every one of
them by a factor of $59/8\approx7.4$.

This last number has a pleasing explanation (Fig.~\ref{fig:shield}d). Gravity
inspects the three-lobed shape of a code from every direction around the line
joining the masses, and the rate depends on the average of the square of what
it finds. A code on the line looks the same from all sides and offers a signal
of $1/4$. The hexagon, seen from an angle $\phi$ to its own plane, offers
$(1+3\cos^2\phi)/4$: four times as much face-on, and never less. Averaging the
square gives exactly $59/8$.

The statement ``within 10\%'' concerns large separations. In the running example, where $d=3.6\,r$, the hexagon
runs at $2.2\%$ of the noiseless speed limit, while the general limit allows
correcting codes up to $15.5\%$: there the hexagon is within a factor of seven
of the best possible code, and the gap narrows to 10\% only as the masses move
apart. Exact correction with discarded error records is one strategy
among many. Above the noise threshold it achieves a rate falling as $1/d^7$,
while Eq.~\eqref{eq:noisy} forbids only rates above $1/d^4$. Whether some other
strategy can do better in between is open.

\subsection*{The bill, in seconds}

Entanglement has to be demonstrated, not just made, and the speed limits also
say how long a demonstration must take. Suppose an experiment ends with a test.
The test may raise a false alarm, reporting entanglement when there is none,
with probability at most $\alpha$; and it detects entanglement with probability
$p$ when the experiment works as intended. We prove that the time for which the
masses interact, averaged over many repetitions of the experiment, must satisfy
\begin{equation}
 \overline T\ \ge\ \frac{1}{C}\left[p\log_2\frac p\alpha+(1-p)\log_2\frac{1-p}{1-\alpha}\right],
 \label{eq:time}
\end{equation}
where $C$ is the speed limit that applies ($C_G$ without a shield). The bracket
counts the bits of evidence that the test provides; for 1\% false alarms and
95\% detection it is 6.0 bits. The bound allows any measurement, feedback and
stopping rule. Without a shield it is essentially the truth: as $\alpha$
becomes small, the least possible time approaches $(p/C_G)\log_2(1/\alpha)$, and
we give protocols that reach it.

Table~\ref{tab:bill} lists what this means for the running example. Without a
shield, no experiment can make its case in less than 100 seconds of
interaction, noise or no noise, and an explicit protocol with smooth finite-energy
wavepackets and purely local measurements does it in 300 seconds
(Supplementary Section~\Sref{sec:evidence}). With a shield, every fixed exactly correcting code with discarded error records needs
at least 648 seconds (Fig.~\ref{fig:bill}a). The hexagon's coupling needs 1,119
seconds to turn two unentangled encoded bits into one Bell pair, where two
unshielded positions need 25 seconds.

\begin{table*}[t]
\centering\small
\begin{tabular}{@{}p{6.2cm}p{4.2cm}p{5.0cm}@{}}
\toprule
 & No shield & Exact error correction\\
\midrule
Limit: least interaction time, for any experiment & 100 s & 648 s (any code)\\[3pt]
An explicit protocol & 300 s (twelve runs of 25 s; no noise) & 19.6 h (hexagon; 63 runs; noise that resolves the positions)\\[3pt]
Time for the coupling to make one Bell pair & 25 s & 1,119 s (hexagon)\\
\bottomrule
\end{tabular}
\caption{\textbf{The bill for the running example.} Two masses of
$10^{-14}$\,kg, $r=125\,\mu$m, $d=450\,\mu$m. The first two rows refer to a
demonstration of entanglement with 1\% false alarms and 95\% detection (96\% for
the 19.6-hour protocol).}
\label{tab:bill}
\end{table*}

Two rules of thumb summarize the bill. In time, shielding against random forces
with the hexagon costs a factor that grows as $(d/r)^4$: 45 in the running
example, about 2,200 at $d=10\,r$. In mass, since every rate grows as the
product of the masses, the lost speed is bought back by making both masses
heavier by a factor that grows as $(d/r)^2$: 6.7 in the running example, 47 at
$d=10\,r$.

\subsection*{When the shield is imperfect}

No real shield is perfect, infinitely fast, or faced only with uniform forces.
Three further results show how the picture bends (Methods and Supplementary
Section~\Sref{sec:control}).

\begin{figure*}[t]
\centering
\includegraphics[width=\textwidth]{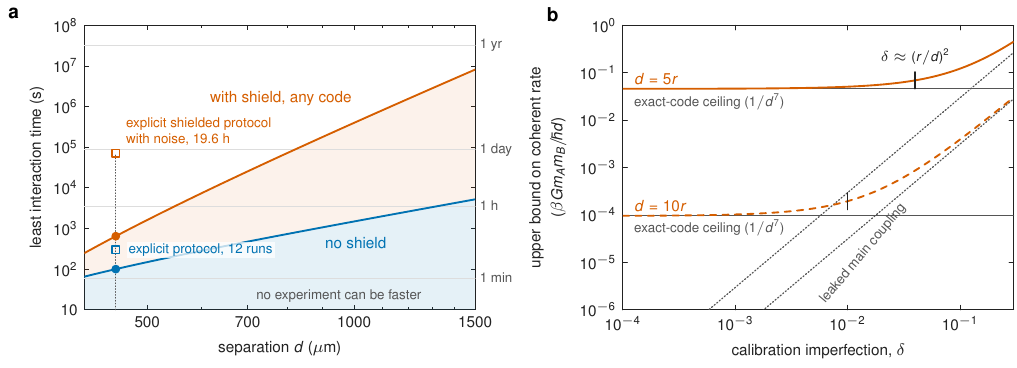}
\caption{\textbf{The bill.} \textbf{a}, Least interaction time needed to
demonstrate entanglement with 1\% false alarms and 95\% detection, for two masses
of $10^{-14}$\,kg in regions of radius $125\,\mu$m, against their separation.
Lines are limits from Eq.~\eqref{eq:time}: no experiment of the stated kind can
be faster. Filled circles: the limits at $450\,\mu$m. Open squares: explicit
protocols, without a shield and without noise (300 s), and with a shield in the
presence of noise (19.6 h). \textbf{b}, An imperfect shield. Horizontal axis:
the calibrated correction imperfection $\delta$ (Methods). Vertical axis: an
upper bound on the coherent entangling rate retained by the code, in units of
$\beta g_0$. For a good shield the bound approaches the flat $1/d^7$ ceiling.
Beyond $\delta\approx(r/d)^2$, its allowance for gravity's main coupling
(dotted lines) takes over.}
\label{fig:bill}
\end{figure*}

First, a leaky shield leaks in both directions. If a code hides the average
position and the spread only approximately, gravity's main coupling leaks back
in (Fig.~\ref{fig:bill}b). The $1/d^7$ law describes
shields whose calibrated correction imperfection $\delta$ (defined in Methods)
is below about $(r/d)^2$. A leakier code keeps some of the
stronger $1/d^3$ coupling, but in the codes we construct it also lets the noise
back in.

Second, the kind of noise matters. A uniform random force is removed completely
by the hexagon code. Noise that can tell the six positions apart, such as
scattering events that reveal where the mass is, is not: it leaves a residual
loss of coherence that no speed of correction removes, and for the continuously corrected hexagon, two equal logical superpositions
become entangled only if the coupling exceeds half the residual rate on each
mass, when those rates are equal.
The residual loss belongs to the hexagon, not to the noise: we construct another
family of codes that approaches a perfect Bell pair for any finite strength of
the Gaussian localization noise considered here, at fixed geometry and bounded average kinetic energy, at the price
of longer interaction times and ever finer control.

Third, correcting at a finite rate is a balancing act. Wait too long between
corrections and errors pile up; correct too often and the imperfections of the
corrections pile up instead. A fully worked example, with noise that resolves
the positions, shows the scale of the task. About 28,000 corrections, one every
0.04 seconds and each with a complete-cycle error below one part in a million, yield a Bell
pair of fidelity at least 0.817; a value above 0.5 certifies entanglement, and
1 is a perfect pair. Repeating this independently 63 times, 19.6 hours of interaction in all, demonstrates
entanglement with 1\% false alarms and 96\% detection, provided that the final
readout is ideal and each run is prepared within trace distance 0.1\% of its target. These
figures are requirements on a future apparatus, not a claim that one exists.

\section*{Discussion}

We set out with three questions and found three speed limits
(Table~\ref{tab:limits}). Left undisturbed, gravity entangles two masses at a
rate that is now known exactly; it falls as the cube of their separation, and
two positions per mass are all it takes. Random forces above a threshold fixed
by the masses and their separation close gravity's main channel, and whatever
is done about it the rate then falls at least as the fourth power. Shielding the
masses by exact, continuously applied error correction removes the noise but
hides the masses from each other, so that the rate falls as the seventh power,
and six positions on a hexagon deliver, at large separation, nearly all that
any such shield can.

\begin{table*}[t]
\centering\small
\begin{tabular}{@{}p{4.6cm}p{5.6cm}p{5.0cm}@{}}
\toprule
Situation & Fastest possible entanglement & What achieves it\\
\midrule
No noise & exactly $\beta g_0\,2u^2/(1-4u^2)$; falls as $1/d^3$ & two positions per mass\\[3pt]
Random forces above the threshold, any strategy & at most $\beta g_0\,[\tfrac92u^3+\dots]$; falls at least as $1/d^4$ & open\\[3pt]
Exact error correction, any code & between $295/32$ and $10.18$ times $\beta g_0u^6$; falls as $1/d^7$ & six positions on a hexagon (lower value)\\
\bottomrule
\end{tabular}
\caption{\textbf{Three speed limits.} Rates are in Bell pairs per unit time;
$g_0=Gm_Am_B/\hbar d$ is the natural gravitational rate of the pair, $u=r/d$ the
ratio of region size to separation, and $\beta=1.9123$. The second and third
rows show the behaviour at large separation; the complete expressions are in
Methods.}
\label{tab:limits}
\end{table*}

One idea runs through all three. Gravity and random forces act on a mass
through its position, so that whatever takes position away from the noise takes
it away from gravity. Strong noise does this crudely, by scrambling the
position until the leading gravitational effect could as well be a classical
message. Error correction does it deliberately, by storing the quantum bit in
features of the mass that a kick cannot reach, and gravity can barely reach
them either. Seen this way, the limits are not peculiar to gravity. They use
only the inverse-distance form of the interaction, and so apply equally to
the electric force between two charges under the same model assumptions. Gravity is simply the case in which the
interaction is so weak that the price decides whether an experiment is
possible.

For those planning experiments, the results amount to a set of signposts. If
the random forces are below the threshold of Eq.~\eqref{eq:threshold}, no shield
is required in principle, and the target is the noiseless limit of
Eq.~\eqref{eq:capacity}, reduced by the noise. If they are above it, the
$1/d^3$ law is gone for good. Exact correction then offers a rate that is immune
to the noise but smaller by roughly $(r/d)^4$, to be won back with heavier
masses or with superpositions that are wider relative to their separation. The
best design also changes shape: without a shield, the useful superposition lies
along the line between the masses; with one, it spreads into a plane.

Our results rest on earlier ones. The imitation of the leading gravitational
effect by measurement and messaging is due to Kafri, Taylor and
Milburn\cite{Kafri2014}, and its noise condition in general form to Li, Miki and
Chen\cite{LiMikiChen2026}; we extend it from the leading effect to the complete
potential and to every strategy. The cancellation under exact correction and
the recoverability of entanglement under noise are due to Wang, Pedernales and
Plenio\cite{WangPedernalesPlenio2026}; we add the limits that hold for all codes
and the code that nearly reaches them. The codes themselves come from quantum
sensing and from quantum spherical codes\cite{LaydenEtAl2019,JainEtAl2024}, and
larger superpositions and other geometries have been explored
before\cite{TillyEtAl2021,NguyenBernards2020}.

The limits also have edges, and it is as well to mark them. They concern
Newton's potential between masses confined to two regions, with ideal local
control. The noise threshold assumes random forces that are uniform, memoryless
and independent for the two masses. The shielded limit assumes fixed codes that
correct exactly and continuously, with error records discarded. Within these
assumptions the limits are proved; beyond them they are reference points.
Noise with memory can change quantum-correlation dynamics and state-evolution
speed limits\cite{PaulsonEtAl2021}.

The most interesting open question lies between two of our limits. Above the
noise threshold, no strategy can beat $1/d^4$, and exact correction reaches
$1/d^7$. Strategies that only detect errors and discard the runs in which one
occurred, or that keep the error record, might do better; deciding this means
counting the failed attempts and following what happens to the runs that
survive. Also open are the exact number in front of the shielded limit, which
lies between $295/32$ and $10.18$, the capacity when the regions nearly touch,
and the effect of a common limit on energy or on the speed of control. Above
all, a physical way to carry out the correction on the motion of a mass, with
the accuracy that our examples demand, has yet to be invented. Gravity's pull
between two specks of matter is the faintest thing anyone has proposed to
entangle with. The limits found here say how quickly that can happen at best,
and what it costs to listen for it through the noise.

\section*{Methods}

\paragraph{Model and capacity.}
The two laboratories contain the masses, confined to balls of radius $r$ with
centre separation $d>2r$, and arbitrary auxiliary systems. The only coupling is
Eq.~\eqref{eq:potential}. Local Hamiltonians, including kinetic and confining
terms, local operations, measurements and classical communication are
unrestricted. The initial state is separable; any entangled resource that a
protocol uses must be generated by the interaction, and its time is counted. The
capacity is the supremum of achievable rates of Bell pairs per unit interaction
time with vanishing error\cite{BennettHarrowLeungSmolin2003}. Protocols with
continuous position are limits of finite protocols using wavepackets of finite
energy; no common bound on energy is imposed along the sequence. We use
$G=6.67430\times10^{-11}$ and $\hbar=1.054571817\times10^{-34}$ in SI units. In
the running example $g_0=0.1406\,{\rm s^{-1}}$ and $u=1/3.6$; the capacity and
the 100-second bound are evaluated with $r=125.001\,\mu$m, which leaves room for
packets of width 1\,nm centred at $125\,\mu$m.

\paragraph{Entanglement rate of a sum of products.}
All bounds rest on one inequality. Suppose that, apart from terms acting on one
laboratory only, the interaction is $H=\sum_\nu A_\nu\otimes B_\nu$ (or an
integral of this form) with $\sum_\nu A_\nu^2\le Q_A^2$ and
$\sum_\nu B_\nu^2\le Q_B^2$. For a pure state of the two laboratories with
Schmidt decomposition $\sum_i\sqrt{p_i}\,|u_i\rangle|v_i\rangle$, the
Cauchy--Schwarz inequality gives
$\sum_j|\langle u_iv_i|H|u_jv_j\rangle|\le Q_AQ_B$, and with
$2\sqrt{p_ip_j}\,|\log_2(p_i/p_j)|\le\beta(p_i+p_j)$ the entanglement entropy
$S$ obeys
\begin{equation}
 \begin{aligned}
 \Bigl|\frac{dS}{dt}\Bigr|&\le\frac{\beta\,Q_AQ_B}{\hbar},\\
 \beta&=\max_{0<q<1/2}2\sqrt{q(1-q)}\,\log_2\frac{1-q}{q}\\
 &=1.9123\ldots
 \end{aligned}
 \label{eq:rate}
\end{equation}
The bound does not depend on the dimensions of the systems or of their
memories\cite{Bravyi2007,VanAcoleyen2013,ChildsLeungVerstraeteVidal2003}. Local
operations with recorded outcomes do not increase the average entropy of the
resulting pure states, and producing $n$ Bell pairs requires entropy $n-o(n)$,
so the capacity is at most $\beta Q_AQ_B/\hbar$. Unbounded local Hamiltonians
and infinite-dimensional systems are treated in Supplementary Section~\Sref{sec:noiseless}.

\paragraph{Noiseless capacity.}
Reflect the axial coordinate of $B$, $\mathbf c=(\mathbf b_\perp,-b_z)$. For a
two-dimensional wavevector $\mathbf k$, define
$f_{\mathbf a}(\mathbf k)=e^{-(d/2-a_z)|\mathbf k|}\bigl(\cos\mathbf k\!\cdot\!\mathbf a_\perp,\ \sin\mathbf k\!\cdot\!\mathbf a_\perp\bigr)$.
With the measure $d^2\mathbf k/(2\pi|\mathbf k|)$,
$\langle f_{\mathbf a},f_{\mathbf c}\rangle=1/|d\,\mathbf e_z+\mathbf b-\mathbf a|$\cite{FrankLieb2010}.
Subtracting the common function $c_0=(f_{r\mathbf e_z}+f_{-r\mathbf e_z})/2$
changes only local terms. We show that
$\|f_{\mathbf a}-c_0\|^2\le\frac14\bigl[\frac1{d-2r}+\frac1{d+2r}-\frac2d\bigr]$
for every $\mathbf a$ in the ball when $d\ge(1+\sqrt5)r$. Equation~\eqref{eq:rate}
then gives $C_G\le\beta\kappa_G$ with
$\kappa_G=\frac{Gm_Am_B}{4\hbar}\bigl[\frac1{d-2r}+\frac1{d+2r}-\frac2d\bigr]
=g_0\,2u^2/(1-4u^2)$. Two packets at $\pm r\mathbf e_z$ on each mass realize the
interaction $\hbar\kappa_G\,Z\otimes Z$ between their labels, whose capacity is
$\beta\kappa_G$\cite{ChildsLeungVerstraeteVidal2003,ChildsLeungVidal2004,BennettHarrowLeungSmolin2003}.
The full proof, the approximation by packets of finite energy, the strict loss
of capacity under a history-wise kinetic-energy budget for equal masses on a
line, and the near-contact advantage of three
positions are in Supplementary Sections~\Sref{sec:noiseless} and \Sref{sec:extensions}.

\paragraph{Noise threshold and the bound for every strategy.}
The noise changes the state $\rho$ of the two masses at the rate
\[
 -\sum_{i=x,y,z}\bigl(\Lambda_A[a_i,[a_i,\rho]]+\Lambda_B[b_i,[b_i,\rho]]\bigr),
\]
with the environment inaccessible. The leading term of the potential is
\[V_{\rm dip}=(Gm_Am_B/d^3)(2a_zb_z-a_xb_x-a_yb_y).\] Consider one component, with
operators $X$ on $A$ and $Y$ on $B$, coupling $\hbar g\,X\otimes Y$ and
eigenvalues $x,\eta$. Over a time $t$ the coupling and the noise multiply the
matrix element between $(x,\eta)$ and $(x',\eta')$ by
$\exp\{t[-ig(x\eta-x'\eta')-\Lambda_A(x-x')^2-\Lambda_B(\eta-\eta')^2]\}$. Let $A$
perform the measurement with operators $M_y\propto\exp[-kt(X-y)^2]$, announce
the outcome $y$, and let $B$ apply $\exp(-igty\,Y)$. Averaged over $y$, this
multiplies the same element by
$\exp\{t[-\frac k2(x-x')^2-\frac{ig}{2}(x+x')(\eta-\eta')-\frac{g^2}{8k}(\eta-\eta')^2]\}$.
Composing it with the mirror-image procedure of strength $k'$ reproduces the
phase exactly, with dampings $\frac k2+\frac{g^2}{8k'}$ and
$\frac{k'}2+\frac{g^2}{8k}$. With $kk'=g^2/4$ these are $k$ and $k'$, which can
be kept below $\Lambda_A$ and $\Lambda_B$ if and only if
$\Lambda_A\Lambda_B\ge g^2/4$; what is left is local
dephasing\cite{Kafri2014,WisemanMilburn1994,LiMikiChen2026}. The strongest
requirement, for the $z$ component, is Eq.~\eqref{eq:threshold}.

During a period in which only the potential and the noise act, all position
operators commute, so the evolution factors into this locally imitable channel,
local phases and the unitary generated by the remainder $W=V-V_{\rm dip}$.
Equation~\eqref{eq:rate} bounds the entropy produced by $W$; local operations
with recorded outcomes, including those that imitate the leading term, do not
increase it on average. The terms of $W$ linear in $\mathbf a$ are
$-(Gm_Am_B/d^2)\,\mathbf a\cdot\mathbf q(\mathbf b)$, where $\mathbf q$ is the
field $d^2(d\,\mathbf e_z+\mathbf b)/|d\,\mathbf e_z+\mathbf b|^3$ without its
constant and linear parts; a constant vector may be subtracted from $\mathbf q$
at the cost of a local term. With
$q_*(u)=\min_{\mathbf c}\max_{|\mathbf b|\le r}|\mathbf q(\mathbf b)-\mathbf c|$,
the same for the terms linear in $\mathbf b$, and the series $S_2$ defined below
for the rest, $B(u)=2u\,q_*(u)+S_2(u)$ with $q_*=\frac94u^2+O(u^3)$. Summing
instead all terms of the expansion~\eqref{eq:expansion} except the leading one
gives the closed form $R(u)=2u^3(3-2u)/[(1-2u)(1-u)]$, and the bound proved is
$C_{\rm noise}\le\beta g_0\min\{B(u),R(u)\}$; $R$ alone gives $32/45$ of the
noiseless capacity at $d=6r$. Evaluating the finite-separation bound gives upper values of
$B(u)/[2u^2/(1-4u^2)]$ of
$0.98$, $0.70$, $0.54$, $0.37$, $0.28$ and $0.13$ for $d/r=4,5,6,8,10$ and $20$.
The conversion of the entropy bound into a bound on Bell pairs delivered to
finite registers, the inclusion of the motion of the masses during the
interaction, and the general statement, in which a noise ratio
$\lambda=\sqrt{\Lambda_A\Lambda_B}/\Lambda_c<1$ adds $2(1-\lambda)u^2$ to
$\min\{B,R\}$, are proved in Supplementary Section~\Sref{sec:noise}; the grey curve of
Fig.~\ref{fig:noise}b shows this bound. Below the threshold, two narrow packets at
$\pm s\,\mathbf e_z$ interact with strength
$\kappa(s)\ge2Gm_Am_Bs^2/(\hbar d^3)$ and each loses coherence at the rate
$\gamma=4\Lambda s^2$; their state is entangled at some time if and only if
$\gamma<2\kappa$, and entangled states of two quantum bits can be
distilled\cite{HorodeckiDistill1997}. The blue curve of Fig.~\ref{fig:noise}b is
the rate of the hashing protocol\cite{DevetakWinter2005} for this state,
maximized over the interaction time; for $d=10\,r$ it reaches zero at about
$0.2\,\Lambda_c$, although the state itself remains entangled at some time for
all $\Lambda<\Lambda_c/[1-4(s/d)^2]$.

\paragraph{Shielded codes.}
Let $\zeta_\phi(\mathbf a)=a_z+i(a_x\cos\phi+a_y\sin\phi)$, so that
$|\zeta_\phi|\le r$, and let $\langle\cdot\rangle_\phi$ denote the average over
$\phi$. For $d>2r$,
\begin{equation}
 \frac{1}{|d\,\mathbf e_z+\mathbf b-\mathbf a|}
 =\frac1d\sum_{j,k\ge0}\binom{j+k}{j}
 \frac{\bigl\langle\zeta_\phi(\mathbf a)^j\,[-\zeta_\phi(\mathbf b)]^k\bigr\rangle_\phi}{d^{\,j+k}},
 \label{eq:expansion}
\end{equation}
with absolute convergence. The function $\zeta_\phi^j$ is a harmonic polynomial
of degree $j$, a combination of multipole moments of order $j$. Let $P$ project
onto a code, which may include local auxiliary systems. The code looks alike
below order $M$ if $P\,h\,P$ is a multiple of $P$ for every harmonic polynomial
$h$ of degree less than $M$. Detection of force errors,
$P\,q_i\,P\propto P$, gives $M=2$; exact correction of force errors, that is,
$P\,q_i\,P\propto P$ and $P\,q_iq_j\,P\propto P$\cite{KnillLaflamme1997}, gives
$M=3$; correction of polynomial errors up to degree $N$ gives $M=2N+1$. Correcting a
set of errors amounts to detecting all of their products, which is why each
corrected order costs four powers of $r/d$ and each detected order two. With
arbitrarily frequent recovery and discarded error records, the encoded systems
evolve under $(P_A\otimes P_B)V(P_A\otimes P_B)$ up to local
terms\cite{WangPedernalesPlenio2026}. Every term of Eq.~\eqref{eq:expansion}
with $j<M$ or $k<M$ is then local, and each remaining term is the real part of a
product of two contractions, which Eq.~\eqref{eq:rate} bounds by $\beta$ per
unit coefficient. Hence the capacity of the encoded interaction is at most
$\beta g_0S_M(u)$, with $S_M(u)=\sum_{j,k\ge M}\binom{j+k}{j}u^{j+k}$.

For the matching codes, take corners $r(\sin\theta_n,0,\cos\theta_n)$,
$\theta_n=n\pi/M$, encoded states the equal superpositions over even and over
odd $n$, and the mirror image at the partner. On a corner,
$\zeta_\phi=r(Ae^{i\theta}+Be^{-i\theta})$ with $A=(1+\cos\phi)/2$ and
$B=(1-\cos\phi)/2$. Averaging $\zeta_\phi^j$ over $M$ equally spaced corners
keeps only terms $e^{im\theta}$ with $m$ a multiple of $M$; for $j<M$ these agree
for the two encoded states, and for $j=M$ they differ in sign and give the
signal $r^M(A^M+B^M)$. The $(M,M)$ term of Eq.~\eqref{eq:expansion} is therefore
an interaction $Z\otimes Z$ of strength
$g_0\binom{2M}{M}\langle(A^M+B^M)^2\rangle_\phi\,u^{2M}=g_0K_Mu^{2M}$, with
$K_M=2\binom{2M}{M}\bigl[\binom{4M}{2M}+\binom{2M}{M}\bigr]/16^M$, and all
further terms add with the same sign. For $M=3$ the signal is
$A^3+B^3=(1+3\cos^2\phi)/4$ and $K_3=295/32$; a code on the axis has the signal
$1/4$ for every $\phi$, which gives $5/4$; $K_2=57/16$ and $K_1=2$. The ceiling
$10.18$, the optimality of $295/32$ among planar codes and among interactions
that are a single product, the bound for codes on the axis, the realization with
smooth wavepackets under white force noise, and the hierarchy for all $M$ are
proved in Supplementary Sections~\Sref{sec:correction} and \Sref{sec:geometry}. At $d=3.6\,r$ every exactly correcting code
has a rate of at most $0.1548\,C_G$ (Supplementary Section~\Sref{sec:correction}); the orange band
of Fig.~\ref{fig:shield}c uses a sharper series from the same section. Evaluating the full potential at $d=3.6\,r$ gives a hexagon coupling 18\%
above the leading term of Eq.~\eqref{eq:hex}.

\paragraph{Time needed for a test.}
Resolve a protocol into pure branches $\psi_h$ with probabilities $w_h$.
Equation~\eqref{eq:rate} gives $\sum_hw_hS(\psi_h)\le C\,\overline T$. Removing
the coherence of each branch in its Schmidt basis gives a separable state at
relative entropy $S(\psi_h)$ from it. Joint convexity of the relative entropy,
and the fact that a measurement cannot increase it, give Eq.~\eqref{eq:time} for
any test whose false-alarm probability is bounded for all separable states of
the complete laboratories. For the noiseless capacity the bound is attained to
leading order by generating $\log_2(1/\alpha)+O(1)$ Bell pairs with probability
close to $p$ and verifying them with local
measurements\cite{ZhuHayashi2019,PallisterLindenMontanaro2018} (Supplementary
Section~\Sref{sec:evidence}). For exactly correcting codes the same argument with the bound
$\beta g_0S_3(u)$ gives the time quoted in the text (Supplementary Sections~\Sref{sec:correction} and \Sref{sec:evidence}).

\paragraph{Imperfect shields and worked examples.}
For each fixed code, apply the position-error calibration channel defined in
Supplementary Section~\Sref{sec:approximate-position-correction}. Let $\delta$
bound the half-diamond distance of its complementary channel from a constant
channel, after encoding. Thus $\delta$ measures how much the calibration
environment can learn about the encoded state. For two codes with defect at
most $\delta$, the coherent capacity of the compressed interaction is at most
\begin{equation}
 \beta\,g_0\left[\frac{16\,\delta^2u^2}{(1-2u)^3}+\frac{16\sqrt2\,\delta\,u^4}{(1-2u)^5}
 +\frac{20\,u^6}{(1-2u)^7}\right],
 \label{eq:approx}
\end{equation}
which is plotted in Fig.~\ref{fig:bill}b. This bound, the codes that attain its first and last terms, the
residual loss of coherence under noise that resolves the positions, the
finite-rate correction example (Gaussian localization noise, 27,986 intervals,
complete-cycle error at most $10^{-6}$, Bell fidelity at least 0.817, 51 passes
in 63 runs) and the twelve-run unshielded protocol are derived in Supplementary
Sections~\Sref{sec:evidence} and \Sref{sec:control}. A recovery of the calibration
channel with worst-case entanglement infidelity $\varepsilon$ has
$\delta\le\sqrt{\varepsilon}$; this does not identify $\delta$ with a physical
cycle-failure probability. In the worked example, coherences separated by $L$
decay at rate $\Gamma(L)=\Lambda_0[1-e^{-L^2/(4r_c^2)}]$, with
$r_c=1.397542486$\,mm and $\Gamma(r)=0.025$\,s$^{-1}$. The complete joint-cycle
error is a half-diamond distance, including drift, reset, leakage and recorded
flags. The Bell fidelity and detection probability assume trusted ideal logical
readout, independent target preparations and initialization within trace
distance $0.001$. All 63 runs are counted, and at least 51 passes are required.

\section*{Acknowledgments}
E.P. acknowledges support by the project skQCI (Slovak Quantum Communication Infrastructure), co-funded by the European Union under the Digital Europe Programme, grant agreement No. 101091548, as part of the EuroQCI initiative.

\ifstandalonesupp\else\clearpage\onecolumngrid\fontsize{11}{14}\selectfont\fi
\setcounter{section}{0}
\setcounter{equation}{0}
\setcounter{figure}{0}
\setcounter{table}{0}
\setcounter{theorem}{0}
\renewcommand{\thesection}{S\arabic{section}}
\renewcommand{\thesubsection}{\thesection.\arabic{subsection}}
\renewcommand{\thesubsubsection}{\thesubsection.\arabic{subsubsection}}
\renewcommand{\theequation}{S\arabic{equation}}
\renewcommand{\thefigure}{S\arabic{figure}}
\renewcommand{\thetable}{S\arabic{table}}
\renewcommand{\thetheorem}{S\arabic{theorem}}
\renewcommand{\theHequation}{supp.\arabic{equation}}
\renewcommand{\theHfigure}{supp.\arabic{figure}}
\renewcommand{\theHtable}{supp.\arabic{table}}
\renewcommand{\theHsection}{supp.\arabic{section}}
\setcounter{tocdepth}{1}
\ifstandalonesupp\else\section*{Supplementary Information}\fi
{\small\begin{tabular}{@{}p{0.87\textwidth}r@{}}
\ref{sec:guide}\quad How to read the Supplement & \pageref{sec:guide}\\[3pt]
\ref{sec:noiseless}\quad The noiseless speed limit & \pageref{sec:noiseless}\\[3pt]
\ref{sec:noise}\quad The noise threshold and the full-potential bound & \pageref{sec:noise}\\[3pt]
\ref{sec:correction}\quad What error detection and correction remove & \pageref{sec:correction}\\[3pt]
\ref{sec:geometry}\quad Optimizing the protected spatial encoding & \pageref{sec:geometry}\\[3pt]
\ref{sec:evidence}\quad The time and finite data needed for evidence & \pageref{sec:evidence}\\[3pt]
\ref{sec:control}\quad Finite recovery, localization and imperfect correction & \pageref{sec:control}\\[3pt]
\ref{sec:extensions}\quad Further geometry, energy and confinement results & \pageref{sec:extensions}\\[3pt]
\end{tabular}}

\clearpage
\section{How to read the Supplement}
\label{sec:guide}
We follow the questions in the article. Section~\ref{sec:noiseless} proves the
noiseless speed limit; Section~\ref{sec:noise} adds force noise and bounds what
any local strategy can retain. Sections~\ref{sec:correction} and
\ref{sec:geometry} explain what an exactly correcting code hides from gravity
and why a hexagon retains nearly the largest possible rate. We then turn rates
into finite records in Section~\ref{sec:evidence}, and examine finite recovery,
localization noise and imperfect correction in Section~\ref{sec:control}.
Section~\ref{sec:extensions} supplies the extensions to other geometries,
energy constraints and finite confinement.

\subsection{The physical model and units}
Each laboratory contains one mass and any local auxiliary systems it needs.
In the three-dimensional model the mass positions are $\mathbf a$ and
$\mathbf b$, measured from the centres of balls of radius $r$. The centres
are separated by $d>2r$ along the unit vector $\mathbf e_z$. The only interaction
between laboratories is
\[
 V(\mathbf a,\mathbf b)=-\frac{Gm_Am_B}{|d\mathbf e_z+\mathbf b-\mathbf a|}.
\]
Here $G$ is Newton's constant and $m_A,m_B$ are the masses. The condition
$d>2r$ keeps the denominator positive throughout the allowed position domains.
Terms that act on only one laboratory can change its local state, but cannot
create entanglement between laboratories. We therefore remove such terms when
bounding the entangling rate, and state explicitly whenever a later estimate
must control them as well.

Two scales recur:
\[
 u=\frac rd<\frac12,\qquad g_0=\frac{Gm_Am_B}{\hbar d}.
\]
The ratio $u$ is dimensionless. The rate $g_0$ has units of inverse seconds.
Consequently $g_0u^2$, $g_0u^3$ and $g_0u^6$ scale respectively as $d^{-3}$,
$d^{-4}$ and $d^{-7}$ when the masses and region sizes are fixed. We sometimes
write $\mathcal G=Gm_Am_B$ to shorten formulas. In the collinear model the
position coordinates are $x,y\in[-r,r]$ and the potential is
$-\mathcal G/(d+y-x)$. A result proved only for that model is identified as
collinear.

\subsection{What a rate, an error and a time mean}
Entanglement is measured in bits: one ideal Bell pair contains one bit of
entanglement. The capacity $C_G$ of the article is also denoted
$C_E^{(3)}$ in the three-dimensional proof. It is the supremum of asymptotic
Bell-pair rates, with vanishing output error, per unit time during which the
interaction acts. The numerical constant $\beta=1.9123\ldots$ converts the
strength $\kappa$ of a two-qubit interaction $\hbar\kappa Z\otimes Z$ into its
capacity $\beta\kappa$; Section~\ref{sec:noiseless} defines the optimization
that gives $\beta$. A Bell-pair preparation time, an asymptotic capacity and
the time needed to reject a classical alternative are distinct quantities.
Section~\ref{sec:evidence} connects them.

Local operations and classical communication, abbreviated LOCC, are free in
the capacity model. The laboratories begin in a separable state. An entangled
resource used later must first be generated, and that interaction time is
counted. Local control is otherwise unrestricted unless a result explicitly
imposes an energy, cycle-error or duration constraint. In particular, a limit
of finite-energy packet protocols does not impose one common energy bound on
the entire sequence.

Logarithms denoted $\log_2$ give information in bits; $\ln$ denotes the natural
logarithm. Later channel-testing sections sometimes use natural relative
entropy and state that convention locally. A factor $\ln2$ converts bits to
natural-log units. Trace distance means $\tfrac12\|\rho-\sigma\|_1$. The
half-diamond distance between channels includes arbitrary reference systems;
it bounds the change of any outcome probability when one channel replaces the
other. An initialization error, a calibration-channel defect and a physical
recovery-cycle error concern different maps and are defined separately where
used.

Interaction times add over all attempts, including failures. Thus the
19.6-hour example is aggregate interaction time across 63 trials. A mean-time
bound allows a random stopping rule and averages over the resulting records;
a hard-time bound limits every record. Neither quantity includes a laboratory
clock time for local operations unless that time is explicitly charged.

\subsection{Which kind of noise is being corrected?}
The white-force model describes independent, spatially uniform random forces
on the masses. A coherence between positions separated by $L$ decays at rate
$\Lambda L^2$, where $\Lambda$ has units ${\rm m^{-2}\,s^{-1}}$. With a
two-sided force spectral density $S_F$ per Cartesian component,
$\Lambda=S_F/(2\hbar^2)$. Its leading error operators are the three position
coordinates. Exact correction requires the Knill--Laflamme conditions for
these coordinates and their pairwise products, as derived in
Section~\ref{sec:correction}.

A finite-correlation-length localization process can resolve the individual
positions within a code and is a different noise model. Its coherence-decay
kernel $\Gamma(L)$ is introduced in Section~\ref{sec:control}. The hexagon
removes the leading uniform-force term but need not remove this whole kernel.
The fixed-mean-energy recovery construction there concerns the specified
Gaussian localization kernel with finite rate and positive correlation length.

The protected-capacity bounds concern fixed codes, arbitrarily frequent ideal
local recovery, discarded recovery records and the retained logical systems.
Their effective Hamiltonian is the potential compressed onto the codes. The
all-strategy noisy bound in Section~\ref{sec:noise} has a wider scope: it also
allows retained quantum recovery systems. Keeping these two settings distinct
explains why the noisy upper rate can fall as $d^{-4}$ while the best fixed,
exactly corrected rate falls as $d^{-7}$.

\section{The noiseless speed limit}
\label{sec:noiseless}
\label{sec:si-noiseless}

We first prove the noiseless capacity for two sufficiently separated
three-dimensional position domains. We bound the entropy production of any
state, including local memories, and then approach that bound with two axial
packets per mass. The geometric step is to represent the inverse distance as
an inner product and bound the radius occupied by these position features.
We subsequently give the simulation argument for collinear supports, where
the entire interaction can be replaced by an Ising interaction at a specified
rate.

\subsection{Three-dimensional position domains}
\label{sec:newton-three-d}

Let each mass occupy a three-dimensional ball of radius $r$, with
centre separation $d>2r$. We denote their displacements from the respective
centres by $\mathbf a$ and $\mathbf b$, and the unit vector from Alice's
centre to Bob's by $\hat z$. Their only quantum interaction is the
Newtonian potential energy
\[
 V(\mathbf a,\mathbf b)=-\frac{\mathcal G}
 {|d\hat z+\mathbf b-\mathbf a|},\qquad
 \mathcal G=Gm_A m_B,\quad |\mathbf a|,|\mathbf b|\le r.
\]
Local confinement is maintained throughout. We retain the ideal-control
capacity convention: local operations, noiseless memories and classical
communication are free, seed entanglement is generated and counted, and
finite protocols are closed in output error. Packet energies are finite
in each approximation, without a common energy or bandwidth bound.
For extended spherical bodies the minimum centre separation $d-2r$ must
also exceed the sum of their physical radii.

Here $G$ is Newton's gravitational constant and $m_A,m_B$ are the masses.
Put $\beta=\max_{0<q<1/2}2\sqrt{q(1-q)}\log_2[(1-q)/q]
=1.91227328895\ldots$, the established Ising entropy-rate constant,
including arbitrary local ancillas\cite{ChildsLeungVerstraeteVidal2003,ChildsLeungVidal2004}.
The variable $q$ is a two-level Schmidt probability. The same maximization
will bound each pair of Schmidt coefficients of an arbitrary state.

\begin{theorem}[Capacity for sufficiently separated equal balls]
\label{thm:newton-three-d}
If $d\ge(1+\sqrt5)r$, then the ideal asymptotic entanglement-generation
capacity per aggregate interaction time is
\begin{equation}
 C_E^{(3)}=\frac{\beta\mathcal G Q^2}{\hbar},\qquad
 Q^2=\frac14\left[\frac1{d-2r}+\frac1{d+2r}-\frac2d\right].
 \label{eq:three-d-capacity}
\end{equation}
Axial two-position encodings approach this supremum. No larger spatial
encoding or local ancillary memory increases it within the stated class.
The separation condition is sufficient; it is not a proved transition
point. This theorem determines entanglement capacity, not reversible
simulation of the entire three-dimensional interaction.
\end{theorem}

\subsection{A real Coulomb feature representation}

To bound every spatial encoding at once, we assign a vector $f_{\mathbf a}$
to each allowed position. We choose the vectors so that their inner products
reproduce the inverse distance. This converts the spatial optimization into
a bound on vector norms.
Reflect Bob's longitudinal coordinate and write
$\mathbf c=(\mathbf b_\perp,-b_z)$, where $\perp$ denotes the two
coordinates perpendicular to the separation axis. On the real Hilbert space with two
components and measure $d\mu(\mathbf k)=d^2\mathbf k/(2\pi|\mathbf k|)$,
where $\mathbf k$ is a transverse wave vector, define
\[
 f_{\mathbf a}(\mathbf k)=e^{-(d/2-a_z)|\mathbf k|}
 \bigl(\cos(\mathbf k\cdot\mathbf a_\perp),
       \sin(\mathbf k\cdot\mathbf a_\perp)\bigr).
\]
The angular integral over $\mathbf k$ produces the zeroth-order Bessel
function $J_0$. For $h>0$ and $u\ge0$, the polar-coordinate identity
$\int_0^\infty e^{-hz}J_0(uz)\,dz=(h^2+u^2)^{-1/2}$ gives
\begin{equation}
 \langle f_{\mathbf a},f_{\mathbf c}\rangle
 =\frac1{\sqrt{(d-a_z-c_z)^2+
                   |\mathbf a_\perp-\mathbf c_\perp|^2}}.
 \label{eq:three-d-features}
\end{equation}
Here $d-a_z-c_z>0$ by $d>2r$. This is a direct Fourier--Laplace identity;
reflection positivity is an established representation principle.
For the Coulomb kernel, see Frank and Lieb, Lemma~2.1 and
Eq.~(2.2)\cite{FrankLieb2010}; we have written its real Fourier form
explicitly to fix the normalization.

The two axial endpoints suggest the centre
$c_0=(f_{r\hat z}+f_{-r\hat z})/2$. Their squared distance from this
centre is $Q^2$ in Eq.~\eqref{eq:three-d-capacity}. We show that every
allowed position lies within that same feature-space radius:
\begin{equation}
 \|f_{\mathbf a}-c_0\|^2\le Q^2\qquad(|\mathbf a|\le r).
 \label{eq:three-d-feature-ball}
\end{equation}
Since $\|f_{\mathbf a}\|^2=(d-2a_z)^{-1}$ and
$\langle f_{r\hat z},f_{-r\hat z}\rangle=d^{-1}$, this is equivalent to
\begin{equation}
 \frac1{d-2a_z}+\frac1d\le
 \frac1{\sqrt{(d-a_z-r)^2+\rho^2}}+
 \frac1{\sqrt{(d-a_z+r)^2+\rho^2}},
 \label{eq:three-d-geometry}
\end{equation}
where $\rho=|\mathbf a_\perp|$. At fixed $a_z$ the right side decreases
with $\rho$, so it suffices to put $\rho^2=r^2-a_z^2$.

We now prove this scalar inequality on the boundary of the physical ball.
Measure lengths in units of $r$ and set $D=d/r$, $t=a_z/r$,
$P=D(D-2t)$ and $\zeta=\sqrt{1+4(1-t^2)/P^2}$. Both sides of
Eq.~\eqref{eq:three-d-geometry} are positive. Squaring and simplifying
reduces it to
\[
 (\zeta^2-1)B(P,\zeta)\ge0,\qquad
 B(P,\zeta)=P^2\zeta^2-2P-4-\frac{2P\zeta}{\zeta+1}.
\]
The two squared denominators on the right are
$P+2\pm2(D-t)$, whose product is $P^2\zeta^2$; this supplies the stated
reduction without any sign ambiguity. Now $\zeta\ge1$ and
$P\ge D(D-2)\ge4$. Moreover
\[
 \partial_\zeta B=2P^2\zeta-\frac{2P}{(\zeta+1)^2}>0,
 \qquad B(P,1)=(P-4)(P+1)\ge0.
\]
This proves Eq.~\eqref{eq:three-d-feature-ball}, including the boundary
$d=(1+\sqrt5)r$.

\subsection{An ancilla-assisted entropy-rate bound}

We use the feature radius to bound matrix elements of the interaction in
any Schmidt basis. Such a bound controls how fast the interaction can change
the Schmidt probabilities, and therefore how fast it can generate
entanglement.
Subtracting $c_0$ from both features changes the potential only by local
Hamiltonians and a scalar. Write the centered interaction as
\[
 H_{\rm nl}=-\mathcal G\int A_\lambda\otimes B_\lambda\,d\mu(\lambda),
 \qquad
 \int A_\lambda^2d\mu\le Q^2I,\quad
 \int B_\lambda^2d\mu\le Q^2I.
\]
Here $A_\lambda$ and $B_\lambda$ multiply a position wavefunction by the
corresponding centred feature component. The index $\lambda$ includes the
transverse wave vector and cosine/sine component, and the factors are real
self-adjoint multiplication operators. Local interaction pictures preserve
the two operator inequalities. They also hold after tensoring arbitrary
local ancillary identities.

For any Schmidt vectors $|u_i\rangle,|v_i\rangle$ set
$h_{ij}=\langle u_i v_i|H_{\rm nl}|u_j v_j\rangle$.
Triangle inequality, Cauchy--Schwarz over $(j,\lambda)$ and completeness
of local bases give the dimension-independent row bound
\begin{align}
 \sum_j|h_{ij}|
 &\le\mathcal G\left[\sum_j\int
  |\langle u_i|A_\lambda|u_j\rangle|^2d\mu\right]^{1/2}
 \left[\sum_j\int
  |\langle v_i|B_\lambda|v_j\rangle|^2d\mu\right]^{1/2}\nonumber\\
 &\le\mathcal G Q^2.
 \label{eq:three-d-row-bound}
\end{align}
An incomplete Schmidt family can be extended to local orthonormal bases.

For squared Schmidt weights $p_i$ with $\sum_i p_i=1$, let
$S=-\sum_i p_i\log_2p_i$ be the entropy of entanglement in bits.
Grouping its derivative into pairs yields
\[
 |\dot S|\le\frac2\hbar\sum_{i<j}\sqrt{p_ip_j}
 \left|\log_2\frac{p_i}{p_j}\right|\,|h_{ij}|.
\]
Each pair transfers probability between two Schmidt components. Applying
the definition of $\beta$ to their normalized probabilities gives the scalar inequality
$2\sqrt{p_ip_j}|\log_2(p_i/p_j)|\le\beta(p_i+p_j)$. Combining it with
Eq.~\eqref{eq:three-d-row-bound} yields
\[
 |\dot S|\le\frac\beta\hbar\sum_i p_i\sum_{j\ne i}|h_{ij}|
 \le\frac{\beta\mathcal G Q^2}{\hbar}.
\]
This is the row-sum entropy method underlying the known Ising/product
capacity proof\cite{ChildsLeungVerstraeteVidal2003}; the Coulomb feature
radius is the additional geometric input. Local instruments cannot
increase average pure-branch entropy, so the rate bounds arbitrary
admissible protocols, including local ancillas, noise and feedback.

The row estimate is uniform in dimension, but we must also connect it to
continuous positions without assuming that their entropy is continuous.
First approximate each ball by a finite
partition with representatives inside the ball. If both position errors
are bounded by $\epsilon_A,\epsilon_B$, then
$\|V-V_h\|\le\mathcal G(\epsilon_A+\epsilon_B)/(d-2r)^2$.
The finite label systems satisfy the feature inequality exactly; cell
multiplicities are local spectator systems. Prove the rate and finite-output
bound there and pass fixed protocol outputs under bounded-perturbation
convergence. This does not assume uniform continuity of entropy on the
full bosonic space. Normal protocol limits retain the same support and
resource-control assumptions as in the collinear result.

To approach the upper bound, it suffices to reproduce the endpoint Ising
interaction on a finite logical register. Take two smooth packets per laboratory
centered at $\pm(r-w)\hat z$, of radius $w$. Identical translated local
wells make their reference profiles stationary. The axial branch
Hamiltonian has Ising coefficient approaching $\mathcal G Q^2/\hbar$.
The difference from its branch-center potential is bounded on the well
rectangles by a quantity tending to zero with $w$; Duhamel's estimate on
the stationary code keeps the full local kinetic Hamiltonians. For each
finite encoded Ising block choose nonzero $w$ sufficiently small, then
take the coding limit. One consistent packet choice is the product of
three smooth one-dimensional bumps, each of halfwidth $w/\sqrt3$.
Its cube lies inside a radius-$w$ ball and its kinetic energy is finite,
scaling as $w^{-2}$. For the same nominal $w$, the initial kinetic energy
is nine times the one-dimensional value in the construction of
Section~\ref{sec:newtonian-packets}.
Seed entanglement
is generated and amortized using the established capacity construction.
This proves Theorem~\ref{thm:newton-three-d} as an operational supremum,
without a uniform energy, control-rate or memory bound.

\subsection{Ising simulation, entanglement capacity and finite-confidence time}
\label{sec:newtonian-capacity}

We now restrict the position coordinates to intervals
$x_A\in[-\ell_A,\ell_A]$ and $x_B\in[-\ell_B,\ell_B]$ on the
centre-to-centre axis, with $d_{\min}=d-\ell_A-\ell_B>0$.
Let $\kappa$ be the magnitude of the endpoint Ising frequency: it is
$\mathcal G/(4\hbar)$ times
$[d-\ell_A-\ell_B]^{-1}+[d+\ell_A+\ell_B]^{-1}
-[d-\ell_A+\ell_B]^{-1}-[d+\ell_A-\ell_B]^{-1}$.
For equal intervals $\ell_A=\ell_B=r$, this is
$\mathcal G Q^2/\hbar$. In this collinear setting we can prove a
stronger statement than equality of capacities: finite interaction
protocols can be simulated in either direction.

The product factors in Eq.~\eqref{eq:newton-centered-integral} share the
same extremal position states. We use this alignment to connect the
confined Newtonian interaction to an Ising interaction. Reversible
simulation of product Hamiltonians, including ancillas, is established by
Childs, Leung and Vidal\cite{ChildsLeungVidal2004}, Theorems 3 and 6. Their
Appendix, Corollary 9, treats a two-term sum with common two-dimensional
extremal spaces and, in their normalization, equal Hilbert--Schmidt
overlaps of the traceless restricted factors on Alice's and Bob's sides;
that simulation allows catalytic entanglement. Here the normalized
factors are already aligned on the endpoint code: all restricted terms
are multiples of one Ising operator with the same sign, so no catalyst is
required. We control the passage from finite position grids to continuous
support below.

For convenience we recall the entropy-rate constant used in this comparison:
\begin{equation}
 \beta=\max_{0<q<1/2}2\sqrt{q(1-q)}\log_2\frac{1-q}{q}
 =1.91227328895\ldots.
 \label{eq:newton-beta}
\end{equation}
Here $Z$ is the Pauli operator with eigenvalues $+1$ and $-1$.
The Hamiltonian $\hbar Z\otimes Z$ has asymptotic entanglement capacity
$\beta$ ebits per unit time; the same constant bounds its instantaneous
entropy production with arbitrary local ancillas\cite{ChildsLeungVerstraeteVidal2003}.
For a finite-dimensional product Hamiltonian $\hbar hX\otimes Y$, the
corresponding rate is $\beta|h|\Delta_X\Delta_Y/4$, where $\Delta_X$ and
$\Delta_Y$ are spectral diameters\cite{ChildsLeungVidal2004}. These rates
use free local control and count net generated entanglement. Initial seed
entanglement is prepared and its cost amortized in the established
capacity construction\cite{BennettHarrowLeungSmolin2003}.

\begin{proposition}[Capacity in the ideal supported model]
\label{prop:newton-capacity}
The confined Newtonian interaction is reversibly equivalent to
$\hbar\kappa Z\otimes Z$ in the closure of finite protocols with
arbitrarily accurate local control. Its asymptotic entanglement-generation
capacity per aggregate interaction time is
\begin{equation}
 C_E=\beta\kappa\quad\text{ebits per unit time}.
 \label{eq:newton-capacity}
\end{equation}
Every finite endpoint-encoding approximation may use smooth packets with
finite kinetic energy. The supremum permits local energy, memory and
control requirements to grow with the requested accuracy and block length.
\end{proposition}

\begin{proof}
We first simulate the potential using Ising interactions. Replace each
position operator by a finite-spectrum step function $X_\delta,Y_\delta$
with values in its allowed interval; $\delta$ labels the mesh refinement.
The smallest allowed separation $d_{\min}$ controls the sensitivity of
the reciprocal distance to these replacements:
\begin{equation}
 \|V(x_A,x_B)-V(X_\delta,Y_\delta)\|
 \le\frac{\mathcal G}{d_{\min}^2}
 \bigl(\|x_A-X_\delta\|+\|x_B-Y_\delta\|\bigr).
 \label{eq:newton-grid-error}
\end{equation}
The approximating potential acts on finitely many cell labels. The
within-cell Hilbert spaces are local spectator systems. On the finite
labels, Eq.~\eqref{eq:newton-centered-integral} has product-strength sum
at most $\kappa$. Its integral can be replaced by a finite convex
decomposition of normalized product matrices with the same budget:
the matrices form a compact set after taking their limiting values, and
their integral belongs to its finite-dimensional convex hull.

Each product term can be simulated at its spectral-diameter cost by an
Ising interaction\cite{ChildsLeungVidal2004}. Time sharing and the product
formula simulate the sum, with total Ising strength-time at most
$\kappa T$. For a fixed finite-duration protocol, replacing the potential
changes its output by a quantity tending to zero with
$T\|V-V_\delta\|/\hbar$. Local maps and ancillary systems do not increase
this channel error. We choose the grid and product-formula accuracy
separately for each finite block. Hence any rate exceeding
$\beta\kappa$ for the continuous potential would exceed the established
Ising capacity on finite output registers. No uniform continuity of
infinite-dimensional entropy is used.

For the reverse simulation, choose two narrow packets at each laboratory,
with centres $\pm(\ell_A-w_A)$ and $\pm(\ell_B-w_B)$ and halfwidths
$w_A,w_B$. Keep their local kinetic Hamiltonians, with identical translated
wells so that the reference packets are stationary. On the full confined
two-well Hilbert space of each laboratory, replace the potential by the
constant centre value on each entire branch rectangle. Its restriction
to the branch qubits is a diagonal Hamiltonian locally equivalent to
$\hbar\kappa_w Z\otimes Z$, where
$\kappa_w=\kappa(\ell_A-w_A,\ell_B-w_B)\to\kappa$.
Writing $r_w=d-\ell_A-\ell_B+w_A+w_B$ for the minimum separation of
packet centres, let $\epsilon_w$ denote the operator-norm error in this
constant-on-branches replacement. The reciprocal-distance estimate gives
\begin{equation}
 \epsilon_w\le
 \frac{\mathcal G(w_A+w_B)}{r_w[r_w-w_A-w_B]}.
 \label{eq:newton-endpoint-error}
\end{equation}
The full kinetic-plus-potential evolutions differ in state norm by at most
$T\epsilon_w/\hbar$. Local logical operations act on the branch labels
and preserve the reference within-well coordinate. The same error bound
therefore holds through any fixed finite Ising protocol.

For each finite block of an Ising capacity protocol, choose nonzero widths
small enough that its error vanishes and $\kappa_w$ approaches $\kappa$.
The explicit packets in Section~\ref{sec:newtonian-packets} have finite
kinetic energy at every such choice. Applying the Ising protocol then
gives rates approaching $\beta\kappa$. Its seed is generated using the
interaction and accounted for before taking the capacity limit. This
proves the reverse inequality and the simulation statement.
\end{proof}

The supremum in Eq.~\eqref{eq:newton-capacity} imposes no common kinetic
energy or control-bandwidth bound. Endpoint localization can require
increasing energy and need not remain within a uniformly nonrelativistic
experimental regime. Equality also assumes that noise can be absent;
local noise remains allowed in the upper-bound class. A fixed compulsory
bath, finite barriers or a specified memory lifetime define further
physical constraints.

\section{The noise threshold and the full-potential bound}
\label{sec:noise}
\label{sec:si-noise}

We now ask how much of the noiseless capacity survives independent random
forces. We first simulate the noisy dipole interaction by local measurements
and classical feedback, then bound the entanglement produced by the remaining
terms of the full potential. The last part converts this instantaneous bound
into a bound on finite Bell registers and constructs entangled outputs below,
and slightly above, the dipole threshold.

\subsection{A Bell-rate bound for the full interaction under white force noise}
\label{analytic:noisy-capacity}

We allow the parties to use the entire position space,
local memories and the quantum systems that carry their recovery syndromes,
without requiring them to remain in a fixed code.
Under sufficiently strong white force noise, the dipole Hamiltonian together
with the noise can be implemented by local operations and classical
communication (LOCC). The higher terms of the Newtonian potential then
bound the rate at which the parties can generate Bell pairs.

We retain the full Newtonian potential $V=-\mathcal G/|d\mathbf e_z+
\mathbf b-\mathbf a|$ on two balls of radius $r$ with centre separation
$d>2r$, using $\mathbf e_z$ for the separation direction. Write
$u=r/d$, $g_0=\mathcal G/(\hbar d)$, and let $\beta$ be the Ising
entropy-rate constant in Eq.~\eqref{eq:newton-beta}.
We allow independent isotropic noise strengths $\Lambda_A,\Lambda_B\ge0$,
so that the generator during a free dwell is $-i[V,\cdot]/\hbar+\mathcal L_{\rm f}$,
where
\begin{equation}
 \mathcal L_{\rm f}(\rho)=
 -\sum_{i=x,y,z}
 \bigl(\Lambda_A[a_i,[a_i,\rho]]+\Lambda_B[b_i,[b_i,\rho]]\bigr).
 \label{analytic:force}
\end{equation}
With the two-sided white force spectral density $S_{F,Q}$ per
Cartesian component, $\Lambda_Q=S_{F,Q}/(2\hbar^2)$. The force
environment is inaccessible, and the free dwells use this Markov
channel on the masses tensored with the identity on local memories.
The double commutator damps a position coherence separated by
$\Delta\mathbf q$ in laboratory $Q$ at rate
$\Lambda_Q|\Delta\mathbf q|^2$; this fixes the noise normalization.

A finite protocol interleaves such dwells with arbitrary normal local
operations and classical communication. The laboratories start in a normal
separable state, with any subsequently generated seed entanglement charged to
the protocol. Ancillas and local instruments are unrestricted within the
usual separable Hilbert spaces and standard Borel outcome spaces. Later
controls and dwell times can depend on the observed outcomes. A protocol can
retain or discard quantum syndromes locally. Its cost is the mean of the sum
$T_{\rm int}$ of the dwell durations, including time spent on failed attempts.
We include limits of finite protocols for which the state on each specified
pair of finite output registers converges in trace norm and the mean costs
converge. Continuous local controls, including local Hamiltonian evolution
during a dwell, are admitted when their outputs and costs arise as such
limits with the same support constraint.

Before deriving the rate, we verify that this protocol definition also
includes kinetic motion during an interaction. The issue is that the kinetic
Hamiltonian is unbounded, whereas the confined potential and force-noise
operators are bounded. For example, let
$K_Q=-\hbar^2\Delta/(2m_Q)$ be the Dirichlet kinetic Hamiltonian on
laboratory $Q$'s ball. On trace class, write $\mathscr A$ for the
generator of the strongly continuous local conjugation group
$\mathcal U_t$ of $K_A+K_B$, and $\mathscr B=-i[V,\cdot]/\hbar+
\mathcal L_{\rm f}$ for the bounded position generator. The full
evolution $e^{t(\mathscr A+\mathscr B)}$ is the strong trace-norm
limit of $(\mathcal U_{t/n}e^{(t/n)\mathscr B})^n$.
In the local interaction picture,
$\mathscr B(s)=\mathcal U_{-s}\mathscr B\mathcal U_s$ is strongly
continuous and uniformly bounded. The interaction-picture product
uses $\mathscr B(s)$ frozen at the start of each step. Its
time-ordered series converges on each trace-class input to the
series with the continuous $\mathscr B(s)$ by termwise dominated
convergence, with norm bounds $(t\|\mathscr B\|)^j/j!$ at order
$j$. This gives the stated limit, also with an identity on any
local reference. Each factor preserves the supported Hilbert
spaces, and the position dwells sum to $t$.

For a finite adaptive protocol the same replacement can be made
at every dwell. In a telescoping comparison, the exact preceding
history gives a conditional input and a duration, and the subsequent
instruments contract the trace norm including their classical
outputs. Dominated convergence over the preceding history then
gives convergence of the joint quantum and classical output. To
handle unbounded durations with finite mean cost, first stop
before exceeding a fixed total time and supply local placeholder
outputs on those histories. The mean cost of the capped protocol
is a bounded function of its record and converges under the
product approximation. Removing the cap recovers the original
output by the vanishing truncation probability and the original
mean cost by monotone convergence.

Write $C_{\Lambda_A,\Lambda_B}$ for the supremum of the asymptotic rates of
producing $L\to\infty$ Bell pairs on specified local registers with fidelity
tending to one, per mean interaction time. The fidelity convention here is
$f=\operatorname{Tr}(\Phi_{2^L}\rho_{\rm out})$, with $\Phi_{2^L}$ the
maximally entangled projector for the two registers. Since local force
dephasing can be added freely to noiseless dwells, $C_{\Lambda_A,\Lambda_B}$
is at most the noiseless capacity. For $d\ge(1+\sqrt5)r$ that capacity is
$C_E^{(3)}=2\beta g_0u^2/(1-4u^2)$ by
Theorem~\ref{thm:newton-three-d}.

To estimate the first terms left after removing the dipole interaction, we
expand the dimensionless Newtonian field $\mathbf F$ through first order.
Its residual $\mathbf q$ begins at quadratic order. Let $\mathbf v$ be a
dimensionless displacement and $D\mathbf F(\mathbf 0)$ the field's
Jacobian matrix at the origin. We define
\begin{align}
 \mathbf F(\mathbf v)&=\frac{\mathbf e_z+\mathbf v}
                          {|\mathbf e_z+\mathbf v|^3},
 &\mathbf q(\mathbf v)&=\mathbf F(\mathbf v)-\mathbf F(\mathbf 0)
                       -D\mathbf F(\mathbf 0)\mathbf v,\nonumber\\
 q_*(u)&=\min_{\mathbf c\in\mathbb R^3}
       \sup_{|\mathbf v|\le u}|\mathbf q(\mathbf v)-\mathbf c|,
 &B(u)&=2u\,q_*(u)+S_2(u),
 \label{analytic:field-bound}
\end{align}
Here $q_*(u)$ is the radius of the smallest ball containing all residual
field values, with its centre $\mathbf c$ varied freely. Subtracting such a
centre will remove only a local Hamiltonian. The tail
$S_2(u)=\sum_{j,k\ge2}\binom{j+k}{j}u^{j+k}$ collects terms that are at
least quadratic in both positions; its general form is given in
Eq.~\eqref{eq:protected-universal-series}. The two contributions
$u q_*(u)$ come from terms linear on one side and at least quadratic on
the other. An alternative bound estimates these terms without centring the
field and uses the elementary function
\begin{equation}
 R(u)=\frac{2u^3(3-2u)}{(1-2u)(1-u)} .
 \label{analytic:elementary-bound}
\end{equation}

\begin{theorem}[Dipole threshold and Bell-pair rate]
\label{analytic:noise-converse}
Set $\Lambda_c=\mathcal G/(\hbar d^3)$,
$\lambda=\sqrt{\Lambda_A\Lambda_B}/\Lambda_c$, and
$\eta=\min\{1,\lambda\}$. Every admissible protocol with output
fidelity $f$ to $L$ Bell pairs satisfies
\begin{equation}
 (1-\sqrt{1-f})L-h_2\bigl(\min\{\sqrt{1-f},1/2\}\bigr)
 \le U\,\mathbb E T_{\rm int},\qquad
 U=\beta g_0\bigl[\min\{B(u),R(u)\}+2(1-\eta)u^2\bigr],
 \label{analytic:output-bound}
\end{equation}
where $h_2$ is binary entropy in bits. In particular,
\begin{equation}
 C_{\Lambda_A,\Lambda_B}\le U,\qquad
 B(u)=\tfrac92 u^3+O(u^4).
 \label{analytic:capacity-bound}
\end{equation}
When $\lambda\ge1$, the upper bound at $d/r=6$ is at most
$32/45$ of the noiseless capacity, and as $u\to0$ the ratio is at most
$\tfrac94 u+O(u^2)$. At any fixed pair of individual ratios
$\Lambda_A/\Lambda_c,\Lambda_B/\Lambda_c$ with $\lambda<1$,
the rate is $\Theta(g_0u^2)$ as $u\to0$.
\end{theorem}

\paragraph{A finite-time bilinear channel.}
Kafri, Taylor and Milburn use measurement and feedback to simulate a
bilinear Hamiltonian with force dephasing\cite{Kafri2014}.
Li, Miki and Chen derive state-independent conditions for bilinear
entanglement generation under white noise\cite{LiMikiChen2026}.
We need an identity for a finite dwell, so that the simulated channel can
be inserted into an adaptive protocol. Let $X,Y$ be bounded
local self-adjoint operators and let $g$ be their real coupling coefficient
in the Hamiltonian divided by $\hbar$. Suppressing identities on the other
laboratory, write
\begin{equation}
 \mathcal K=-ig[X\otimes Y,\cdot]
 -\Lambda_A[X,[X,\cdot]]-\Lambda_B[Y,[Y,\cdot]] .
 \label{analytic:bilinear}
\end{equation}
If $\Lambda_A\Lambda_B\ge g^2/4$, then $\exp(t\mathcal K)$ is LOCC
for each $t\ge0$. For $g=0$ it is local dephasing, and for $t=0$ it
is the identity. Otherwise choose positive measurement strengths $k,k'$ as
specified below, and let Alice apply
$M_y=(2kt/\pi)^{1/4}\exp[-kt(X-y)^2]$ with real outcome $y$.
The instrument is normalized since $\int M_y^\dagger M_y\,dy=I$.
She sends $y$ to Bob, who applies $\exp(-igtyY)$. In the
spectral representation of $X,Y$, use $(x,e)$ and $(x',e')$ for the
two sets of coordinates of a matrix element. Completing the square
about $(x+x')/2$ and integrating over $y$ gives its multiplier
\begin{equation}
 \exp\left[t\left(-\frac{k}{2}(x-x')^2
 -\frac{ig}{2}(x+x')(e-e')
 -\frac{g^2}{8k}(e-e')^2\right)\right].
 \label{analytic:gaussian-multiplier}
\end{equation}
The reverse instrument, with measurement strength $k'$,
interchanges the two spectral variables. The two phases add to
$-igt(xe-x'e')$, and the two damping coefficients are
$k/2+g^2/(8k')$ and $k'/2+g^2/(8k)$. Taking $kk'=g^2/4$ makes
these coefficients $k,k'$. The assumed noise product allows
$k\le\Lambda_A$ and $k'\le\Lambda_B$. Independent random local phases
with variances $2t(\Lambda_A-k)$ and $2t(\Lambda_B-k')$ complete the
required noise. The proof applies equally to discrete and continuous spectra
by the spectral calculus.
Thus the product condition is precisely what allows both measurements and
feedback kicks to fit within the available local dephasing strengths.

For $X=Y=Z$, the Pauli matrix, the same threshold is necessary for the
channel to preserve separability for all times. Local Hadamards rotate
the input $|++\rangle$ to $|00\rangle$ and the Hamiltonian to
$\hbar g X_{\rm P}\otimes X_{\rm P}$, with $X_{\rm P}$ the Pauli
bit flip. The block of the partial transpose on $|01\rangle,|10\rangle$
is
\[
 t\begin{pmatrix}2\Lambda_B&ig\\-ig&2\Lambda_A\end{pmatrix}+O(t^2).
\]
Its determinant is negative to order $t^2$ whenever
$\Lambda_A\Lambda_B<g^2/4$. The output is then entangled for sufficiently
small positive $t$.

\paragraph{Separating the full Newtonian potential.}
The angular expansion derived in Eq.~\eqref{eq:protected-universal-expansion}
organizes the potential by degrees $(j,k)$ in Alice's and Bob's positions.
Terms with either degree zero are local. The lowest term with both degrees
nonzero, $(1,1)$, is the dipole interaction
\begin{equation}
 V_{\rm dip}=\hbar\sum_{i=x,y,z}g_i a_i b_i,\qquad
 g_z=2\Lambda_c,\quad g_x=g_y=-\Lambda_c .
 \label{analytic:dipole}
\end{equation}
For each component, $\Lambda_A\Lambda_B\ge(\eta g_i)^2/4$.
Apply the bilinear channel construction to the three components with the
couplings $\eta g_i$, using $\Lambda_A,\Lambda_B$ for each component's
dephasing. All position multipliers commute. Consequently, for any
decomposition $V=\eta V_{\rm dip}+W+V_{\rm loc}$ with position
multipliers $W,V_{\rm loc}$ and $V_{\rm loc}$ local, the exact dwell
channel factors as the LOCC channel for $\eta V_{\rm dip}$ with noise,
conjugation by $\exp(-itW/\hbar)$, and local position phases. Each
factor uses the full dwell duration $t$ in this identity. The instrument
outcomes used in the LOCC representation are auxiliary variables in the
proof. The parties' subsequent controls depend on their own recorded
outcomes.

After simulating the fraction $\eta$ of the dipole interaction locally, only
$W$ can increase entanglement. We bound $W$ using a row sum that is uniform
in local ancillas. For
any paired orthonormal families $|u_j\rangle,|v_j\rangle$ on the two
laboratories, let $w_{ij}=\langle u_iv_i|W|u_jv_j\rangle$. If
$W=\int A_\nu\otimes B_\nu\,d\mu(\nu)$ with self-adjoint factors and
positive measure $\mu$, then
\begin{equation}
 \sup_i\sum_j |w_{ij}|
 \le \left\|\int A_\nu^2\,d\mu(\nu)\right\|^{1/2}
      \left\|\int B_\nu^2\,d\mu(\nu)\right\|^{1/2}.
 \label{analytic:rows}
\end{equation}
Indeed Cauchy--Schwarz over $(j,\nu)$ bounds the sum of the products of
matrix elements, and Bessel's inequality bounds each sum of squares.
Signs can be included in either self-adjoint factor. Row bounds add
under summation. In particular, a complex tensor product of contractions
has a real part with row bound one by
Eq.~\eqref{eq:protected-complex-rate-factors}. These estimates also apply
to norm-convergent decompositions with summable row bounds by taking
the limit in each entry and using Fatou's lemma on the row sums.

The residual dipole $(1-\eta)V_{\rm dip}$ has row bound at most
$2(1-\eta)\hbar\Lambda_c r^2$. To see this, put $\sqrt{|g_i|}$ in each
position factor, place the sign of $g_i$ on one side and use
$\sum_i|g_i|a_i^2\le2\Lambda_c r^2 I$ and the analogous bound on
Bob's side. For the higher terms, the strip $j=1,k\ge2$ of
Eq.~\eqref{eq:protected-universal-expansion} is
$-(\mathcal G/d)(\mathbf a/d)\cdot\mathbf q(\mathbf b/d)$.
This follows by differentiating $V$ in $\mathbf a/d$ at zero and
subtracting the constant and linear terms of $\mathbf F$.
Subtracting a constant $\mathbf c$ from $\mathbf q$ adds only a local
term. The squared position factors sum to at most $u^2 I$, so the
strip has row bound $(\mathcal G/d)u q_*(u)$ with the minimizing
centre. The strip $j\ge2,k=1$ is related by
$(\mathbf a,\mathbf b)\mapsto(-\mathbf b,-\mathbf a)$ and has the same
bound. The terms with $j,k\ge2$ contribute at most
$(\mathcal G/d)S_2(u)$ by the real and imaginary factor decomposition.
This gives the row bound $(\mathcal G/d)B(u)$ beyond the dipole.

Alternatively, bound each degree separately before choosing any
centres. Write $S_1(u)=\sum_{j,k\ge1}\binom{j+k}{j}u^{j+k}$ for
the corresponding series including the dipole. The entire series with
$j,k\ge1$ except $(1,1)$ contributes
at most $(\mathcal G/d)[S_1(u)-2u^2]$. Summing the unrestricted
binomial series and removing the two axes gives
\[
 S_1(u)-2u^2
 =\frac1{1-2u}-\frac2{1-u}+1-2u^2
 =R(u).
\]
The subtracted axes are the purely local terms, and $2u^2$ is the
dipole contribution. The two decompositions can differ by local position terms, and we
choose the one giving the smaller bound at the specified separation.
At $u=1/6$ the ratio of $R(u)$ to $2u^2/(1-4u^2)$ is $32/45$.

To obtain the leading coefficient of $B(u)$, expand
$\mathbf q$ uniformly on the ball. With $\mathbf v_\perp$ the transverse
part of $\mathbf v$, its quadratic part is
\[
 \mathbf q_2(\mathbf v)
 =\bigl(3v_z^2-\tfrac32|\mathbf v_\perp|^2\bigr)\mathbf e_z
 -3v_z\mathbf v_\perp .
\]
The polar and equatorial boundary points have images
$3u^2\mathbf e_z$ and $-\tfrac32u^2\mathbf e_z$, so any enclosing ball
has radius at least $9u^2/4$. The centre $\tfrac34u^2\mathbf e_z$
attains that radius. In fact, with $s=|\mathbf v|^2/u^2$ and
$x=v_z^2/u^2$, the squared distance to this centre in units of $u^4$
is
\[
 \bigl(\tfrac92 x-\tfrac32 s-\tfrac34\bigr)^2+9x(s-x).
\]
Here $s$ is the squared radial coordinate and $x$ its longitudinal
contribution, so $0\le x\le s\le1$. The expression is convex in $x$ at fixed $s$ and takes values at most $(9/4)^2$
at $x=0$ and $x=s$ for $0\le s\le1$. The uniform Taylor remainder is
$O(u^3)$ and changes the enclosing radius by at most its supremum
norm. Thus $q_*(u)=9u^2/4+O(u^3)$, and $S_2(u)=O(u^4)$.
This determines the leading coefficient of the bound $B(u)$.

\paragraph{Finite-separation bounds.}
An explicit upper bound on $q_*(u)$ suffices for the finite-separation values.  Fix any real number $c$ and use the axial centre $c\mathbf e_z$.
We first reduce the spatial supremum to the boundary, then control the error
of replacing that boundary by a finite grid. Each component of $\mathbf q$
is harmonic on the open unit ball, since
$\mathbf F(\mathbf v)=-\nabla_{\mathbf v}|\mathbf e_z+\mathbf v|^{-1}$.
Consequently $|\mathbf q-c\mathbf e_z|^2$ is subharmonic, and its maximum
over $|\mathbf v|\le u<1$ occurs on the sphere.  Rotational symmetry
reduces this maximum to
$\mathbf v(\theta)=u(\sin\theta,0,\cos\theta)$, $0\le\theta\le\pi$.

Here is an explicit discretization allowance.  For
$\mathbf f(\mathbf w)=\mathbf w/|\mathbf w|^3$ with $\mathbf w\ne0$ and a unit vector
$\mathbf h$, put $s=(\mathbf h\cdot\mathbf w)^2/|\mathbf w|^2$.
Here $D^2\mathbf f[\mathbf h,\mathbf h]$ is the second directional
derivative, and $s\in[0,1]$ is the squared cosine of the angle between
$\mathbf h$ and $\mathbf w$. Direct differentiation gives
\[
 |D^2\mathbf f(\mathbf w)[\mathbf h,\mathbf h]|^2
 =|\mathbf w|^{-8}(45s^2-18s+9)\le36|\mathbf w|^{-8}.
\]
The same bound applies to the bilinear operator norm: after fixing a
unit output direction, the scalar Hessian is a real symmetric matrix,
whose norm is attained on repeated unit inputs.  Integrating this
Hessian bound on the segment from $0$ to $\mathbf v$ bounds the change
$D\mathbf F(\mathbf v)-D\mathbf F(0)=D\mathbf q(\mathbf v)$ and yields
\[
 \|D\mathbf q(\mathbf v)\|
 \le 6|\mathbf v|\int_0^1(1-t|\mathbf v|)^{-4}\,dt
 =2[(1-|\mathbf v|)^{-3}-1].
\]
The spherical curve has speed $u$, and every angle is at most
$\pi/(2N)$ from a grid point when $[0,\pi]$ is divided into $N$
equal intervals. Multiplying these two factors by the derivative bound
gives the allowance added to the sampled maximum:
\begin{equation}
 q_*(u)\le Q_N(u,c):=
 \max_{0\le j\le N}
 |\mathbf q(u\sin(j\pi/N),0,u\cos(j\pi/N))-c\mathbf e_z|
 +\frac{\pi u}{N}[(1-u)^{-3}-1].
 \label{analytic:finite-field-radius}
\end{equation}
The maximization is now finite. The added term covers every position
between grid points, so a sampled value alone is never used as an upper
bound. Any centre $c$ gives a valid bound;
minimizing the sampled maximum is only a way of choosing it.  A
truncation of $S_2(u)$ through total degree $L$ has positive remainder
at most $(2u)^{L+1}/(1-2u)$.  Thus replacing $q_*$ by $Q_N$ and adding
this remainder gives the upper value
$B_{N,L}(u)=2uQ_N(u,c)+\sum_{j,k\ge2,\,j+k\le L}\binom{j+k}{j}u^{j+k}
+(2u)^{L+1}/(1-2u)$.

For $N=8192$ and $L=200$, the following choices give the stated upper
values. The last column is rounded upwards; in particular it gives the
$54\%$ and $28\%$ limits used in the article.
\begin{center}
\begin{tabular}{rrr}
\toprule
$d/r$ & Centre $c$ & Upper value of $B_{N,L}(u)/[2u^2/(1-4u^2)]$\\
\midrule
4 & 0.091627292805921 & 0.975513\\
5 & 0.051088656324545 & 0.695735\\
6 & 0.032421075970972 & 0.538447\\
8 & 0.016317001649121 & 0.368872\\
10 & 0.009773379542487 & 0.279646\\
20 & 0.002140940355088 & 0.125568\\
\bottomrule
\end{tabular}
\end{center}
After division by the noiseless capacity, the bound on the noisy-capacity
ratio can consequently use
\[
 \min\!\left\{1,
 \frac{\min\{B_{N,L}(u),R(u)\}
       +2[1-\min\{1,\lambda\}]u^2}
      {2u^2/(1-4u^2)}\right\},
\]
where the cap at one uses the noiseless capacity in its stated
separation range.

\paragraph{From the dwell bound to finite Bell registers.}
Write $H_*=\hbar U/\beta$ for the bound on the rows of $W$.
For a pure state $|\psi\rangle$ of the complete laboratories, we first
show that evolution under $W$ for time $t$ increases its entropy of
entanglement by at most $Ut$. To handle infinite Schmidt rank,
take the reduced state $\sigma$ on Alice's side and use a smooth entropy
whose derivative remains bounded at a zero eigenvalue:
\[
 S_\epsilon(\sigma)=\operatorname{Tr}f_\epsilon(\sigma),
 \qquad f_\epsilon(p)=-p\log_2[(p+\epsilon)/(1+\epsilon)],
 \qquad \epsilon>0 .
\]
This quantity increases to the entropy $S(\sigma)$ as $\epsilon$
decreases to zero. The trace-norm derivative of $\sigma$ exists since
$W$ is bounded. The trace derivative of $S_\epsilon$ is
$\operatorname{Tr}(f_\epsilon'(\sigma)\dot\sigma)$, which follows by
expanding the logarithm in powers of $(I-\sigma)/(1+\epsilon)$.
With the factor $\sigma$ in $f_\epsilon(\sigma)$ the series and its
trace derivatives converge, since $\|(I-\sigma)/(1+\epsilon)\|<1$
and $\dot\sigma$ is trace class. Thus we can evaluate the derivative
in any Schmidt bases at the current time, with squared Schmidt weights
$p_i$. We compare the regularized derivative difference with the logarithmic
difference of the ordinary entropy. The curvature bound below makes that
comparison possible, and the second inequality supplies the Ising constant:
\[
 |f_\epsilon''(p)|=\frac{p+2\epsilon}{(p+\epsilon)^2\ln 2}
                  \le \frac1{p\ln 2}\quad(p>0),\qquad
 2\sqrt{p_ip_j}|\log_2(p_i/p_j)|\le\beta(p_i+p_j).
\]
In the derivative, diagonal elements of $\dot\sigma$ corresponding to
zero Schmidt weights vanish. Pairing the other terms by $(i,j)$ gives
\begin{align}
 \left|\frac{d}{dt}S_\epsilon(\sigma)\right|
 &\le\frac{2}{\hbar}\sum_{i<j}\sqrt{p_ip_j}
             |f_\epsilon'(p_i)-f_\epsilon'(p_j)|\,|w_{ij}|
 \nonumber\\
 &\le\frac{\beta}{\hbar}\sum_i p_i\sum_{j\ne i}|w_{ij}|
 \le U .
 \label{analytic:regularized-entropy}
\end{align}
Integrating the curvature inequality between $p_i$ and $p_j$ bounds
$|f_\epsilon'(p_i)-f_\epsilon'(p_j)|$ by
$|\log_2(p_i/p_j)|$ when both weights are positive. The row bound also
gives absolute convergence of the paired series.
Integrate the inequality and take $\epsilon\downarrow0$ to obtain the
entropy increment bound for any finite initial entropy.

An initial normal separable state has a probability-measure
decomposition into pure product states\cite{HolevoShirokovWerner2005}.
We keep this ensemble and refine it through the finite protocol.
Each simulated LOCC factor has a pure-state refinement, indexed by its
Gaussian outcomes or local phase variables. Local control instruments
also have such refinements. One can use a local Stinespring dilation
and resolve the discarded environment in a basis. For continuous
outcomes, the normal instrument dilation resolves the pointer as a
direct integral, with an additional basis resolution in its
multiplicity spaces\cite{Ozawa1984}. Only the prescribed outcomes are
available to the controls. All retained systems remain in their
laboratory. A pure branch can therefore be followed at every step and
its average gives the specified state's statistics.

A local step on Alice's side leaves Bob's reduced state unchanged on
averaging. Concavity of entropy then bounds the average of the pure
output entropies by the pure input entropy. This remains valid for
continuous outcomes: in an eigenbasis of Bob's input state, each
conditional entropy is at most the Shannon entropy of its diagonal,
and concavity of Shannon entropy bounds the average by the entropy of
the input diagonal. The same argument applies to a local step on
Bob's side. The initial branch entropies are zero. Applying
Eq.~\eqref{analytic:regularized-entropy} to each remainder unitary and
averaging over the actual and auxiliary histories bounds the final
average pure-branch entropy by $U\,\mathbb E T_{\rm int}$. In particular,
if the cost is finite, the entropy is finite almost surely at every
step. Dwell durations may be chosen from the recorded history since
the increment inequality holds conditionally for every chosen duration.

At extraction, resolve every system being discarded in a local basis,
again solely for the entropy proof. This gives a pure ensemble on the
two finite Bell registers with average entropy still at most
$U\,\mathbb E T_{\rm int}$. Let $D=2^L$. A pure output branch with
fidelity $f_\omega$ to $\Phi_D$ has a reduced state at trace distance
at most $\sqrt{1-f_\omega}$ from $I_D/D$. Denote the total variation
distance between its eigenvalue distribution and uniform by $z_\omega$.
A coupling of the two distributions with disagreement probability
$z_\omega$ bounds their entropy difference by
$h_2(z_\omega)+z_\omega\log_2 D$: condition on agreement and bound
the entropy on disagreement by $\log_2 D$. Thus the branch entropy
is at least $L-h_2(z_\omega)-z_\omega L$. The mean of $z_\omega$ is at
most $\sqrt{1-f}$ by concavity of the square root and linearity of
fidelity with the pure target. Concavity of $h_2$ and its maximum at
$1/2$ now give Eq.~\eqref{analytic:output-bound}. The proof uses entropy
of pure branches and the specified finite registers throughout.

For a countable adaptive protocol with almost-sure stopping and
finite mean cost, truncate after a finite number of steps, supplying
local placeholder outputs on branches still running. The output
trace distance is at most the probability of truncation and the
mean cost converges by monotone convergence. For any admitted
normal protocol limit, apply Eq.~\eqref{analytic:output-bound} to
the fixed finite output registers. Their fidelities converge, and
the inequality passes to the limit with the mean cost. Applying the
finite-output bound to every member of an asymptotic Bell-generation
sequence yields the rate converse in
Eq.~\eqref{analytic:capacity-bound}.

\paragraph{Unprotected states below and near the dipole threshold.}
The upper bound removes the leading rate at the threshold. To determine
whether that threshold is also the end of entanglement generation, we now
examine a two-position protocol for the full potential.
Take two positions $\pm s\mathbf e_z$ in each ball with $0<s<r$,
and use $|++\rangle$ on the two position labels, where
$|+\rangle=(|0\rangle+|1\rangle)/\sqrt2$ assigns equal amplitude to
both positions. After removal of
local phases, the point-mode Hamiltonian is $\hbar\kappa Z_AZ_B$ and
the off-diagonal elements on side $Q$ decay at rate $\gamma_Q$, with
\begin{align}
 \kappa&=\frac{\mathcal G}{4\hbar}
 \left[\frac1{d-2s}+\frac1{d+2s}-\frac2d\right]
 =\frac{2\Lambda_c s^2}{1-4(s/d)^2},&
 \gamma_Q&=4\Lambda_Qs^2.
 \label{analytic:packet-coupling}
\end{align}
The Pauli partial-transpose calculation after
Eq.~\eqref{analytic:gaussian-multiplier} applies with strengths
$\gamma_Q/4$. It gives entanglement at small positive time when
$\sqrt{\gamma_A\gamma_B}<2\kappa$, equivalently
$\lambda<1/[1-4(s/d)^2]$.

A fixed positive margin below the dipole threshold gives a rate
of order $g_0u^2$. To see this, take $s=\alpha r$ for a fixed
$0<\alpha<1$ and write $\lambda_Q=\Lambda_Q/\Lambda_c$.
The ratios $\gamma_Q/\kappa=2\lambda_Q[1-4(s/d)^2]$ can be increased
to $2\lambda_Q$ by adding local dephasing. At fixed
$\lambda_A,\lambda_B$ with $\lambda_A\lambda_B<1$,
choose a small positive dimensionless time $\tau=\kappa t$
for which this padded two-qubit state is entangled. Its state is
independent of $u$, so it has a fixed positive two-way distillation
yield $D_{\lambda_A,\lambda_B}$ in ebits per copy. This uses the
distillability theorem for entangled two-qubit states\cite{HorodeckiDistill1997}:
local filtering and a finite number of recurrence steps reach
positive hashing yield with nonzero total success fraction.
Independent preparation and distillation give a rate at least
$D_{\lambda_A,\lambda_B}\kappa/\tau=\Omega(g_0u^2)$.
The noiseless capacity supplies the matching upper order.
For equal noise, this lower bound can be taken uniformly over
$0\le\lambda\le1-\delta$ with any fixed $0<\delta<1$ by padding to the
same ratios $2(1-\delta)$. Equation~\eqref{analytic:capacity-bound}
also shows that $\lambda(u)\to1$ forces a rate $o(g_0u^2)$,
including approach from below.

At each finite separation the full axial coupling in
Eq.~\eqref{analytic:packet-coupling} exceeds its dipole
approximation. For equal noise, $\gamma_A=\gamma_B=\gamma$, the
negative-partial-transpose criterion from
Eq.~\eqref{eq:protected-entanglement-condition} is
$|\sin(2\kappa t)|>\sinh(\gamma t)$. For a concrete example, take
$s/d=1/10$, $r/d>1/10$, and
$\Lambda_A=\Lambda_B=(51/50)\Lambda_c$. Then
$\gamma/\kappa=1.9584$ exactly. At $\kappa t=1/10$,
\[
 \sin(0.2)>\sinh(0.19584),\qquad
 \lambda_{\min}(\rho^{T_B})
 =\frac{e^{-0.19584}}2\bigl[\sinh(0.19584)-\sin(0.2)\bigr]
 \simeq-6.47\times10^{-4}.
\]
This produces unprotected entanglement above the dipole threshold.
The strict inequality also follows from the alternating lower bound
$\sin x\ge x-x^3/6$ at $x=0.2$ and the upper bound
$\sinh y\le y+y^3/[6(1-y^2/20)]$ at $y=0.19584$, obtained by
bounding the ratios of successive series terms beginning with the
cubic term.

Both arguments admit finite-energy packets. Centre disjoint, smooth,
compactly supported packets at the two positions in each laboratory.
On the union of the packet supports, replace each position
operator by its packet-centre value and $V$ by the corresponding
position-label potential. The two generators converge in
trace-class operator norm, also after tensoring an identity on
references. Indeed $\|[H,\cdot]\|_{1\to1}\le2\|H\|$,
the difference between two bounded double commutators is bounded
by $4(\|X\|+\|X'\|)\|X-X'\|$, and the position and potential
supremum errors vanish as the widths decrease. Both dwell channels
preserve the packet-support subspace, so the integrated generator
error bounds the trace-norm error at any fixed dwell time. The
two support regions can be decoded locally into a position label
and an internal spatial register. Tracing out the latter gives
convergence to the two-qubit point-mode state. For the example at
$r=d/6$, one explicit choice is support radius $w=10^{-6}d$.
The potential error is at most $2\mathcal G w/(d-2r)^2$ and each
coordinate error is at most $w$. The corresponding trace-norm
generator error is at most $(429/25)g_0 w/d$. Since
$\kappa=g_0/48$ here, the dwell lasts $24/(5g_0)$ and the
trace-norm output error is less than $8.3\times10^{-5}$.
Partial transposition preserves the Hilbert--Schmidt norm, so the
change in the minimum partial-transpose eigenvalue is bounded by
this trace-norm error. The series inequalities above and
$e^{-0.19584}\ge 1-0.19584$ bound the point-mode eigenvalue above by
$-6\times10^{-4}$, proving strict negativity for these smooth
packets. In a distillation block, widths may be chosen small enough to make the
total approximation error vanish along the sequence of blocks.
Each packet has finite kinetic energy under the stated
ideal-control convention.

These unprotected packet dwells can also be approached if physical
evolution includes the Dirichlet kinetic Hamiltonians. After each
short physical dwell $h$, undo its local kinetic unitary with the
free local controls. The resulting map is
$F_h=\mathcal U_{-h}e^{h(\mathscr A+\mathscr B)}$.
Its interaction-picture integral equation gives
$(F_h-I)/h\to\mathscr B$ strongly on trace class, with a uniform
bound on these difference quotients for small $h$.
This convergence is uniform on the compact trajectory
$\{e^{s\mathscr B}\rho:0\le s\le t\}$ for any fixed input $\rho$.
Comparing $F_h$ with $e^{h\mathscr B}$ at successive points of that
trajectory gives $F_{t/n}^{\,n}\rho\to e^{t\mathscr B}\rho$
by contractivity and telescoping. The physical dwell times sum to
$t$. Thus the strict packet witness persists after sufficiently
many finite local pulses. A finite distillation block is
approximated by choosing the pulse count for its desired error.

Thus the dipole threshold removes the order $g_0u^2$ contribution to
the capacity. The ideal exact force-recovery protocol retains a rate
of order $g_0u^6$, and above the threshold the all-protocol upper
bound is of order $g_0u^3$. The optimal subleading rate is open.

\section{What error detection and correction remove}
\label{sec:correction}
\label{sec:si-correction}

We now restrict attention to a code that is restored after each short
interaction, with the quantum error record discarded. We derive the
projected Hamiltonian, identify which position moments exact correction
removes, and bound the rate left in the full Newtonian series. Alternating
polygon codes attain the resulting distance exponent. We then separate
moment detection from correction and verify that smooth packets approach
the corrected dynamics even when kinetic evolution is retained.

\subsection{Noise protection and the geometry of the retained interaction}
\label{sec:protected-gravity}

Random forces couple to the position of each mass. To correct these forces,
we must encode a logical state in distributions whose first and second
position moments agree. The Newtonian interaction then distinguishes the
logical states through higher moments. We first bound the rate that
survives polynomial position-error correction in three dimensions.
We then show that protection changes the useful geometry: a planar code
produces a larger protected entangling rate than
every axial code, as proved in Section~\ref{sec:si-geometry}. Here we
establish the moment cancellations and recovery limit needed for that
comparison.

\subsection{Protected logical dynamics}

Let $\mathcal G=Gm_Am_B$, and let the two allowed position balls have radius
$r$ and centres separated by $d>2r$. Local position operators are measured
from their respective centres. We first consider independent white force
noise,
\begin{equation}
 \mathcal L_{\rm force}(\rho)
 =-\Lambda\sum_{Q=A,B}\sum_{i=x,y,z}
 [q_{Q,i},[q_{Q,i},\rho]],
 \label{eq:protected-force-generator}
\end{equation}
Here $q_{Q,i}$ is Cartesian position component $i$ in laboratory $Q$,
and $\Lambda\ge0$ is the force-dephasing strength defined in
Section~\ref{sec:si-noise}. Let $P_Q$ denote the orthogonal projector
onto a fixed local protected subspace, including any
local ancillary system used in the encoding. This code corrects the infinitesimal
error span $\{I,q_{Q,x},q_{Q,y},q_{Q,z}\}$ precisely when the
Knill--Laflamme conditions hold:
\begin{equation}
 P_Qq_{Q,i}P_Q=\mu_{Q,i}P_Q,
 \qquad
 P_Qq_{Q,i}q_{Q,j}P_Q=\mu_{Q,ij}P_Q.
 \label{eq:protected-KL}
\end{equation}
The numbers $\mu_{Q,i}$ and $\mu_{Q,ij}$ are independent of the logical
state. Thus the first and second moments cannot reveal which state the
code stores, and the off-diagonal logical matrix elements of these
observables vanish. These operator conditions apply to an arbitrary logical
dimension, not only to two chosen codewords.
We recover locally after every dwell and reset or discard the quantum
syndrome systems. In the limit of vanishing dwell, the retained logical
system evolves under
\begin{equation}
 H_L=(P_A\otimes P_B)V(P_A\otimes P_B),
 \qquad
 V=-\frac{\mathcal G}{|d\mathbf e_z+\mathbf b-\mathbf a|},
 \label{eq:protected-Hamiltonian}
\end{equation}
Here $\mathbf a=\mathbf q_A$ and $\mathbf b=\mathbf q_B$ are the
local positions and $\mathbf e_z$ is the separation direction. The logical
evolution is specified up to local Hamiltonians. This follows either by expanding one recovered
cycle to first order or by the local recovery limit used in
Ref.~\cite{WangPedernalesPlenio2026}. Exact correction here concerns the
infinitesimal error span; finite dwells are treated below.

The protected entangling rate is the ideal ancilla-assisted capacity of
$H_L$, allowing local controls within the protected logical systems.
The bounds below apply to every fixed subspace satisfying
Eq.~\eqref{eq:protected-KL}, and hence also to time sharing between such
subspaces. They concern the retained logical output with quantum syndromes
reset, rather than entanglement stored in discarded error spaces. An axial
code means that both masses are confined to the line joining the two
centres, with each local coordinate in $[-r,r]$.

\subsection{A universal rate bound for polynomial position-error correction}
\label{sec:protected-universal}

Correcting more position errors removes more of the gravitational
interaction. Wang and colleagues identify the surviving higher Newtonian
order for a particular oscillator encoding\cite{WangPedernalesPlenio2026}.
We now bound the entangling rate of every code in the three-dimensional
balls, including codes with ancillary systems. A family of planar codes
then shows that the distance exponent in this bound is attainable.

Let $N\ge1$ be an integer. At each laboratory, suppose that the fixed
code corrects all Cartesian position monomials of total degree at most
$N$. Products of these errors span the polynomials of degree at most
$2N$, so the Knill--Laflamme conditions are equivalent to
\begin{equation}
 P_Qp(\mathbf q_Q)P_Q=c_{Q,p}P_Q,
 \qquad \deg p\le2N,\quad Q=A,B.
 \label{eq:protected-polynomial-KL}
\end{equation}
We retain the projected logical dynamics in
Eq.~\eqref{eq:protected-Hamiltonian}, with quantum syndromes reset.
All gravitational seed preparation is included in the interaction time.
Let $C_{\rm prot}^{(N)}(d,r)$ be the supremum of the ideal capacities over
these fixed local codes and their finite-protocol limits. Local controls
preserve the codes, and arbitrary local memories are allowed. Finite
packet realizations have finite kinetic energy, with no common bound on
energy or recovery bandwidth along an approximating sequence.

Write $M=2N+1$ for the first degree not required to be scalar by correction,
$u=r/d<1/2$, and let $\beta$ have its value in
Eq.~\eqref{eq:newton-beta}. The series $S_M$ bounds all interaction
terms of degree at least $M$ on both sides; $K_M$ will be the coefficient
attained by the alternating polygon construction. We define them by
\begin{align}
 S_M(u)&=\sum_{j,k\ge M}\binom{j+k}{j}u^{j+k},
 \label{eq:protected-universal-series}\\
 K_M&=\frac{2\binom{2M}{M}}{16^M}
       \left[\binom{4M}{2M}+\binom{2M}{M}\right].
 \label{eq:protected-polygon-coefficient}
\end{align}

\begin{theorem}[Universal protected rate and its distance scaling]
\label{thm:protected-universal}
For $d>2r$, the protected capacities satisfy
\begin{equation}
 \frac{\beta\mathcal G}{\hbar d}K_Mu^{2M}
 \le C_{\rm prot}^{(N)}(d,r)
 \le\frac{\beta\mathcal G}{\hbar d}S_M(u).
 \label{eq:protected-universal-capacity}
\end{equation}
The upper bound applies to each code of arbitrary logical dimension,
including arbitrary local ancillary systems. The series obeys
\begin{equation}
 S_M(u)\le\min\left\{
 \frac{(2u)^{2M}}{1-2u},\,
 \frac{\binom{2M}{M}u^{2M}}{(1-2u)^2}\right\},
 \qquad
 S_M(u)=\binom{2M}{M}u^{2M}+O(u^{2M+1}).
 \label{eq:protected-universal-series-bounds}
\end{equation}
Consequently, for fixed $N$ and $r$, the optimal protected rate scales as
$\mathcal G r^{4N+2}/(\hbar d^{4N+3})$ as $d/r\to\infty$,
up to positive constants depending on $N$. The lower bound holds at
every finite separation $d>2r$ as a supremum of finite-energy packet
constructions.
\end{theorem}

\begin{proof}
We first express the potential as a convergent sum of separated
position factors. In this form a corrected local moment removes an entire
class of bipartite terms. For the transverse angle $0\le\phi<2\pi$, define
\[
 \zeta_\phi(\mathbf a)
 =a_z+i(a_x\cos\phi+a_y\sin\phi),\qquad
 |\zeta_\phi(\mathbf a)|\le r,
\]
and use $\langle\cdot\rangle_\phi$ for uniform angular averaging.
The complex coordinate combines the longitudinal position with its
projection onto a transverse direction. Expanding its exponentials will
separate Alice's and Bob's powers. The Laplace--Bessel identity gives
\begin{align}
 \frac1{|d\mathbf e_z+\mathbf b-\mathbf a|}
 &=\int_0^\infty e^{-dt}
   \left\langle e^{t\zeta_\phi(\mathbf a)}
                  e^{-t\zeta_\phi(\mathbf b)}\right\rangle_\phi dt
   \nonumber\\
 &=\frac1d\sum_{j,k\ge0}\binom{j+k}{j}
   \frac{\left\langle\zeta_\phi(\mathbf a)^j
                 [-\zeta_\phi(\mathbf b)]^k\right\rangle_\phi}
        {d^{j+k}}.
 \label{eq:protected-universal-expansion}
\end{align}
Here $t$ is the nonnegative Laplace integration variable, with units of
inverse length, and $j,k$ are the powers on Alice's and Bob's sides.
For the first equality, the angular average is
$e^{t(a_z-b_z)}J_0(t|\mathbf a_\perp-\mathbf b_\perp|)$,
whose Laplace transform is the inverse distance. The bound $d>2r$
allows both the integral and the series to converge absolutely in
operator norm.

Put $A_j=P_A(\zeta_\phi/r)^jP_A$ and
$B_k=P_B(-\zeta_\phi/r)^kP_B$, acting on their respective code spaces.
They are contractions, meaning operators of norm at most one, because
$|\zeta_\phi|\le r$ and compression cannot increase the norm.
Equation~\eqref{eq:protected-polynomial-KL}
makes all terms with $j<M$ or $k<M$ local or scalar. Moreover, replacing
$\phi$ by $\phi+\pi$ conjugates the multiplication factors. Thus the
remaining interaction is the sum in
Eq.~\eqref{eq:protected-universal-expansion} with $j,k\ge M$ and
Hermitian factors $\operatorname{Re}(A_j\otimes B_k)$.

To bound their entangling rate, we use real separated operators. For
any contractions $A,B$, write
\begin{equation}
 \operatorname{Re}(A\otimes B)
 =A_R\otimes B_R-A_I\otimes B_I,
 \qquad
 A_R^2+A_I^2=\frac{A^\dagger A+AA^\dagger}{2}\le I,
 \label{eq:protected-complex-rate-factors}
\end{equation}
where $A_R=(A+A^\dagger)/2$, $A_I=(A-A^\dagger)/(2i)$,
and similarly for $B$. For a pure state of the complete laboratories,
write $S_2$ for its von Neumann entanglement entropy with logarithms in
base two; the subscript denotes the logarithm base, not a R\'enyi order.
The expectation values below are in this state. The product-Hamiltonian moment bound
\eqref{eq:newton-moment-rate} gives
\begin{align}
 |\dot S_2|
 &\le\frac\beta\hbar\left[
 \sqrt{\langle A_R^2\rangle\langle B_R^2\rangle}
 +\sqrt{\langle A_I^2\rangle\langle B_I^2\rangle}\right]
 \nonumber\\
 &\le\frac\beta\hbar
 \sqrt{\langle A_R^2+A_I^2\rangle
        \langle B_R^2+B_I^2\rangle}
 \le\frac\beta\hbar.
 \label{eq:protected-complex-entropy-rate}
\end{align}
The middle step is Cauchy--Schwarz. Compression need not preserve
normality of $A$ or $B$; the contraction identity above still holds.
Summing the positive coefficients of the separated expansion proves
the upper rate in Eq.~\eqref{eq:protected-universal-capacity}.
This step uses the product structure and second moments, rather than
the operator norm of an unrestricted bipartite Hamiltonian.

For arbitrary normal states and code spaces, we use the entropy
regularization described in Section~\ref{sec:newton-fixed-energy}.
The regularized Schmidt-pair logarithms have smaller magnitude, so
Eq.~\eqref{eq:protected-complex-entropy-rate} is uniform in the
regularization parameter. Concavity controls averages over local
measurement outcomes, and Fatou's lemma gives the final entropy bound.
The interaction series has summable operator norms. Every finite
initially separable protocol therefore has average final pure-branch
entropy at most $\beta\mathcal G S_M(u)\mathbb E T_{\rm int}/(\hbar d)$.
Producing $L$ Bell pairs with vanishing error requires at least $L-o(L)$
entropy on their finite output registers. This proves the capacity
bound and its extension to normal limits of finite protocols with
convergent output probabilities and interaction costs. The same bound
also holds under time sharing between admissible codes.

For the series estimates, summing the full binomial row at total degree
$n$ gives $2^n$, proving the first bound in
Eq.~\eqref{eq:protected-universal-series-bounds}. At degree $2M+n$,
each of the $n+1$ retained binomial coefficients is at most
$2^n\binom{2M}{M}$: the largest binomial coefficient grows by at most
a factor two when its row index increases by one. Summing
$(n+1)(2u)^n$ proves the second bound. A truncation after total degree
$L$ has remainder at most $(2u)^{L+1}/(1-2u)$.

We next construct codes that attain the distance exponent. Take the
same orientation of a regular $2M$-gon of radius $a<r$ at both laboratories,
\begin{equation}
 \mathbf a_j=\mathbf b_j
 =a(\sin\theta_j,0,\cos\theta_j),\qquad
 \theta_j=\pi j/M,\quad j=0,\ldots,2M-1,
 \label{eq:protected-general-polygon}
\end{equation}
where $|j\rangle$ labels the orthogonal position mode at vertex $j$.
We form the two logical states by assigning alternate vertices to each:
\begin{equation}
 |0_L\rangle=\frac1{\sqrt M}\sum_{j\ {\rm even}}|j\rangle,
 \qquad
 |1_L\rangle=\frac1{\sqrt M}\sum_{j\ {\rm odd}}|j\rangle.
 \label{eq:protected-general-polygon-code}
\end{equation}
Polynomial moment cancellation and interleaved spatial code principles
have the antecedents in Refs.~\cite{LaydenEtAl2019,JainEtAl2024}.
Here a polynomial of degree less than $M$ has angular Fourier
frequencies below $M$, so its averages over the two parity classes agree.
The codewords have disjoint support, making their off-diagonal moments
zero. Thus they satisfy Eq.~\eqref{eq:protected-polynomial-KL} for all
three coordinates.

We evaluate the full logical coupling to show that its higher orders
reinforce the leading term. Let
$A(\phi)=(1+\cos\phi)/2$, $B(\phi)=(1-\cos\phi)/2$.
On a polygon vertex,
$\zeta_\phi/a=A(\phi)e^{i\theta_j}+B(\phi)e^{-i\theta_j}$.
Let $Z=|0_L\rangle\langle0_L|-|1_L\rangle\langle1_L|$.
For a nonnegative integer $q$, the coefficient of this logical $Z$ in
the $q$th compressed power is
\begin{equation}
 d_q(\phi)=
 \sum_{\substack{0\le l\le q\\q-2l\equiv M\ ({\rm mod}\ 2M)}}
 \binom ql A(\phi)^{q-l}B(\phi)^l.
 \label{eq:protected-polygon-moments}
\end{equation}
This follows by summing the signed root-of-unity phases over the
$2M$ vertices. The coefficient vanishes for $q<M$ and for even $q$,
and is nonnegative for odd $q$. In particular,
$d_M=A(\phi)^M+B(\phi)^M$.

Since Bob's factor in Eq.~\eqref{eq:protected-universal-expansion}
contains $(-1)^k$, every surviving nonlocal term has the same sign.
Writing the compressed Hamiltonian, modulo local terms, as
$\hbar J_M Z_AZ_B$, we obtain
\begin{equation}
 J_M=\frac{\mathcal G}{\hbar d}
 \sum_{j,k\ge M}\binom{j+k}{j}(a/d)^{j+k}
 \langle d_jd_k\rangle_\phi
 \ge\frac{\mathcal G}{\hbar d}K_M(a/d)^{2M}.
 \label{eq:protected-polygon-full-lower}
\end{equation}
All terms in this expression are nonnegative, so the lower bound holds
for every $d>2a$. To evaluate its leading coefficient, use
\[
 \left\langle[A(\phi)^M+B(\phi)^M]^2\right\rangle_\phi
 =\frac{2}{16^M}\left[\binom{4M}{2M}+\binom{2M}{M}\right].
\]
This average is the squared size of the first logical moment, so
multiplying it by $\binom{2M}{M}$ gives $K_M$. The first contribution is the average of
$2\cos^{4M}(\phi/2)$; the cross term is the average of
$2\sin^{2M}\phi/4^M$. This proves
Eq.~\eqref{eq:protected-polygon-coefficient}. Because both nonzero
degrees are odd, the next total degree is $2M+2$. Hence
\begin{equation}
 J_M=\frac{\mathcal G K_M a^{2M}}{\hbar d^{2M+1}}
 +O\!\left(\frac{\mathcal G a^{2M+2}}{\hbar d^{2M+3}}\right).
 \label{eq:protected-polygon-asymptotic}
\end{equation}
The established Ising capacity is $\beta J_M$, including the
amortized cost of generating its seed entanglement.

Finally, we replace the point modes by normalizable states. Choose
$a<r$ and an identical smooth radial packet at every vertex, with
support radius
$w<\min\{r-a,a\sin[\pi/(2M)]\}$. The packets are disjoint and lie
within the allowed balls. Each translated packet moment is a polynomial
in its centre of no higher degree, so the parity cancellation and all
Knill--Laflamme conditions remain exact. The Coulomb kernel is harmonic
on either packet support. Averaging it against a normalized radial
density therefore returns its value at the centre; applying this mean-value
identity successively to both packets preserves each compressed
potential entry. Each codeword has finite kinetic energy. Taking
$a\uparrow r$ with nonzero widths tending to zero proves the lower
supremum in Eq.~\eqref{eq:protected-universal-capacity}. These packets
realize the projected Hamiltonian; maintaining their spatial modes and
correcting their internal error sectors are part of the stipulated
ideal local control.
\end{proof}

For $N=1$, $M=3$ and $K_3=295/32$, so this family recovers the hexagon
coefficient used below. The reflected orientation used there exchanges
Bob's two logical labels and leaves the capacity unchanged. The general upper coefficient is
$\binom63=20$. The cubic refinement in
Section~\ref{sec:protected-cubic-ceiling} reduces it to
$10.17758296\ldots$, placing the planar construction within $10.41\%$
of the optimal leading coefficient over all three-dimensional codes. At the finite separation
$d/r=3.6$, $S_3(1/3.6)=0.0345517560525961\ldots$. Comparing with the
bare capacity in Eq.~\eqref{eq:three-d-capacity}, every protected
three-dimensional code has rate at most $0.154792$ times that capacity.

The entropy proof also bounds the interaction time required by a
statistical test. Let $p$ be the protocol's acceptance probability and
$\alpha$ its largest acceptance probability on a separable output. Write
$d_{\rm bin}(p\|\alpha)=p\ln(p/\alpha)+(1-p)\ln[(1-p)/(1-\alpha)]$
for their binary relative entropy. If every interaction
in an initially separable protocol obeys the correction conditions,
a final retained-logical test of all-separable size $0<\alpha<p\le1$
satisfies
\begin{equation}
 d_{\rm bin}(p\|\alpha)
 \le\frac{\beta\mathcal G}{\hbar d}S_M(u)\ln2\,
        \mathbb E T_{\rm int}.
 \label{eq:protected-universal-confidence}
\end{equation}
The separable-comparator argument includes every recorded outcome.
This concerns the fixed-code projected dynamics; entanglement exported
into quantum syndrome memories requires a different capacity problem.

\subsection{The first retained multipole under moment detection}
\label{analytic:moment-ladder}

A code that detects linear force errors has the same first moments in
both logical states. It can still distinguish second moments, which
appear in the quadrupole--quadrupole part of the interaction. Exact
correction of linear errors also makes their products scalar and
therefore first permits a cubic moment. We quantify this difference
for every moment order in the projected Hamiltonian of
Eq.~\eqref{eq:protected-Hamiltonian}.

Let $M\ge1$ be an integer. At each laboratory, require
$P_Q h(\mathbf q_Q)P_Q$ to be scalar on the code for every real
harmonic polynomial $h$ of degree less than $M$. A harmonic polynomial
satisfies $\nabla^2h=0$. The same compression condition applies
to complex harmonic polynomials by linearity. Let $C_M^{\rm proj}$
be the supremum of the ideal entanglement-generation capacity of
these projected Hamiltonians, with free local operations, ancillary
systems, and seed preparation counted in the time.
Here $g_0=\mathcal G/(\hbar d)$ and $u=r/d$, as in the noisy-capacity
section. Unlike the preceding correction theorem, this condition allows
even $M$: it specifies the moments that are invisible within the code,
without asserting that a noisy evolution can be corrected.

\begin{theorem}[Capacity by order of detected moments]
\label{analytic:ladder}
For every integer $M\ge1$ and $0<u<1/2$,
\begin{equation}
 \beta g_0K_M u^{2M}\le C_M^{\rm proj}\le\beta g_0S_M(u),
 \qquad
 K_M=\frac{2\binom{2M}{M}}{16^M}
             \left[\binom{4M}{2M}+\binom{2M}{M}\right].
 \label{analytic:ladder-bound}
\end{equation}
The upper bound holds for each admissible code, irrespective of
logical dimension. The lower bound is a supremum of finite-energy
packet constructions at every $d>2r$.
\end{theorem}

\begin{proof}
Use the expansion in Eq.~\eqref{eq:protected-universal-expansion}
with $\zeta_\phi(\mathbf a)=a_z+i(a_x\cos\phi+a_y\sin\phi)$.
The coefficient vector of this linear form has complex bilinear
square zero: $1-\cos^2\phi-\sin^2\phi=0$. For $j\ge2$, taking a
Laplacian of the power multiplies that square by
$j(j-1)\zeta_\phi^{j-2}$. The constant and linear cases are harmonic
directly, so $\zeta_\phi^j$ is harmonic for every $j$.
Consequently every term with $j<M$ or $k<M$ becomes local or scalar
upon compression. Each remaining compressed factor
$P_Q(\pm\zeta_\phi/r)^jP_Q$ is a contraction. Pairing
$\phi$ with $\phi+\pi$ takes the real part of the separated
product, with row bound one by
Eqs.~\eqref{eq:protected-complex-rate-factors} and \eqref{analytic:rows}.
The sum of the row bounds is $(\mathcal G/d)S_M(u)$.
Apply the entropy and extraction argument above with unitary
dwells to obtain the upper bound.

For the lower bound, place the same regular $2M$-gon of radius
$a<r$ in the $xz$ plane in each laboratory, with vertices
$a(\sin\theta_n,0,\cos\theta_n)$ and $\theta_n=\pi n/M$,
$n=0,\ldots,2M-1$. The logical states are equal superpositions
over the even and odd vertices, respectively. A polynomial of
degree less than $M$ on the vertices has angular frequencies below
$M$, so its two parity-class averages agree. The two logical
states have disjoint support, making all position-polynomial
off-diagonal elements vanish.

On a vertex, $\zeta_\phi/a=A e^{i\theta_n}+B e^{-i\theta_n}$ with
$A=(1+\cos\phi)/2$ and $B=(1-\cos\phi)/2$.
The logical $Z$ coefficient of the compressed $q$th power of
$\zeta_\phi/a$ is the signed Fourier average
\[
 d_q(\phi)=
 \sum_{\substack{0\le l\le q\\q-2l\equiv M\ ({\rm mod}\ 2M)}}
 \binom ql A^{q-l}B^l .
\]
Every summand is nonnegative, and a term can occur only when
$q\ge M$ and $q\equiv M\pmod 2$. In the full Newtonian series the
sign of Bob's surviving terms is therefore always $(-1)^M$.
Thus the magnitude of the logical Ising frequency receives a
nonnegative contribution from every surviving degree,
\[
 |J_M|=g_0\sum_{j,k\ge M}\binom{j+k}{j}(a/d)^{j+k}
                  \langle d_j d_k\rangle_\phi .
\]
The lowest term, at $j=k=M$, uses $d_M=A^M+B^M$.
Integration of the powers of $\cos(\phi/2)$ and $\sin\phi$ gives
$\langle d_M^2\rangle_\phi=
2[\binom{4M}{2M}+\binom{2M}{M}]/16^M$. This term alone gives
$|J_M|\ge g_0 K_M(a/d)^{2M}$, and the projected Ising Hamiltonian
has capacity $\beta|J_M|$.

Replace each vertex by the translate of the same smooth radial
packet with support radius
$0<w<\min\{r-a,a\sin[\pi/(2M)]\}$. The averaged polynomial at a
vertex is a polynomial in its centre of no higher degree, so
the moment cancellation remains exact. The Coulomb kernel is
harmonic separately in each position throughout the packet
supports. Integrating over a normalized radial density gives
its value at the centre by the harmonic mean-value identity.
Applying that identity to each packet in turn preserves all
compressed potential entries. The packets have finite kinetic
energy and $a\uparrow r$ gives the lower supremum.
\end{proof}

Since $S_M(u)=\binom{2M}{M}u^{2M}+O(u^{2M+1})$, the optimal
projected capacity scales as $\mathcal G r^{2M}/(\hbar d^{2M+1})$
at fixed $r$ and large $d$. For $M=1$ the exact leading coefficient
in units of $\beta g_0u^2$ is two. For $M=2$ it lies between
$57/16$ and $6$, with a projected rate of order $d^{-5}$. The
four-vertex construction encodes one logical state at the
two axial vertices and the other at the two transverse vertices.
This projected result requires only $P_Qq_{Q,i}P_Q\propto P_Q$.
A noisy implementation must also account for disturbance and
any rejected outcomes. If a code corrects all position errors
through degree $N$, Knill--Laflamme tests their products
through degree $2N$ and $M=2N+1$. The exponent becomes $4N+3$
in distance, as in Theorem~\ref{thm:protected-universal}.
In particular, correcting linear errors yields the $d^{-7}$ law
under the ideal recovery convention.

\paragraph{Exact force recovery with supported kinetic evolution.}
The projected Hamiltonian becomes a physical limiting dynamics only if
the noise disappears under the recovered cycles. We show this directly
by differentiating one encoded dwell followed by a decoder. This also
identifies why the kinetic and force terms contribute no logical evolution.
For $M=3$, the smooth translated packets used in the polygon
construction obey the full Knill--Laflamme conditions
in Eq.~\eqref{eq:protected-KL}. Here we verify that they approach
the ideal projected Ising rate for every finite pair of force
strengths even when the kinetic Hamiltonians act during each
dwell. In each laboratory let $E_0=I$ and $E_i=q_i$ for the
three position coordinates. The relations
$P_Q E_i^\dagger E_j P_Q=G_{ij}P_Q$ identify the span of the
error images with a logical qubit tensored with a syndrome
space. Here $G_{ij}$ is the scalar Gram matrix of these error images.
Explicitly, $E_i|\psi_L\rangle$ maps isometrically to
$|\psi\rangle|e_i\rangle$, where
$\langle e_i|e_j\rangle=G_{ij}$. Extend this isometry to the
whole local Hilbert space using an additional orthogonal
syndrome space. Trace out the syndrome to obtain the decoder
$\mathcal D_Q$. A code vector is mapped to its logical vector
and the fixed unit syndrome $|e_0\rangle$, so every vector
orthogonal to the code has zero component in that syndrome.
The decoder therefore annihilates cross operators between
the code and its complement.

Let $\mathcal E$ be the joint encoding channel and $\mathcal D$
the product decoder. We use $\mathscr A$ for the kinetic generator and
$\mathscr B=-i[V,\cdot]/\hbar+\mathcal L_{\rm force}$ for the bounded
position generator, as in Section~\ref{sec:si-noise}; $\tau$ is the
duration of one dwell. Every code matrix unit belongs to the
trace-class generator domain because its ket and bra are smooth
and supported strictly inside the balls. The expansion on these
matrix units gives
\[
 \mathcal D e^{\tau(\mathscr A+\mathscr B)}\mathcal E
 = {\rm id}+\tau\mathscr G_L+o(\tau),\qquad
 \mathscr G_L=-\frac{i}{\hbar}[PVP,\cdot],
\]
where $P=P_A\otimes P_B$ and $PVP$ is represented on the logical
qubits. The remainder is $o(\tau)$ in diamond norm since the
logical input is finite dimensional. The cross-operator
cancellation compresses each Hamiltonian in the derivative.
Equal radial packet profiles have equal kinetic diagonal
matrix elements, and disjoint supports give zero off-diagonal
ones. Thus $P_QK_QP_Q$ is scalar. For each force term,
decoding $q_i\mathcal E_Q(\rho)q_i$ gives $G_{ii}\rho$,
and decoding either one-sided multiplication by $q_i^2$ also
gives $G_{ii}\rho$ by compression. These three terms cancel
in the double commutator. These calculations hold with
identities on memories.

Repeating the decoded cycle with a fresh local encoding at
the start of each dwell converges in diamond norm to
$e^{t\mathscr G_L}$ as $\tau=t/n\to0$, by telescoping against
the ideal logical evolution and summing $n\,o(t/n)$.
The syndrome is reset at each cycle and the charged time is
$t$. For any finite Ising-capacity block a finite number of
ideal local recoveries therefore approximates its output to
any prescribed accuracy. The projected coupling is $|J_3|$
for the inward hexagons and tends to $g_{\rm hex}$ of
Eq.~\eqref{eq:protected-hexagon-g} as their centres approach
the radius $r$. The rate $\beta g_{\rm hex}$ is attainable
as a supremum in this control model.

\section{Optimizing the protected spatial encoding}
\label{sec:geometry}
\label{sec:si-geometry}

Exact force correction leaves cubic moments as the first possible logical
signal. We now optimize that signal over spatial encodings. We first bound
all three-dimensional codes, then determine the leading optimum for planar
codes and for codes whose leading interaction is a single product. Finally,
we compare the best axial coefficient with the hexagon and give a strict
advantage at finite separation. Throughout, capacities refer to the retained
logical Hamiltonian with the correction convention of
Section~\ref{sec:si-correction}.

\subsection{A tighter three-dimensional bound for linear-force protection}
\label{sec:protected-cubic-ceiling}

For codes correcting linear position errors, the first retained
interaction is cubic in each laboratory. We can bound this term more
sharply by removing a linear function before estimating its size.
This subtraction changes the protected Hamiltonian only by local terms.
We seek separate subtraction coefficients $a_0$ and $b_0$ for the
longitudinal and transverse components. The following choice makes the
squared-error bound factor into nonnegative terms; $s_0$ will identify a
latitude where equality holds, and $\gamma$ is the resulting radius in
units of $r^3$:
\begin{align}
 a_0&=\frac{1310-47\sqrt{730}}{140},&
 b_0&=\frac{165-6\sqrt{730}}{14},\nonumber\\
 s_0&=\frac{117-4\sqrt{730}}{35},&
 \gamma&=1-a_0=0.713357657814227335\ldots.
 \label{eq:protected-cubic-constants}
\end{align}
Only the cubic factor is improved by this subtraction. Let
$\gamma_3=\gamma$ and $\gamma_j=1$ for $j\ge4$, and replace the
universal series $S_3$ by the correspondingly weighted series
\begin{equation}
 \widetilde S_3(u)=\sum_{j,k\ge3}\binom{j+k}{j}
                    \gamma_j\gamma_k u^{j+k},\qquad 0<u<\tfrac12.
 \label{eq:protected-cubic-series}
\end{equation}

\begin{proposition}[Three-dimensional cubic ceiling]
\label{prop:protected-cubic-ceiling}
For every fixed code in the two radius-$r$ balls satisfying the exact
linear-position correction conditions, with arbitrary logical dimension
and local ancillary systems, the ideal entangling capacity of the retained
logical Hamiltonian obeys
\begin{equation}
 C_L\le\frac{\beta\mathcal G}{\hbar d}\widetilde S_3(r/d).
 \label{eq:protected-cubic-ceiling}
\end{equation}
In particular, uniformly over these codes,
\begin{equation}
 C_L\le\frac{\beta\mathcal G r^6}{\hbar d^7}
 \left[20\gamma^2+O(r/d)\right],\qquad
 20\gamma^2=10.17758295924400515\ldots.
 \label{eq:protected-cubic-leading}
\end{equation}
Quantum syndromes are reset or discarded, as in the protected-Hamiltonian
model. This is not a bound on all outputs of a finite-rate recovery
protocol that retains entangled syndrome memories.
\end{proposition}

\begin{proof}
For $t_\phi=a_x\cos\phi+a_y\sin\phi$, write
$q_\phi=a_z+i t_\phi$ and
\[
 f_\phi(\mathbf a)=q_\phi^3
                  -r^2(a_0a_z+i b_0t_\phi).
\]
Here $t_\phi$ is a transverse coordinate, not a dwell time, and
$f_\phi$ is the cubic feature after the permitted linear subtraction.
Angular averages are normalized over $[0,2\pi]$. On $|\mathbf a|=r$,
put $s=a_z^2/r^2$. Using
$\langle t_\phi^{2}\rangle=(r^2-a_z^2)/2$,
$\langle t_\phi^{4}\rangle=3(r^2-a_z^2)^2/8$ and
$\langle t_\phi^{6}\rangle=5(r^2-a_z^2)^3/16$, direct expansion gives
\begin{equation}
 \gamma^2-r^{-6}\langle|f_\phi(\mathbf a)|^2\rangle_\phi
 =\frac5{16}(1-s)(s-s_0)^2\ge0.
 \label{eq:protected-cubic-certificate}
\end{equation}
Since $0\le s\le1$, the right side is nonnegative. This proves the
desired root-mean-square bound on the sphere, with equality at $s=1$
and $s=s_0$. We still need the same bound at interior positions.
For each $\phi$, both parts of $f_\phi$ are harmonic: in the orthogonal
coordinates $(a_z,t_\phi)$ they are
$a_z^3-3a_zt_\phi^2-r^2a_0a_z$ and
$3a_z^2t_\phi-t_\phi^3-r^2b_0t_\phi$.
For a real harmonic function $h$, $\nabla^2h^2=2|\nabla h|^2\ge0$.
Applying this to both parts shows that
$\langle|f_\phi|^2\rangle_\phi$ is subharmonic, meaning its Laplacian
is nonnegative.
The maximum principle extends the sphere bound to the entire ball.

Set $f_{3,\phi}=f_\phi$ and $f_{j,\phi}=q_\phi^j$ for $j\ge4$.
For a local code projector $P$, put
$F_{j,\phi}=P f_{j,\phi}P$ on its code space.
We next transfer the position-wise bound to arbitrary logical states.
Compression of the self-adjoint real and imaginary parts gives
\begin{equation}
 \left\langle (\operatorname{Re}F_{j,\phi})^2+
                  (\operatorname{Im}F_{j,\phi})^2\right\rangle_\phi
 \le P\langle|f_{j,\phi}|^2\rangle_\phi P
 \le\gamma_j^2r^{2j}I.
 \label{eq:protected-cubic-sos}
\end{equation}
The first inequality follows from
$(PXP)^2\le PX^2P$ for self-adjoint $X$. It does not assume that the
compressed complex operator is normal. The $j\ge4$ bound uses
$|q_\phi|\le r$. Tensoring arbitrary local ancillary identities
preserves these inequalities.

Use the absolutely convergent angular expansion of the Newtonian kernel,
\[
 \frac1{|d\mathbf e_z+\mathbf b-\mathbf a|}
 =\frac1d\sum_{j,k\ge0}\binom{j+k}{j}d^{-(j+k)}
   \left\langle q_\phi(\mathbf a)^j[-q_\phi(\mathbf b)]^k
   \right\rangle_\phi.
\]
Exact correction makes every compressed polynomial of degree at most two
scalar. Terms with $j<3$ or $k<3$ are therefore local. Replacing a cubic
factor by $f_{3,\phi}$ also changes only local terms, because its
subtraction is linear. The factor $(-1)^k$ and the sign in
$\operatorname{Re}(A\otimes B)=A_R\otimes B_R-A_I\otimes B_I$
do not affect the following estimate.

For any real separated interaction
$H=\int\sum_\nu A_{\nu,\phi}\otimes B_{\nu,\phi}\,d\phi/(2\pi)$
with
$\int\sum_\nu A_{\nu,\phi}^2d\phi/(2\pi)\le Q_A^2I$
and the analogous bound $Q_B^2I$, its matrix elements in paired Schmidt
bases satisfy
\[
 \sum_j|\langle u_iv_i|H|u_jv_j\rangle|\le Q_AQ_B.
\]
This follows by Cauchy--Schwarz over $(j,\nu,\phi)$ and completeness of
the local bases. The Schmidt-pair entropy inequality then gives
$|\dot S_2|\le\beta Q_AQ_B/\hbar$.
Applying this once to both real and imaginary terms of each $(j,k)$
angular average, using Eq.~\eqref{eq:protected-cubic-sos}, bounds its
rate by $\beta\gamma_j\gamma_k r^{j+k}/\hbar$.
Summing the positive coefficient bounds proves
Eq.~\eqref{eq:protected-cubic-ceiling}. No factor of two is required.

The estimates are independent of the number of Schmidt coefficients.
For normal infinite-dimensional encodings, the same argument applies to
the regularized entropy with derivative logarithms
$\log_2[(p_i+\varepsilon)/(p_j+\varepsilon)]$.
Their magnitudes are bounded by the unregularized logarithms.
Concavity gives average monotonicity under local instruments, and
Fatou's lemma passes the integrated estimate to finite output-register
entanglement as $\varepsilon\downarrow0$. This gives the same
ideal-control capacity bound with arbitrary local memories.
Finally, the first term of the series is $20\gamma^2u^6$; its positive
tail is uniformly bounded by the corresponding geometric tail in
$2u$, proving Eq.~\eqref{eq:protected-cubic-leading}.
\end{proof}

The planar code already has leading coefficient $295/32=9.21875$.
It therefore achieves at least
\[
 \frac{295/32}{20\gamma^2}=0.9057897181399907\ldots
\]
of the optimal three-dimensional protected rate in the far-separation
limit. Thus its leading rate is at least $90.57\%$ of the unrestricted
protected-code optimum under these assumptions. This brackets the
optimum; it does not identify the optimal coefficient or prove that the
planar code is globally optimal over three-dimensional codes.

To evaluate the full-distance bound, we can sum the correction to the
universal series explicitly. Only terms with $j=3$ or $k=3$ change;
their intersection at $(3,3)$ must be counted once. If
$S_3(u)=\sum_{j,k\ge3}\binom{j+k}{j}u^{j+k}$, then
\begin{equation}
 \widetilde S_3(u)=S_3(u)
 -2(1-\gamma)u^3[(1-u)^{-4}-1-4u-10u^2]
 +20(1-\gamma)^2u^6.
 \label{eq:protected-cubic-closed}
\end{equation}
Truncation through total degree $L\ge6$ has remainder at most
$(2u)^{L+1}/(1-2u)$. The finite-distance bound is exact as an inequality;
the $90.57\%$ comparison concerns its leading far-field coefficient.
\subsection{The exact leading optimum for planar codes}
\label{sec:protected-planar-optimum}

The hexagon's leading coefficient can be optimized over all planar
encodings. The comparison allows arbitrary logical dimension, local
ancillas and noncommuting compressed moments. We first establish the
planar result, then show that the same optimum applies in three
dimensions whenever the leading logical interaction is a single product.

Normalize positions by the support radius $r$. For an encoding isometry
$W_Q$ from the logical space into the physical laboratory, write
$\Phi_Q(f)=W_Q^\dagger f(\mathbf x_Q)W_Q$, where
$|\mathbf x_Q|\le1$ and local ancillary identities are understood.
The force-error correction conditions are
\[
 \Phi_Q(x_i)=\mu_{Q,i}I,\qquad
 \Phi_Q(x_ix_j)=S_{Q,ij}I.
\]
The coefficients $\mu_{Q,i}$ and $S_{Q,ij}$ are scalar first and second
moments. Put $\Omega=\mathcal G r^6/(\hbar d^7)$ for the frequency
scale of the first retained term. To keep its angular components together,
we use a real Hilbert-space feature $F$ represented by the complex function
\begin{equation}
 f_\phi(x,y,z)=\sqrt{20}\,[z+i(x\cos\phi+y\sin\phi)]^3,
 \qquad
 \langle F(\mathbf x),F(\mathbf y)\rangle
 =\operatorname{Re}\langle f_\phi(\mathbf x)
                          \overline{f_\phi(\mathbf y)}\rangle_\phi.
 \label{eq:protected-planar-feature}
\end{equation}
The inner product takes the real part of the normalized angular average.
Its seven real coordinates $F_\ell$ in an orthonormal basis are harmonic cubic
polynomials. Reflect Bob's normalized position by
$R(x,y,z)=(x,y,-z)$. The cubic-by-cubic term of the Newtonian expansion is
\begin{equation}
 H_6=-\hbar\Omega\sum_\ell
       \Phi_A(F_\ell)\otimes\Phi_B(F_\ell\circ R)
       \quad\text{modulo local terms}.
 \label{eq:protected-planar-H6}
\end{equation}
The subscript six denotes total position degree three on each side.
The coefficient $20=6!/(3!)^2$ follows from
Eq.~\eqref{eq:protected-universal-expansion}. Subtracting any polynomial
of degree at most two from a feature changes only local terms after
compression.

\begin{proposition}[Planar optimality at leading order]
\label{prop:protected-planar-optimum}
Suppose each laboratory's position support lies in an affine plane
inside its unit ball. The planes may have different orientations and
offsets. Under the fixed-code projected dynamics,
\begin{equation}
 C(H_6)\le\beta\Omega K_{\rm pl},
 \qquad K_{\rm pl}=\frac{295}{32}.
 \label{eq:protected-planar-optimum}
\end{equation}
The hexagon attains this coefficient in the ideal position-mode model.
No larger planar code or local ancillary system increases it.
\end{proposition}

\begin{proof}
We first consider a plane through the ball's centre. Rotation about the
separation axis allows us to parameterize its unit circle as
\[
 \mathbf n(\theta)
 =(\sqrt{1-t}\cos\theta,\sin\theta,\sqrt t\cos\theta),
 \qquad 0\le t\le1.
\]
Here $\theta$ runs around the circle, and $t$ fixes the plane's tilt:
$t=0$ is transverse to the separation axis, while $t=1$ contains it.
Only the first and third angular harmonics occur in a cubic feature, so
\begin{equation}
 F(\mathbf n(\theta))
 =L_c\cos\theta+L_s\sin\theta+U\cos3\theta+V\sin3\theta.
 \label{eq:protected-plane-fourier}
\end{equation}
The vectors $L_c,L_s,U,V$ are coefficients in the feature Hilbert space.
The first two terms define a linear feature $L_1$ on the plane and can
therefore be subtracted without changing the nonlocal logical interaction.
The third-harmonic vectors $U,V$ contain the part we must bound.
To evaluate the four coefficient vectors, put
$a_\phi=\sqrt t+i\sqrt{1-t}\cos\phi$ and $b_\phi=i\sin\phi$.
Their representing functions are
\begin{align*}
 L_c&:\ \frac{3\sqrt{20}}4(a_\phi^3+a_\phi b_\phi^2),&
 L_s&:\ \frac{3\sqrt{20}}4(a_\phi^2b_\phi+b_\phi^3),\\
 U&:\ \frac{\sqrt{20}}4(a_\phi^3-3a_\phi b_\phi^2),&
 V&:\ \frac{\sqrt{20}}4(3a_\phi^2b_\phi-b_\phi^3).
\end{align*}
Angular averaging gives $U\perp V$ and $L_c\perp L_s$, and
\begin{align}
 \|U\|^2&=\frac5{64}(5t^3+45t^2+60t+8),&
 \|V\|^2&=\|U\|^2-\frac{25t^3}{64},\nonumber\\
 \|L_c\|^2&=\frac{45}{64}(5t^3-11t^2+4t+8),&
 \|L_s\|^2&=\frac{45}{64}(5t^2-12t+8).
 \label{eq:protected-plane-fourier-norms}
\end{align}
These identities can be obtained by expanding the displayed functions
and using
\[\langle\cos^{2m}\phi\sin^{2n}\phi\rangle_{\phi}
=(2m)!(2n)!/[4^{m+n}m!n!(m+n)!],\] an odd power averages to zero.

The first polynomial in Eq.~\eqref{eq:protected-plane-fourier-norms}
increases on $[0,1]$, so $\|U\|^2,\|V\|^2\le K_{\rm pl}$.
For the linear coefficients,
$5t^3-11t^2+4t+8\le8+4t-6t^2\le26/3$ and
$5t^2-12t+8\le8$. Thus
$\sup_\theta\|L_1(\mathbf n(\theta))\|^2\le195/32<K_{\rm pl}$.
Bounding the boundary circle alone is not sufficient because subtracting
$L_1$ destroys cubic homogeneity. For an interior point $\rho\mathbf n$,
with $0\le\rho\le1$, the cubic and linear degrees give
\[
 F(\rho\mathbf n)-L_1(\rho\mathbf n)
 =\rho^3[F(\mathbf n)-L_1(\mathbf n)]
   -\rho(1-\rho^2)L_1(\mathbf n).
\]
The triangle inequality therefore bounds its norm by
$[\rho^3+\rho(1-\rho^2)]\sqrt{K_{\rm pl}}
=\rho\sqrt{K_{\rm pl}}\le\sqrt{K_{\rm pl}}$ throughout the disk.

For an affine plane, write $\mathbf x=\mathbf c+\mathbf v$, where
$\mathbf c$ is its point closest to the origin. Its available disk has
radius $R_0=\sqrt{1-|\mathbf c|^2}\le1$. Since $F$ is cubic,
$F(\mathbf c+\mathbf v)-F(\mathbf v)$ has degree at most two in
$\mathbf v$ and therefore compresses to a scalar vector. Subtract this
vector polynomial and the linear term $R_0^2L_1(\mathbf v)$.
The remaining feature $G$ satisfies
$\|G\|^2\le R_0^6K_{\rm pl}\le K_{\rm pl}$ on the support.
The factor $R_0^6$ is the square of the cubic rescaling from a unit disk
to a disk of radius $R_0$. The zero-radius case is local and follows by continuity.

Let $A_\ell=\Phi_A(G_{A,\ell})$ and
$B_\ell=\Phi_B(G_{B,\ell})$, with the reflected Bob plane treated
in the same way. The compression inequality gives
\[
 \sum_\ell A_\ell^2
 \le\Phi_A\!\left(\sum_\ell G_{A,\ell}^2\right)
 \le K_{\rm pl}I,
 \qquad \sum_\ell B_\ell^2\le K_{\rm pl}I.
\]
Indeed, for each real multiplication operator $g$,
$W^\dagger g^2W-(W^\dagger gW)^2
=W^\dagger g(I-WW^\dagger)gW\ge0$.
Applying Eq.~\eqref{eq:newton-moment-rate} to each separated term of
$H_6$ and then Cauchy--Schwarz over $\ell$ bounds the entropy rate by
\[
 \beta\Omega\sum_\ell
 \sqrt{\langle A_\ell^2\rangle\langle B_\ell^2\rangle}
 \le\beta\Omega
 \sqrt{\left\langle\sum_\ell A_\ell^2\right\rangle
        \left\langle\sum_\ell B_\ell^2\right\rangle}
 \le\beta\Omega K_{\rm pl}.
\]
No commutativity of the compressed moments was used. The normal-protocol
and finite-register limit argument of
Theorem~\ref{thm:protected-universal} proves the capacity statement.

For the vertical plane $t=1$, the alternating hexagon code has a
logical cubic feature proportional to $U$, with $\|U\|^2=295/32$.
Its leading Hamiltonian is an Ising interaction of that strength, so
its capacity is $\beta\Omega K_{\rm pl}$. This proves equality in the
planar position-mode model. The identical radial packets described
above preserve the compressed harmonic features at their centre values
and approach the same coefficient with finite energy in each
approximation; their supports are thin three-dimensional regions.
\end{proof}

We have optimized the degree-six term. To bound the full potential at a
finite separation, retain the universal estimate for all higher terms and
replace its degree-six coefficient $20$ by $K_{\rm pl}$. Thus the planar comparison also
has the finite-distance upper bound
\begin{equation}
 C_{\rm planar}(d,r)\le\frac{\beta\mathcal G}{\hbar d}
 \left[K_{\rm pl}u^6+S_3(u)-20u^6\right],\qquad u=r/d.
 \label{eq:protected-planar-full-bound}
\end{equation}
In particular, separation-dependent choices of planes and codes cannot
increase the leading coefficient. Equation~\eqref{eq:protected-planar-optimum}
determines the leading optimum; optimizing the full finite-distance
capacity is a further problem.

\subsection{Single-product logical interactions in three dimensions}
\label{sec:protected-product-optimum}

We can remove the planar-support restriction when the leading logical
interaction is a single product Hamiltonian. The proof bounds the
gravitational energy difference between two local logical states, then
uses the established capacity formula for a product interaction.

\begin{proposition}[Leading optimum for a product interaction]
\label{prop:protected-product-optimum}
Under the force-error correction conditions, suppose that the leading
logical Hamiltonian satisfies $H_6=X\otimes Y$ modulo local terms,
with bounded Hermitian factors $X,Y$. Positions may occupy the full
three-dimensional balls and the logical systems may have arbitrary
dimension and local ancillas. Then
\begin{equation}
 C(H_6)\le\beta\Omega\frac{295}{32}.
 \label{eq:protected-product-optimum}
\end{equation}
The hexagon attains this coefficient.
\end{proposition}

\begin{proof}
Choose any two normalized local logical states and let $\mu_0,\mu_1$
be their position probability measures on the unit ball. Their
half-difference $\nu=(\mu_0-\mu_1)/2$ annihilates every polynomial of
degree at most two, and its total variation is at most one. Let
$\iota(\mathbf x)=-\mathbf x$ and
$\nu_o=(\nu-\iota_*\nu)/2$, where $\iota_*\nu$ is the reflected
measure. We keep only this odd part because $F$ is odd:
$\int F\,d\nu_o=\int F\,d\nu$. The positive and negative Jordan
parts are the nonnegative measures whose difference is $\nu_o$; they
are antipodal images. If the positive part has mass
$m>0$, its normalized probability measure $\pi$ therefore satisfies
\[
 \int\mathbf x\,d\pi=0,\qquad
 \int F\,d\nu=2m\int F\,d\pi,\qquad 2m\le1.
\]
The first identity follows because $\nu_o$ annihilates linear functions;
the other two express its feature difference and total-variation bound.
The case $m=0$ has zero feature difference. Thus it suffices to bound
the feature of a zero-mean probability measure.

We move this measure to the sphere without changing its linear or
harmonic cubic moments. An interior point $\mathbf x$ is replaced by
its Poisson harmonic measure
\[
 d\omega_{\mathbf x}(\mathbf n)
 =\frac{1-|\mathbf x|^2}{4\pi|\mathbf n-\mathbf x|^3}\,dS(\mathbf n),
 \qquad |\mathbf n|=1;
\]
a boundary point is left unchanged. The harmonic representation formula
preserves both types of moments. Write the resulting symmetric tensor
as $T_{ijk}=\mathbb E[n_in_jn_k]$. Since
$\mathbb E\mathbf n=0$ and $|\mathbf n|=1$, it is trace-free:
$\sum_iT_{iik}=0$.

For any unit vector $\mathbf u$, the degree-three Chebyshev inequality
gives
\[
 |T(\mathbf u,\mathbf u,\mathbf u)|
 =\left|\mathbb E\left[(\mathbf u\cdot\mathbf n)^3
                    -\frac34(\mathbf u\cdot\mathbf n)\right]\right|
 \le\frac14.
\]
Banach's theorem for real symmetric tensors identifies the norm over
three independent unit vectors with this diagonal norm
\cite{FriedlandLim2018}. Hence
$|T(\mathbf u,\mathbf v,\mathbf w)|\le1/4$ for all unit
$\mathbf u,\mathbf v,\mathbf w$.

We now collect four tensor norms that determine the gravitational feature.
Define $(T_z)_{ij}=T_{ijz}$ and $(T_{zz})_i=T_{izz}$; the subscript $z$
contracts an index with the separation direction. We write $\|\cdot\|_F$
for the Frobenius norm, the square root of the sum of squared components.
The symmetric matrix $T_z$ is trace-free and has eigenvalues in
$[-1/4,1/4]$. Maximizing their squared sum under these constraints
gives $\|T_z\|_F^2\le1/8$, attained by eigenvalues
$1/4,0,-1/4$. The same multilinear bound gives
$\|T_{zz}\|^2\le1/16$ and $T_{zzz}^2\le1/16$. Finally,
\[
 \|T\|_F^2=\mathbb E\,T(\mathbf n,\mathbf n,\mathbf n)\le\frac14.
\]
In the last identity, contracting $T_{ijk}=\mathbb E[n_in_jn_k]$
with itself gives the expectation shown, and the diagonal tensor bound
controls its value. Expanding Eq.~\eqref{eq:protected-planar-feature} and using the
trace-free identities gives the positive quadratic form
\begin{align}
 Q(T):=\left\|\mathbb EF(\mathbf n)\right\|^2
 &=\frac52\|T\|_F^2+\frac{75}{2}\|T_z\|_F^2
   +\frac{225}{4}\|T_{zz}\|^2+\frac{25}{4}T_{zzz}^2
 \nonumber\\
 &\le\frac{295}{32}.
 \label{eq:protected-cubic-tensor-form}
\end{align}
One direct way to verify the identity is to expand
$20\langle|T(\mathbf v_\phi,\mathbf v_\phi,\mathbf v_\phi)|^2\rangle_{\phi}$
with $\mathbf v_\phi=(i\cos\phi,i\sin\phi,1)$ and use the angular
moments following Eq.~\eqref{eq:protected-plane-fourier-norms}.
Since $2m\le1$, the same norm bound holds for $\int F\,d\nu$.
It also holds after reflecting Bob's measure by $R$.

Apply this result to half-differences on the two laboratories. Let $E_{uv}$
be the expectation of $H_6$ in the chosen product of local states $u,v$,
with $u,v\in\{0,1\}$. The quarter rectangular difference
$(E_{00}+E_{11}-E_{01}-E_{10})/4$ cancels all local energies and has magnitude at most
$\hbar\Omega\sqrt{Q(T_A)Q(T_B)}\le\hbar\Omega(295/32)$.
For $H_6=X\otimes Y$ modulo local terms, choose states approaching the
maximum and minimum of each factor's spectrum. This rectangle equals
$\Delta_X\Delta_Y/4$, where $\Delta_X,\Delta_Y$ are the spectral
diameters. The product-Hamiltonian capacity
$\beta\Delta_X\Delta_Y/(4\hbar)$\cite{ChildsLeungVidal2004}
therefore proves Eq.~\eqref{eq:protected-product-optimum}.

The uniform triangle with vertices
$(0,0,1)$ and $(\pm\sqrt3/2,0,-1/2)$ has
$T_{zzz}=1/4$, $T_{zxx}=-1/4$ and its symmetric permutations, with
all remaining components zero. It saturates every tensor bound above.
This triangle and its negative are the two parity classes of the
hexagon, proving attainment.
\end{proof}

The scalar rectangle bound does not determine the capacity of a sum of
noncommuting product interactions. Indeed, force-error correction does
not make the compressed cubic moments commute. On seven equally spaced
points of a circle, the Fourier modes $0$ and $3$ span an exact
correcting code: position polynomials of degree at most two have only
Fourier frequencies $-2,\ldots,2$ and hence compress to scalar matrices.
Nevertheless, the compressed operators $\cos3\theta$ and $\sin3\theta$
are $X/2$ and $-Y/2$ in this basis. Thus any improvement over the
hexagon requires a nonplanar support on at least one side and a
nonproduct leading interaction. The optimum over that remaining class
is not fixed by the two propositions.

\subsection{A bound for every protected axial code}

We next compare the planar optimum with codes confined to the
centre-to-centre axis. For this geometry we can bound each retained
position power more tightly than by its maximum magnitude. We remove the
first three Chebyshev terms from each position monomial. Their
compression is scalar by Eq.~\eqref{eq:protected-KL}; the remaining terms
therefore bound the part that can entangle the logical systems.

\begin{theorem}[Protected axial rate]
\label{thm:protected-axial}
Put $u=r/d<1/2$. Let $e_k$ be the sum of the Chebyshev coefficients
left in the monomial $t^k$ after removing degrees zero, one and two;
$F(u)$ sums the resulting bipartite bounds. Explicitly,
\begin{align}
 e_k&=0\quad (k=0,1,2),\nonumber\\
 e_k&=
 \begin{cases}
 1-2^{1-k}\binom{k}{(k-1)/2},&k\ge3\text{ odd},\\
 1-2^{-k}\binom{k}{k/2}-2^{1-k}\binom{k}{k/2-1},
       &k\ge4\text{ even},
 \end{cases}\label{eq:protected-ek}\\
 F(u)&=\sum_{k,l\ge3}\binom{k+l}{k}e_ke_lu^{k+l}.
 \label{eq:protected-F}
\end{align}
Every protected axial code, of arbitrary logical dimension and with
arbitrary local ancillary systems, has entangling rate
\begin{equation}
 C_{\rm ax}^{\rm prot}\le
 \frac{\beta\mathcal G}{\hbar d}F(u),
 \qquad \beta=1.912273288953718\ldots.
 \label{eq:protected-axial-rate}
\end{equation}
If $F_M$ retains terms of total degree at most the truncation order $M$, then
\begin{equation}
 0\le F(u)-F_M(u)\le\frac{(2u)^{M+1}}{1-2u}.
 \label{eq:protected-F-tail}
\end{equation}
The leading constant is sharp:
\begin{equation}
 \sup C_{\rm ax}^{\rm prot}
 =\frac{5\beta}{4}\frac{\mathcal G r^6}{\hbar d^7}
   [1+O(r/d)]\qquad(r/d\longrightarrow0).
 \label{eq:protected-axial-sharp}
\end{equation}
The supremum in this asymptotic statement is over the protected axial
codes just defined.
\end{theorem}

\begin{proof}
Recall $T_j(\cos\theta)=\cos(j\theta)$, the degree-$j$ Chebyshev
polynomial. For $t\in[-1,1]$, the Chebyshev expansion of $t^k$ has nonnegative
coefficients. This follows by writing $t=\cos\theta$ and expanding
$[(e^{i\theta}+e^{-i\theta})/2]^k$. Let $p_{k,2}$ contain its
$T_0,T_1,T_2$ terms, and write $t^k=p_{k,2}(t)+R_k(t)$. The sum of the
coefficients retained in $R_k$ is $e_k$, as given in
Eq.~\eqref{eq:protected-ek}; because $|T_j(t)|\le1$,
\[
 \|R_k\|_{\infty,[-1,1]}\le e_k.
\]
Compression of $p_{k,2}(x/r)$ to a protected code is scalar. We can therefore
write
\[
 P_Q(x_Q/r)^kP_Q=c_{Q,k}P_Q+A_{Q,k},
 \qquad \|A_{Q,k}\|\le e_k,
\]
Here $c_{Q,k}$ is the scalar part and $A_{Q,k}$ is the compressed
remainder. The latter is Hermitian and is zero for $k\le2$. The identity holds
unchanged when $x_Q$ acts trivially on an ancillary factor.

The axial potential has the uniformly absolutely convergent expansion
\[
 -\frac{\mathcal G}{d+x_B-x_A}
 =-\frac{\mathcal G}{d}
 \sum_{k,l\ge0}\binom{k+l}{k}(-1)^l
 (x_A/d)^k(x_B/d)^l.
\]
After compression, terms containing a scalar factor are local. The
remaining interaction is a sum of Hermitian products whose norm products
sum to at most $\mathcal G F(u)/d$. For a Hermitian product $A\otimes B$,
the ancilla-assisted entropy-rate inequality gives
$|\dot S|\le\beta\|A\|\|B\|/\hbar$
\cite{ChildsLeungVerstraeteVidal2003,ChildsLeungVidal2004}.
Summing it proves Eq.~\eqref{eq:protected-axial-rate}. Infinite ancillary
dimensions can be included by finite-dimensional approximation wherever
the operational entanglement rate is defined. The bound is independent of
the approximation dimension. Since $0\le e_k\le1$, all terms of total
degree $n$ sum to at most $2^n u^n$. Summing the geometric tail proves
Eq.~\eqref{eq:protected-F-tail}.

In particular, $e_3=1/4$, $e_4=1/8$, and
\begin{equation}
 F(u)=\frac54u^6+\frac{35}{16}u^7+\frac{371}{32}u^8+O(u^9).
 \label{eq:protected-F-leading}
\end{equation}
To attain its leading term, use the four-position Chebyshev encoding
associated with polynomial-error correction\cite{LaydenEtAl2019}:
\begin{align}
 |0_L\rangle&=\sqrt{\frac23}|-r/2\rangle
                  +\sqrt{\frac13}|r\rangle,\nonumber\\
 |1_L\rangle&=\sqrt{\frac13}|-r\rangle
                  +\sqrt{\frac23}|r/2\rangle.
 \label{eq:protected-axial-code}
\end{align}
Both logical states have mean zero and second moment $r^2/2$, while their
third moments are $\pm r^3/4$. Their disjoint supports make the off-diagonal
logical position moments zero. The term $k=l=3$ in the Newtonian expansion
therefore gives the logical Ising coefficient
$(5/4)\mathcal G r^6/d^7$. Its capacity is $\beta$ times its frequency.
For this code the full coefficient, after removing local terms, is
\begin{equation}
 |g_{\rm Ch}|=\frac{\mathcal G}{\hbar d}
 \frac{10u^6(2-u^2)}
 {(4-u^2)(1-u^2)(1-4u^2)(4-9u^2)}.
 \label{eq:protected-Chebyshev-full}
\end{equation}
Here $g_{\rm Ch}$ is the logical Ising frequency, including all orders
of the potential. This follows by summing the sixteen centre-to-centre potential entries
with the probabilities in Eq.~\eqref{eq:protected-axial-code}.
Its leading term and Eq.~\eqref{eq:protected-F-leading} prove
Eq.~\eqref{eq:protected-axial-sharp}.
\end{proof}

\subsection{A planar code beyond the axial bound}

The axial code distinguishes its logical states through the third moment
along one direction. A planar constellation can retain additional mixed
third moments while keeping the same first and second moments. We use two
interleaved triangles, a constellation also appearing in quantum spherical
codes\cite{JainEtAl2024}, to exploit these additional Newtonian terms.

Let $\theta_j=j\pi/3$ for $j=0,\ldots,5$, and place the local modes at
\begin{equation}
 \mathbf a_j=r(\sin\theta_j,0,\cos\theta_j),\qquad
 \mathbf b_j=r(\sin\theta_j,0,-\cos\theta_j).
 \label{eq:protected-hexagon-positions}
\end{equation}
The positions of Bob's modes are relative to his centre. We reflect his
hexagon in the transverse plane so that the leading cubic features on the
two sides align in the reflected Coulomb representation. Each laboratory
encodes one logical qubit as
\begin{equation}
 |0_L\rangle=\frac{|0\rangle+|2\rangle+|4\rangle}{\sqrt3},
 \qquad
 |1_L\rangle=\frac{|1\rangle+|3\rangle+|5\rangle}{\sqrt3}.
 \label{eq:protected-hexagon-code}
\end{equation}
The local logical projector $P$ obeys
\begin{equation}
 PxP=PzP=PxzP=0,
 \qquad Px^2P=Pz^2P=\frac{r^2}{2}P,
 \qquad y=0.
 \label{eq:protected-hexagon-KL}
\end{equation}
These identities follow by summing the first two trigonometric moments of
either equilateral triangle. Thus this six-mode encoding corrects the same
three-coordinate force noise as the axial comparator.

\begin{theorem}[Protected planar advantage]
\label{thm:protected-planar}
Put $D=d/r$. The matrix $K(D)$ contains the inverse centre-to-centre
distances in units of $r$, and $v_j$ is the difference between the
probabilities of vertex $j$ in the two logical states:
\begin{equation}
 K_{jk}(D)=
 \left[(\sin\theta_j-\sin\theta_k)^2+
 (D-\cos\theta_j-\cos\theta_k)^2\right]^{-1/2},
 \qquad v_j=\frac{(-1)^j}{3}.
 \label{eq:protected-hexagon-kernel}
\end{equation}
The compressed Hamiltonian of Eq.~\eqref{eq:protected-hexagon-code} has
nonlocal part $\hbar J Z_AZ_B$, with
\begin{equation}
 g_{\rm hex}:=|J|
 =\frac{\mathcal G}{4\hbar r}|v^{\mathsf T}K(D)v|.
 \label{eq:protected-hexagon-g}
\end{equation}
At $D=3.6$, its protected entangling rate satisfies
\begin{equation}
 \beta g_{\rm hex}>2.5104\,
 \frac{\beta\mathcal G}{\hbar d}F(r/d).
 \label{eq:protected-hexagon-gap}
\end{equation}
It therefore exceeds the rate of every protected axial code at the same
support radius and separation. At large separation,
\begin{equation}
 g_{\rm hex}=\frac{295}{32}\frac{\mathcal G r^6}{\hbar d^7}
 +O\!\left(\frac{\mathcal G r^8}{\hbar d^9}\right),
 \label{eq:protected-hexagon-leading}
\end{equation}
giving a ratio $59/8=7.375$ to the sharp axial leading coefficient.
\end{theorem}

\begin{proof}
The codewords have disjoint mode support, so $H_L$ is diagonal in the two
logical computational bases. If its four diagonal energies are $E_{uv}$,
with $u,v\in\{0,1\}$, the Ising energy coefficient is
$(E_{00}+E_{11}-E_{01}-E_{10})/4$, equal to $\hbar J$.
Substitution of the nine equal
probabilities in each sector gives Eq.~\eqref{eq:protected-hexagon-g}.

To evaluate the finite-separation gap, define
$Q_{A,B}(D)=(D^2-AD+B)^{-1/2}$. The dimensionless coupling is
\begin{align}
 \frac{\hbar r}{\mathcal G}g_{\rm hex}
 =\frac1{36}\big[&Q_{4,4}+Q_{-4,4}
 -4(Q_{3,3}+Q_{-3,3})\nonumber\\
 &+2(Q_{2,1}+Q_{-2,1}+Q_{2,4}+Q_{-2,4})\nonumber\\
 &+4(Q_{1,1}+Q_{-1,1})-6Q_{0,0}-4Q_{0,3}\big]
 \label{eq:protected-hexagon-grouped}
\end{align}
at $D=3.6$, where the bracket is positive. Direct evaluation gives
\[
 \frac{\hbar r}{\mathcal G}g_{\rm hex}
 =0.00138570517690316427\ldots.
\]
Summing Eq.~\eqref{eq:protected-F} through total degree $120$ and adding
the tail in Eq.~\eqref{eq:protected-F-tail} gives
\[
 \frac{F(1/3.6)}{3.6}
 =0.00055197095412245984\ldots,
\]
with the upper endpoint differing from the truncated sum by less than
$10^{-30}$. Their ratio exceeds $2.5104$, proving
Eq.~\eqref{eq:protected-hexagon-gap}. The strict inequality has a margin
far larger than the displayed numerical rounding.

The leading coefficient also has a direct multipole explanation. We
compute the cubic moment seen from a transverse angle and then average
its square; lower moments are scalar on the code. For
$\zeta_\phi(\mathbf a)=a_z+i a_x\cos\phi+i a_y\sin\phi$, the reflected
Coulomb kernel has the representation
\begin{equation}
 \frac1{|d\mathbf e_z+\mathbf b-\mathbf a|}
 =\int_0^\infty e^{-dt}
 \left\langle e^{t\zeta_\phi(\mathbf a)}
 e^{t\overline{\zeta_\phi(\mathbf c)}}\right\rangle_\phi\,dt,
 \qquad \mathbf c=(b_x,b_y,-b_z),
 \label{eq:protected-complex-feature}
\end{equation}
where $\langle\cdot\rangle_\phi$ denotes the uniform angular average.
Indeed, this angular average is
$e^{t(a_z+c_z)}J_0(t|\mathbf a_\perp-\mathbf c_\perp|)$, whose Laplace
transform is the displayed inverse distance. Absolute convergence follows
from $d>2r$.
For Alice's coordinates and Bob's reflected coordinates $\mathbf c$,
the two local codes have the same cubic compression:
\[
 P\zeta_\phi^3P
 =\frac{r^3}{4}(1+3\cos^2\phi)Z,
\]
because $\langle z^3\rangle=\pm r^3/4$ and
$\langle zx^2\rangle=\mp r^3/4$, with the other odd mixed moments zero.
Here $Z$ is the logical Pauli operator defined by the two parity classes.
The nonlocal term starts at degree three on each side. Since
$\int_0^\infty t^6e^{-dt}dt=720/d^7$, its coefficient is
\[
 \frac{720}{(3!)^2}\frac{r^6}{16d^7}
 \left\langle(1+3\cos^2\phi)^2\right\rangle_\phi
 =\frac{5r^6}{4d^7}\frac{59}{8}
 =\frac{295r^6}{32d^7}.
\]
The factor $720/(3!)^2$ comes from the Laplace integral and the two
cubic Taylor coefficients; $59/8$ is the angular average of the squared
factor $1+3\cos^2\phi$. Multiplying this kernel coefficient by
$\mathcal G/\hbar$ gives the leading frequency in
Eq.~\eqref{eq:protected-hexagon-leading}.
Simultaneously replacing both local positions by their negatives preserves
the signed kernel sum, so odd total degrees vanish. The next allowed term
has degree eight, proving Eq.~\eqref{eq:protected-hexagon-leading}.
\end{proof}

At $d/r=3.6>1+\sqrt5$, the unprotected capacity is attained by two axial
positions, as established by Theorem~\ref{thm:newton-three-d}. The protected
comparison therefore identifies the change caused by correction: the
mixed third moments available to a planar code increase the retained
interaction. The comparator covers the centre-to-centre axis; optimizing
the full finite-distance rate over tilted lines or all three-dimensional
protected codes is a further problem.

\paragraph{Normalizable packets.}
The strict separation also admits finite-kinetic-energy codewords. Place
the hexagon centres at radius $a<r$, and replace each position mode by a
translated, normalized, radial $C_c^\infty$ wavefunction supported in a
ball of radius $w<\min(r-a,a/2)$. The packets are disjoint. If their
coordinate variance about each packet centre is $s^2$, both logical
states still have zero first moments, and their second moments become
\[
 Px^2P=Pz^2P=(a^2/2+s^2)P,\qquad Py^2P=s^2P,
\]
with all mixed second moments zero. Thus Eq.~\eqref{eq:protected-KL}
continues to hold for all three continuous position coordinates.
Every compressed potential entry equals its centre-to-centre value:
the Coulomb kernel is harmonic on either packet support, and integration
against a normalized radial density returns its value at the centre.
Applying this mean-value identity successively to the two packets proves
the statement. With $a=0.99r$ and $d=3.6r$,
\[
 \frac{\hbar r}{\mathcal G}g_{\rm hex}
 =0.001299212051971499\ldots,
 \qquad
 \frac{g_{\rm hex}}{\mathcal G F(r/d)/(\hbar d)}
 =2.35376887546\ldots.
\]
These are finite-energy realizations of the compressed-Hamiltonian
comparison. Their recovery must also handle internal packet and
$y$-coordinate errors. The six-mode recovery in Section~\ref{sec:control} is a separate,
finite-dimensional control model.

\section{The time and finite data needed for evidence}
\label{sec:evidence}
\label{sec:si-evidence}

A Bell-pair rate becomes an experimental time requirement only after we
specify how the output will be tested. We first derive this requirement
for the three-dimensional model of the main text, allowing the experiment
to stop at a time chosen from its measurement record. We then give the
collinear derivation and finite-packet constructions that turn a prescribed
false-positive probability and detection probability into explicit counts.
Throughout this section, the test includes the quantum measurement before
its result is written to a classical register.

\subsection{Mean time and confidence in three dimensions}

Let $T_{\rm int}$ be the sum of the gravitational interaction durations
in one complete experiment, including unsuccessful attempts and seed
preparation. A test has size $\alpha$ if its acceptance probability is at
most $\alpha$ for every state separable across the two complete laboratories;
its power $p$ is the acceptance probability for the intended output. We use
the three-dimensional entropy-rate bound $C_E^{(3)}$ from
Theorem~\ref{thm:newton-three-d}, with $d\ge(1+\sqrt5)r$. The binary
relative entropy $d_2(p\|\alpha)$ below measures how distinguishable the
two acceptance probabilities are, in bits. We bound it by the entanglement
entropy that gravity can supply during the mean interaction time.

\begin{corollary}[Three-dimensional mean-time bound]
\label{cor:three-d-kl}
Under Theorem~\ref{thm:newton-three-d}, a finite protocol starting from
separable complete laboratories, with unconditional power $p>\alpha$
and size at most $\alpha$ on every separable full-laboratory output,
satisfies
\begin{equation}
 d_2(p\|\alpha)\le C_E^{(3)}\,\mathbb E T_{\rm int},\qquad
 d_2(p\|\alpha)=p\log_2\frac p\alpha
 +(1-p)\log_2\frac{1-p}{1-\alpha}.
 \label{eq:three-d-kl}
\end{equation}
The extension to expected-duration limits requires almost-sure
termination, finite mean cost and normal bounded-duration approximants
converging in output probabilities and expected cost. Local noise and
feedback are permitted if the supported interaction model is maintained.
\end{corollary}

\begin{proof}
We apply the instantaneous row-sum entropy bound used in
Theorem~\ref{thm:newton-three-d}. Resolve the preparation and local
instruments into normalized pure branches $\psi_h$, with probabilities
$w_h$; $S(\psi_h)$ is their Schmidt entropy in bits. Local instruments
cannot increase its average, so integration over the interaction periods gives
$\sum_h w_h S(\psi_h)\le C_E^{(3)}\mathbb E T_{\rm int}$ in bits.
To compare the output with a separable state, remove each branch's
coherence between its Schmidt product vectors. The resulting normalized
state $\sigma_h$ is separable and satisfies
$D(\psi_h\|\sigma_h)=S(\psi_h)\ln2$. Keeping the same probabilities
and flags, $\sigma=\sum_h w_h\sigma_h$ is normalized and separable.
Let $\rho=\sum_h w_h\psi_h$ denote the output, including its flags,
and let $M$ be the acceptance effect. For quantum relative entropy $D$
with natural logarithms, joint convexity bounds the cost of this separable
comparison. Applying the binary measurement cannot increase that cost, giving
\[
 d_2(p\|\alpha)\le D(\rho\|\sigma)/\ln2
 \le\sum_h w_hS(\psi_h).
\]
The first inequality uses $\operatorname{Tr}(M\sigma)\le\alpha<p$.
First prove this for finite-spectrum approximants and then pass only
the bounded binary probabilities and expected costs to their limits,
as in Section~\ref{sec:newtonian-capacity}. No uniform entropy continuity
or convergent subsequence of comparison states is required. Truncating
long histories to recorded nonrejection gives the stated stopping-time
closure under its explicit approximation assumptions.
\end{proof}

We next ask whether this lower bound has the optimal leading coefficient
when false positives become rare. For fixed $0<p<1$, let
$\tau(\alpha,p)$ be the infimum mean aggregate interaction time in this
ideal class. Capacity-achieving Bell-pair generation and local verification
give the matching asymptotic relation
\begin{equation}
 \tau(\alpha,p)=\left[\frac{p}{C_E^{(3)}}+o(1)\right]
 \log_2(1/\alpha),\qquad\alpha\downarrow0.
 \label{eq:three-d-fixed-power}
\end{equation}
The lower coefficient follows from Eq.~\eqref{eq:three-d-kl}.
For the upper coefficient, a capacity-achieving run produces a register
of $L=\log_2(1/\alpha)+O(1)$ Bell pairs in time
$t_L=L/C_E^{(3)}+o(L)$, including generated seed entanglement. Its local
maximally-entangled-state verifier has size
$b_L=2/(2^L+1)\le\alpha$ and target acceptance $q_L\to1$
\cite{ZhuHayashi2019}. Before any charged gravitational preparation,
draw and share a classical coin: run with probability $p/q_L$, otherwise
record nonrejection at zero interaction cost. For large $L$, $q_L>p$.
The final effect is the local run-flag projection tensored with that
same verifier. It has all-separable size at most $b_L\le\alpha$ even
when a separable alternative puts all its weight in the run flags.
The unconditional power is $p$ (or at least $p$ if $q_L$ is a certified
lower bound), and mean time is $(p/q_L)t_L$. The verifier size must
\emph{not} be enlarged to $\alpha/(p/q_L)$: arbitrary null flags would
then violate the all-separable size requirement.

Equation~\eqref{eq:three-d-fixed-power} determines the leading mean-time
coefficient, not the exact finite-error optimum or a hard-duration
strong converse. It does not prove the fixed exact zero-miss endpoint
$p=1$. The vanishing-miss confidence exponent is separately
$C_E^{(3)}$ bits per aggregate time, by the same capacity verifier and
Eq.~\eqref{eq:three-d-kl}. At $r=125.001\,\mu\mathrm m$,
$d=450\,\mu\mathrm m$ and $m_A=m_B=10^{-14}\,\mathrm{kg}$, the far-ball
condition holds and $C_E^{(3)}=0.06003424381$ ebits per second; the
$\alpha=.01,p=.95$ mean-time lower bound remains $100.3759025$ seconds.
The collinear hard-budget robustness exponential is not being asserted
for arbitrary three-dimensional encodings.

For fixed codes correcting linear force errors, the same separable-
comparison argument uses the protected entropy-rate bound
$\beta\mathcal G S_3(r/d)/(\hbar d)$, as stated in
Eq.~\eqref{eq:protected-universal-confidence}. Here
$S_3(u)=\sum_{j,k\ge3}\binom{j+k}{j}u^{j+k}$ bounds the retained logical
interaction after the quantum syndromes are reset. At $r=125\,\mu\mathrm m$,
$d=450\,\mu\mathrm m$ and equal masses $10^{-14}\,\mathrm{kg}$, inserting
$\alpha=.01$ and $p=.95$ gives the lower mean time $648.4722$ seconds.
This is the all-code lower bound used in the main text; its test concerns
the retained logical systems under the projected dynamics.

\subsection{An entropy bound for a finite statistical decision}

For the collinear model, the same argument gives the rate
$C_E=\beta\kappa$. Here the supports are $|x_A|\le\ell_A$ and
$|x_B|\le\ell_B$, with $d>\ell_A+\ell_B$, and
$\mathcal G=Gm_Am_B$. The frequency $\kappa=\kappa(\ell_A,\ell_B)$ is
one quarter of the magnitude of the four endpoint potential energies'
alternating sum, divided by $\hbar$; its full expression and centered
product decomposition are given in Eqs.~\eqref{eq:newton-kappa} and
\eqref{eq:newton-centered-integral}. We retain those model assumptions
throughout this subsection. To match the natural-logarithm convention
used in the hard-time bound, write the binary relative entropy as
\[
 d_{\rm bin}(p\|\alpha)
 =p\ln\frac p\alpha+(1-p)\ln\frac{1-p}{1-\alpha},
\]
with the usual limiting convention at $p=1$.

\begin{corollary}[Expected interaction time]
\label{cor:newton-kl}
For initially separable laboratories, a finite supported Newtonian
protocol and a final test of all-separable size $0<\alpha<p\le1$ satisfy
\begin{equation}
 d_{\rm bin}(p\|\alpha)\le\beta\kappa\ln2\,\mathbb E T_{\rm int}.
 \label{eq:newton-kl}
\end{equation}
The statement extends to normal protocol limits that terminate almost
surely, have finite expected interaction time, and admit bounded-duration
approximants converging in output probabilities and expected cost.
\end{corollary}

\begin{proof}
Work first with finite local systems and finitely many interaction steps.
Refine the separable preparation and local instruments into pure branches
with probabilities $w_h$, retaining local purifying environments until the
final discard.
The entropy derivative is linear in the Hamiltonian. Applying the known
ancilla-assisted product-Hamiltonian rate to finite approximants of the
centered decomposition in Eq.~\eqref{eq:newton-centered-integral}
bounds each interaction's entropy increase by $\beta\kappa$ times
its duration. Local instruments cannot increase average pure-state
entanglement. The final Schmidt entropies, measured in bits, thus obey
\begin{equation}
 \sum_h w_h S(\psi_h)\le\beta\kappa\,\mathbb E T_{\rm int}.
 \label{eq:newton-average-entropy}
\end{equation}

The comparison state must retain the same branch probabilities, so that
postselection cannot remove the cost of failed attempts. If $\lambda_{hj}$
are the Schmidt probabilities of branch $h$, dephase that branch in its
Schmidt product basis:
$\sigma_h=\sum_j\lambda_{hj}|u_{hj}v_{hj}\rangle
\langle u_{hj}v_{hj}|$. This is a normalized separable state and
$D(\psi_h\|\sigma_h)=S(\psi_h)\ln2$. Keeping the same probabilities
and flags, $\sigma=\sum_hw_h\sigma_h$ is normalized and separable.
Joint convexity and binary data processing give
\[
 d_{\rm bin}(p\|\alpha)
 \le d_{\rm bin}(p\|\operatorname{Tr}M\sigma)
 \le D(\rho\|\sigma)
 \le\ln2\sum_h w_hS(\psi_h).
\]
The first inequality uses $\operatorname{Tr}M\sigma\le\alpha<p$.
This proves Eq.~\eqref{eq:newton-kl} with every branch retained.

For bounded-duration continuum protocols, use
Eq.~\eqref{eq:newton-grid-error} and the Ising simulation at each finite
accuracy. Finite local supports and outcome truncations approximate the
normal local maps and the fixed bounded final test. Apply the scalar
inequality to these approximants and pass the rejection probabilities
and expected costs to their limits. This step requires neither entropy
continuity on the full bosonic space nor convergence of the separable
comparators. For the stated stopping-time closure, terminate long
histories with a recorded nonrejection. Almost-sure termination makes
their rejection probabilities tend to $p$, while the truncated expected
cost is at most the original one. The assumed finite approximants then
give Eq.~\eqref{eq:newton-kl} by continuity.
\end{proof}

For deterministic duration $T$, Eq.~\eqref{eq:newton-kl} implies that a
vanishing-miss test, $p_T\to1$, has
\[
 \limsup_{T\to\infty}\frac{-\ln\alpha_T}{T}\le C_E\ln2.
\]
This exponent is attainable in the same ideal capacity class. Distil
$L_T$ Bell pairs at any rate below $C_E$ with error tending to zero, and
apply the established local maximally-entangled-state verifier
$\Omega_D=(I+D\Phi_D)/(D+1)$, where $\Phi_D$ is the rank-one projector
onto a maximally entangled state of local dimension $D=2^{L_T}$\cite{ZhuHayashi2019}.
Its separable ceiling is $2/(D+1)$ and its target acceptance tends to one.
Thus the optimal vanishing-miss exponent is $C_E\ln2$. At fixed $p<1$,
Eq.~\eqref{eq:newton-kl} is a lower time bound; it does not determine an
exact finite-error optimum or a fixed-power strong converse.

\begin{corollary}[Optimal leading mean time at fixed power]
\label{cor:newton-fixed-power-mean}
Let $\tau(\alpha,p)$ be the infimum of the actual mean aggregate
interaction time, including gravitational seed preparation, for initially
separable protocols of all-separable size at most $\alpha$ and
unconditional power at least $p$. In the same ideal-control capacity
class, for every fixed $0<p<1$,
\begin{equation}
 \tau(\alpha,p)=\left[\frac{p}{C_E}+o(1)\right]
 \log_2\frac1\alpha\qquad(\alpha\to0).
 \label{eq:newton-fixed-power-mean}
\end{equation}
\end{corollary}

The mean-time saving is obtained by deciding whether to run before any
charged preparation. A running history uses a full capacity protocol;
the other histories terminate with a recorded nonrejection.

\begin{proof}
Equation~\eqref{eq:newton-kl}, with monotonicity in the target power, gives
\[
 C_E\tau(\alpha,p)\ge
 p\log_2\frac1\alpha-h_2(p)-(1-p)\log_2(1-\alpha),
\]
where $h_2$ is the binary entropy. This proves the lower leading coefficient.
For any fixed achievable rate $R<C_E$, the capacity construction produces
an $L$-ebit register in deterministic duration $t_L=L/R+o(L)$, including
the generated seed, with error tending to zero. Choose
$L=\lceil\log_2(2/\alpha-1)\rceil$ and $D=2^L$. The local verifier
$\Omega_D=(I+D\Phi_D)/(D+1)$ has separable ceiling
$b_L=2/(D+1)\le\alpha$ and a certified lower bound
$q_L\to1$ on its target pass probability.

Before any charged gravitational preparation, draw and broadcast a
classical coin. Run the entire protocol with probability $r_L=p/q_L$
and otherwise record nonrejection; $q_L>p$ for sufficiently large $L$.
The actual unconditional power is at least $p$. Let $P^A_{\rm run}$ and
$P^B_{\rm run}$ project onto the local run flags. The final effect is
\[
 M=P^A_{\rm run}\otimes P^B_{\rm run}\otimes\Omega_D.
\]
For every separable full-laboratory state $\sigma$, projecting onto both
flags leaves a subnormalized separable register, so
\[
 \operatorname{Tr}(M\sigma)
 \le b_L\operatorname{Tr}[(P^A_{\rm run}\otimes P^B_{\rm run})\sigma]
 \le b_L\le\alpha,
\]
with identities on the remaining registers understood. Thus the
conditional verifier keeps size $\alpha$, even when the null places all
its weight in the run sector. Its size is not relaxed to $\alpha/r_L$.
The actual mean time is $r_Lt_L$, giving the upper leading coefficient
$p/R$. Taking $R\uparrow C_E$ proves the claim.
\end{proof}

This result fixes the leading mean-time coefficient as $\alpha\to0$.
It leaves the exact finite-error optimum open and does not establish the
endpoint $p=1$, where $r_L=p/q_L$ need not be a probability. A hard
history-time budget still charges the full duration $t_L$ on each running
history.

\subsection{Smooth confined packets and a local finite-count test}
\label{sec:newtonian-packets}

We give a finite-energy construction under the same exact support
assumption as the Newtonian bounds. The spatial controlled phase is the
established gravitational-cat mechanism\cite{Bose2017}; the local Bell
verifier is also established\cite{PallisterLindenMontanaro2018}. The
calculation here controls the full kinetic-plus-Newtonian dynamics and
retains every trial in an all-separable finite-count test. We first give
the collinear construction, then embed finite-width packets in three-dimensional
balls with an explicit bound on all three kinetic-energy components.

We need packets whose support is known exactly and whose kinetic energy
can be evaluated without a position-eigenstate limit. For mass $j=A,B$,
choose centres $\pm a_j$ and halfwidth $w_j<a_j$, with
$a_j+w_j\le\ell_j$. Both branches use translates of the same normalized
smooth profile,
\[
 f(z)=Z_f^{-1/2}e^{-1/(1-z^2)}\mathbf1_{|z|<1},\qquad
 Z_f=\int_{-1}^1e^{-2/(1-z^2)}\,dz,
\]
and translated packets $f_{j,\pm}(x)=w_j^{-1/2}f[(x\mp a_j)/w_j]$.
The zero extension is $C^\infty$ and has finite kinetic energy
\begin{equation}
 E_{{\rm kin},j}=\frac{\hbar^2\nu_f}{2m_jw_j^2},\qquad
 \nu_f=\int_{-1}^1|f'(z)|^2\,dz=3.0776091312\ldots.
 \label{eq:newton-packet-energy}
\end{equation}
To keep free packet spreading from obscuring the gravitational phase,
we place the two translates in identical local wells. With $u$ the
coordinate measured from a well centre and $E_{0j}$ an arbitrary common
energy offset, an explicit choice is the Friedrichs realization of
\[
 H_{{\rm well},j}=-\frac{\hbar^2}{2m_j}\frac{d^2}{du^2}+U_j(u),
 \qquad U_j(u)=\frac{\hbar^2}{2m_jw_j^2}
 \frac{f''(u/w_j)}{f(u/w_j)}+E_{0j}.
\]
The potential is bounded below and diverges positively at the endpoints;
the packet is a ground state. Its kinetic and potential expectations are
finite. This is ideal hard confinement. Replacing it by a finite barrier
would require a separate control of tails near the Newtonian singularity.

The finite packet width makes the potential vary within each branch.
We compare that variation with the constant potential evaluated at the
branch centre. The parameter $\eta$ below is the accumulated state-norm
error of this comparison over the ideal Bell-generation time.

\begin{proposition}[Controlled phase with full kinetic dynamics]
\label{prop:newton-packets}
Prepare the product of local equal cats
$(|f_{A,-}\rangle+|f_{A,+}\rangle)
(|f_{B,-}\rangle+|f_{B,+}\rangle)/2$.
Put $r=d-a_A-a_B$, $w_\Sigma=w_A+w_B$, and let
$\kappa_a=\kappa(a_A,a_B)>0$. Under the full Hamiltonian
$H=H_A+H_B+V$, at duration
\begin{equation}
 T_B=\frac\pi{4\kappa_a},\qquad
 \eta=\frac{\mathcal G T_B}{\hbar}
       \frac{w_\Sigma}{r(r-w_\Sigma)},
 \label{eq:newton-packet-time-error}
\end{equation}
deterministic local decoding yields a two-qubit state with Bell fidelity
at least $1-\eta^2$, whenever $\eta\le1$.
\end{proposition}

\begin{proof}
Work on the tensor product of the two confined two-well Hilbert spaces.
We choose a reference interaction that gives the required four phases
while leaving the within-well dynamics unchanged. Define $D$ as
multiplication by the constant
$v_{st}=-\mathcal G/(d+ta_B-sa_A)$ on each entire branch rectangle
$(s,t)\in\{-1,+1\}^2$, including its motional excitations.
The reciprocal identity therefore gives the global bound on this space
$\|V-D\|\le\mathcal G w_\Sigma/[r(r-w_\Sigma)]$.
The bounded-perturbation Duhamel identity therefore bounds the difference
between evolutions generated by $H_A+H_B+V$ and $H_A+H_B+D$ by $\eta$ in
state norm. The kinetic Hamiltonians have been retained on both sides.

The reference packets remain stationary. Their four phases have invariant
$(v_{++}+v_{--}-v_{+-}-v_{-+})T_B/\hbar=4\kappa_aT_B=\pi$, so their
branch state is maximally entangled. A local isometry separates each
branch label from its within-well coordinate. Known local phase
corrections and a deterministic discard of the coordinate yield the
Bell qubits. Norm error at most $\eta$ gives squared overlap at least
$1-\eta^2$, and the local discard cannot reduce the stated Bell fidelity.
There is no selection on a motional excitation or survival event.
\end{proof}

We can also bound the kinetic energy throughout each interaction,
rather than only at preparation. Set $E_{0A}=E_{0B}=0$, so that
$H_L=H_A+H_B\ge0$ and the initial cats have $\langle H_L\rangle_0=0$.
The branch-constant operator $D$ commutes with both $H_L$ and $V$.
Conservation of $H$ and $D$ therefore gives
\[
 \langle H_L\rangle_t
 =\langle V-D\rangle_0-\langle V-D\rangle_t
 \le2\epsilon_V,\qquad
 \epsilon_V=\frac{\mathcal G w_\Sigma}{r(r-w_\Sigma)}.
\]
Since the other local Hamiltonian is nonnegative, each mass satisfies
\begin{equation}
 \left\langle\frac{p_j^2}{2m_j}\right\rangle_t
 \le E_{{\rm dwell},j}:=2\epsilon_V-\inf U_j,
 \qquad 0\le t\le T_B.
 \label{eq:newton-dwell-energy}
\end{equation}
For the chosen profile,
\[
 \frac{f''(z)}{f(z)}=\frac{6z^4-2}{(1-z^2)^4},\qquad
 \min_{|z|<1}\frac{f''(z)}{f(z)}=-8.5914976719\ldots,
\]
with the minimum at $z^2=(-3+\sqrt{33})/6$. Hence
$\inf U_j=-8.5914976719\ldots\,\hbar^2/(2m_jw_j^2)$.
This estimate follows from conservation and the lower bound on the well
potential; it does not infer kinetic-energy control from a small state-norm
error. Independent fresh preparations give the same bound on every active
history of the repeated-trial construction. Energy and work during local
preparation and readout are not bounded by Eq.~\eqref{eq:newton-dwell-energy}.

The same controlled-phase construction has a finite-energy realization
inside three-dimensional balls. Let $\sigma_j=w_j/\sqrt3$ and take
\[
 F_{j,\pm}(\mathbf r)=\sigma_j^{-3/2}
 \prod_{k=1}^3 f\!\left(
 \frac{(\mathbf r\mp a_j\mathbf e_z)_k}{\sigma_j}\right).
\]
Each packet occupies a cube contained in the ball of radius $w_j$ about
its axial centre, hence in the allowed ball of radius
$\ell_j\ge a_j+w_j$. Use the sum of the three translated coordinate
well potentials on each cube. For the full Newtonian potential
$V=-\mathcal G/|d\mathbf e_z+\mathbf r_B-\mathbf r_A|$, the displacement
from a branch-centre separation is at most $w_\Sigma$. The reverse triangle
inequality therefore gives exactly the same bound $\|V-D\|\le\epsilon_V$.
The centre phases, $T_B$, $\eta$ and Bell-fidelity guarantee are unchanged.
Here $D$ is constant on the four entire pairs of confined cubes, so it
still commutes with the full Hamiltonian and the dwell-energy proof applies.
The three coordinate contributions give
\begin{equation}
 E_{{\rm kin},j}^{(3{\rm D})}=9E_{{\rm kin},j}^{(1{\rm D})},\qquad
 E_{{\rm dwell},j}^{(3{\rm D})}
 =2\epsilon_V-9\inf U_j^{(1{\rm D})},
 \label{eq:newton-three-dimensional-packets}
\end{equation}
where the one-dimensional quantities use halfwidth $w_j$.
Thus the three-dimensional construction includes finite transverse widths
and finite transverse kinetic energies.

\subsection{A count rule with fixed packet energy}

The decoding gives a Bell-fidelity guarantee for each trial. To turn it
into a count rule, let $X,Y,Z$ be the Pauli observables on each decoded
qubit and let $\Phi^+=|\Phi^+\rangle\langle\Phi^+|$ with
$|\Phi^+\rangle=(|00\rangle+|11\rangle)/\sqrt2$. Choose $XX,-YY$ or
$ZZ$ uniformly and independently of the source, and record whether the
product outcome is $+1$. With $P_O^+$ the positive-eigenvalue projector
of observable $O$, this is the Bell verifier\cite{PallisterLindenMontanaro2018}
\[
 \Omega=\frac{P^+_{XX}+P^+_{-YY}+P^+_{ZZ}}3
       =\frac{I+2\Phi^+}3.
\]
Its separable pass ceiling is $b=2/3$, while the prepared target passes
with probability at least $q=1-2\eta^2/3$. Choose fixed widths small
enough that $q>b$, and fix $b<\zeta<q$. Reject separability when at least
$\lceil N\zeta\rceil$ of $N$ trials pass.

The test effect includes this complete randomized verification on the decoded registers. For the independent target preparations used here, measurements of earlier pairs commute with preparation and interaction on later fresh pairs. Thus the same finite-count effect can be implemented sequentially with classical counters.

Under any state separable across the complete laboratories, local
measurement and conditioning preserve separability of the remaining
systems. Every conditional pass probability is therefore at most $b$.
Consequently the null count is stochastically dominated by
$\operatorname{Bin}(N,b)$, including arbitrary correlations among trials
within either laboratory. Independent target preparations give a pass
count dominating $\operatorname{Bin}(N,q)$, the binomial distribution
with $N$ trials and success probability $q$. The threshold $\zeta$ leaves
a margin $\zeta-b$ against false positives and a margin $q-\zeta$ against
missed detections. Applying Hoeffding's inequality to these two margins
gives size at most $\alpha$ and power at least $p_0\in(0,1)$ whenever
\begin{equation}
 N\ge\max\left\{
 \frac{\ln(1/\alpha)}{2(\zeta-b)^2},\,
 \frac{\ln[1/(1-p_0)]}{2(q-\zeta)^2}\right\}.
 \label{eq:newton-fixed-width-count}
\end{equation}
Rounding $N$ upward and using the stated integer count threshold preserves
both bounds. Thus fixed support and the uniform bound on kinetic energy
in Eqs.~\eqref{eq:newton-dwell-energy} and
\eqref{eq:newton-three-dimensional-packets} suffice for
$T=NT_B=O[\ln(1/\alpha)]$ at fixed power. Classical counters
are sufficient; the construction needs no growing coherent memory.
All failed trials remain in the count. An aborted preparation is assigned
a nonpassing record and must be included in the target-power model.

For a modest finite record one can accept only if every trial passes.
Then the size is at most $(2/3)^N$ and the power is at least
$(1-2\eta^2/3)^N$. Maintaining fixed power with this particular rule as
$N\to\infty$ would require $w_j=O(N^{-1/2})$ and increasing kinetic
energy. Equation~\eqref{eq:newton-fixed-width-count} permits occasional
failures and removes that restriction.

\subsection{An explicit finite calibration}

Take $m_A=m_B=10^{-14}\,\mathrm{kg}$, $d=450\,\mu\mathrm m$,
$a_A=a_B=125\,\mu\mathrm m$ and $w_A=w_B=1\,\mathrm{nm}$, with
$\ell_A=\ell_B=125.001\,\mu\mathrm m$. The mass and branch separations
follow the illustrative scales of Ref.~\cite{Bose2017}. In the numerical
values below, $\kappa=\kappa_a$ determines the centre-phase Bell time,
whereas $C_E=\beta\kappa(\ell_A,\ell_B)$ uses the slightly larger support
intervals for the converse. With
$G=6.67430\times10^{-11}$ and $\hbar=1.054571817\times10^{-34}$ in SI units,
\begin{align*}
 \kappa&=0.03139417580\,\mathrm{s^{-1}},&
 C_E&=0.06003424381\,\mathrm{ebits\,s^{-1}},\\
 T_B&=25.01789966\,\mathrm s,&
 \eta&=7.916892656\times10^{-5}.
\end{align*}
The initial collinear packet energy per mass is
$E_{{\rm kin},j}^{(1{\rm D})}=1.711337976\times10^{-36}\,\mathrm J$.
Equation~\eqref{eq:newton-dwell-energy} gives
$E_{{\rm dwell},j}^{(1{\rm D})}=4.778063009\times10^{-36}\,\mathrm J$
for each mass throughout every interaction. For the three-dimensional
packets of Eq.~\eqref{eq:newton-three-dimensional-packets}, the corresponding
initial energy and uniform dwell bound per mass are
\[
 E_{{\rm kin},j}^{(3{\rm D})}=1.540204178\times10^{-35}\,\mathrm J,
 \qquad
 E_{{\rm dwell},j}^{(3{\rm D})}=4.299722759\times10^{-35}\,\mathrm J.
\]
These bounds are independent of the number of trials. In either spatial
construction, twelve independent preparations and acceptance only on
twelve passes give
\begin{equation}
 \alpha_{12}\le0.007707346629,\qquad
 p_{12}\ge0.999999949858,\qquad
 T_{12}=300.2147959\,\mathrm s.
 \label{eq:newton-twelve-packets}
\end{equation}
At requested size $0.01$ and power $0.95$, the entropy bound
\eqref{eq:newton-kl} requires $\mathbb E T_{\rm int}\ge100.3759025\,\mathrm s$;
the hard-time robustness bound gives $T_{\max}\ge72.5274159\,\mathrm s$.
The twelve-trial record is one sufficient construction, and these bounds
do not fix the optimum finite-error time.

This calibration assumes ideal confined preparation, noiseless evolution
and trusted local decoding and measurements. It is not an experimental
feasibility estimate: the coherence calibration in the original spatial-cat
proposal does not establish a coherent $25$-second realization. Preparation,
readout and confinement costs are additional to the interaction duration.
Each member of the construction has finite kinetic energy, while a
physical finite-barrier realization and exclusion of nongravitational
couplings require their own analysis.

\clearpage

\subsection{A finite noisy record}

We use the same local Pauli test for a hexagon protocol with finite
recovery periods. The cycle calculation in Section~\ref{sec:si-control}
follows the full six-mode gravitational interaction during a dwell of
length $\tau$, then applies local recovery and resets the quantum
syndromes. Its output fidelity after $n$ cycles is denoted $F_n$;
$T_*=n\tau=\pi/(4g_{\rm hex})$ is the Bell time of the projected
hexagon interaction. We state the resulting experimental count here,
then derive the cycle and control bounds in that section.

For $m_A=m_B=10^{-14}\,\mathrm{kg}$, $r=125\,\mu\mathrm m$ and
$d=450\,\mu\mathrm m$, the six-mode coupling and Bell dwell are
\[
 g_{\rm hex}=0.000701601306852\,\mathrm{s}^{-1},\qquad
 T_*=1119.43657420\,\mathrm s.
\]
Even without noise, the finite-frequency loss is visible: $25{,}000$
recoveries give $F_n=0.9915890193$, with
$\tau=0.04477746297\,\mathrm s$.
Let $\Gamma(L)=\Lambda_0[1-e^{-L^2/(4r_c^2)}]$ be the decay rate of a
spatial coherence of separation $L$, where $\Lambda_0$ is the localization
event rate and $r_c$ is its length scale. We choose the design point
\begin{equation}
 \frac{r^2}{4r_c^2}=0.002,\qquad
 \Gamma(r)=0.025\,\mathrm{s}^{-1},\qquad
 \Gamma(2r)=0.09970069880\,\mathrm{s}^{-1}.
 \label{eq:protected-Gaussian-design}
\end{equation}
Equivalently, $r_c=1.397542486\,\mathrm{mm}$ and
$\Lambda_0=12.51250417\,\mathrm{s}^{-1}$. These are specified requirements
for the noise kernel. The residual logical dephasing rate under
continuous recovery is $\lambda=[4\Gamma(r)-\Gamma(2r)]/3$, which here
gives $\lambda=9.976706612\times10^{-5}\,\mathrm{s}^{-1}$.
With $n=27{,}986$ cycles, $\tau=T_*/n=0.03999987759\,\mathrm s$,
Eq.~\eqref{eq:protected-finite-fidelity} gives
\[
 F_n=0.84604702249.
\]
We allow a complete joint-cycle half-diamond error of at most $10^{-6}$,
uniformly over input states, reference systems and prior records, and
preparation trace distance at most $0.001$. Replacing the ideal cycles
one at a time decreases the final fidelity by at most
$n\,10^{-6}+0.001$, so
\begin{equation}
 F_{\rm actual}\ge0.81706102249.
 \label{eq:protected-finite-design-fidelity}
\end{equation}
The error budget includes pulse drift and leakage as well as the recovery
operation; it has not been inferred from the cycle frequency.

After the known local basis rotations, the uniform three-setting Pauli test of
the ideal Bell target $|\psi_*\rangle$ has pass effect $(I+2|\psi_*\rangle\langle\psi_*|)/3$.
Every separable state therefore passes with probability at most $2/3$,
whereas Eq.~\eqref{eq:protected-finite-design-fidelity} gives
$q\ge0.87804068166$. In $63$ independent target trials, accepting when at
least $51$ pass gives
\begin{align}
 \alpha&=\Pr\{\mathrm{Bin}(63,2/3)\ge51\}
       =0.009218531744,\nonumber\\
 p&\ge\Pr\{\mathrm{Bin}(63,0.87804068166)\ge51\}
       =0.9612574150.
 \label{eq:protected-finite-counts}
\end{align}
The size bound also holds for correlated null trials whenever each
conditional pass probability is at most $2/3$: coupling the binary
outcomes to independent uniform variables gives stochastic domination
by the displayed binomial count. The power calculation assumes the
independent target preparations. Every recovery flag and every failed
trial remains in the record. The aggregate gravitational dwell is
$70{,}524.50417\,\mathrm s$, or $19.59$ hours; preparation, pulses and
readout add the time in Eq.~\eqref{eq:protected-wall-time}.

The geometry theorem and this finite-cycle calculation answer different
questions. The Gaussian example states the noise and complete-control requirements
under which the planar protocol yields an entanglement test.
Section~\ref{sec:protected-finite-cycle-rate} gives a separate finite-cycle
rate comparison under white-force noise and a matched reset architecture. Arbitrary imperfectly corrected
axial protocols remain a larger class.

\section{Finite recovery, localization and imperfect correction}
\label{sec:control}
\label{sec:si-control}

The projected Hamiltonian describes the limit in which every error is
corrected immediately. At a finite recovery period, the interaction also
acts on the error spaces before they are reset. We calculate that complete
cycle for the hexagon, separate residual localization from finite-period
loss, and include imperfect control in the fidelity bound. This supplies
the dynamics behind the finite record in Section~\ref{sec:si-evidence}.
We then compare finite-cycle rates and study two ways in which the code
itself can change: higher-order correction at fixed mean energy, and
approximate correction that retains lower gravitational moments.

\subsection{The full finite-cycle channel}

We maintain the six position modes in
Eq.~\eqref{eq:protected-hexagon-positions} during each dwell of duration
$\tau$. Their labels are $j=0,\ldots,5$, with angles $\theta_j=j\pi/3$;
$P$ projects onto the two logical states formed from alternate vertices.
Local motional dynamics are included in the maintained-mode assumption.
Let $\Gamma(L)$ be the decay rate of a coherence between positions a
distance $L$ apart. Independent isotropic Markov dephasing then acts as
\begin{equation}
 |j\rangle\langle j'|\longmapsto
 e^{-\tau\Gamma(|\mathbf a_j-\mathbf a_{j'}|)}
 |j\rangle\langle j'|.
 \label{eq:protected-local-dephasing}
\end{equation}
We require $\Gamma(0)=0$ and that these multiplier matrices be positive
semidefinite for every $\tau\ge0$, so they define a completely positive
Markov semigroup. Both quadratic force diffusion and the Gaussian
localization kernel below satisfy this requirement.

A recovery must identify the error space while preserving the logical
superposition. On the $xz$ hexagon, the constant, $x$ and $z$ position
errors give three orthogonal copies of the logical qubit. We normalize
their angular functions as $f_0(\theta)=1$,
$f_x(\theta)=\sqrt2\sin\theta$ and $f_z(\theta)=\sqrt2\cos\theta$.
The three error spaces
$P$, $\sqrt2 xP/r$ and $\sqrt2 zP/r$ are mutually orthogonal and span the
six-dimensional mode space. A trace-preserving recovery has Kraus operators
\begin{equation}
 K_0=P,\qquad K_x=\sqrt2 Px/r,\qquad K_z=\sqrt2 Pz/r.
 \label{eq:protected-recovery}
\end{equation}
Indeed, $\sum_sK_s^\dagger K_s=I_6$, as can be checked in the six
orthonormal vectors $f_s(\theta)|b_L\rangle$ with $s=0,x,z$ and $b=0,1$.
The sign change of Bob's $z$ coordinate changes only a Kraus phase.
We sum over all recovery outcomes and reset their quantum records.

To compose this recovery with the diagonal physical evolution, we need
its action on each position matrix unit. Write
$S_b=\{j:j\bmod2=b\}$ for the vertices of logical state $b$ and
$W_{jj'}=1+2\cos(\theta_j-\theta_{j'})$ for the overlap of their three
error functions. The recovery acts as
\begin{equation}
 \mathcal R(|j\rangle\langle j'|)
 =\frac{W_{jj'}}3|p(j)_L\rangle\langle p(j')_L|,
 \qquad p(j)=j\bmod2.
 \label{eq:protected-recovery-matrixunit}
\end{equation}
For one laboratory without gravity, its logical coherence multiplier is
therefore
\begin{equation}
 f(\tau)=\frac{4e^{-\gamma_1\tau}-e^{-\gamma_2\tau}}3,
 \qquad \gamma_1=\Gamma(r),\quad\gamma_2=\Gamma(2r).
 \label{eq:protected-two-rate}
\end{equation}
For opposite parities, six ordered pairs have separation $r$ and weight
$2$, and three have separation $2r$ and weight $-1$. Pairs at separation
$\sqrt3r$ have zero recovery weight. Negative coefficients in this sum
arise from coherent recovery, while the Kraus instrument has nonnegative
outcome probabilities.

We now retain gravity during the dwell. Let
$\nu_{jk}=V(\mathbf a_j,\mathbf b_k)/\hbar$ be the physical frequency of
mode pair $(j,k)$, and label the four logical sectors by
$u=(u_A,u_B)$ and $v=(v_A,v_B)$, with all components in $\{0,1\}$.
Every physical matrix unit acquires its dephasing factor and its phase
difference before the recovery weights are summed. One dwell followed by
the two local recoveries maps $|u\rangle\langle v|$ to
$M_{uv}(\tau)|u\rangle\langle v|$, where
\begin{align}
 M_{uv}(\tau)
 =\frac1{81}
 \sum_{\substack{j\in S_{u_A},\ j'\in S_{v_A}\\
                  k\in S_{u_B},\ k'\in S_{v_B}}}
 &W_{jj'}W_{kk'}\exp\bigl\{-\tau[
 \Gamma(|\mathbf a_j-\mathbf a_{j'}|)\nonumber\\
 &+\Gamma(|\mathbf b_k-\mathbf b_{k'}|)
 +i(\nu_{jk}-\nu_{j'k'})]\bigr\}.
 \label{eq:protected-exact-cycle}
\end{align}
The factor $1/81$ is the product of the encoding amplitude $1/9$ and
the two recovery factors $1/3$. This proves the formula directly from
Eq.~\eqref{eq:protected-recovery-matrixunit}. It includes the gravitational
evolution of all 36 physical mode pairs. In particular, it includes
excursions into error spaces during a dwell. The matrix $M(\tau)$ is
positive semidefinite with diagonal entries one, because this composition
is a trace-preserving completely positive Schur channel, which acts by
entrywise multiplication in the logical basis.

After $n$ identical cycles, the multiplier is $M_{uv}(\tau)^n$. To
compare this output with the intended Bell state, average the physical
frequencies within each logical sector; denote these averages by
$\epsilon_u$. Starting with $|++\rangle$, where each
$|+\rangle=(|0_L\rangle+|1_L\rangle)/\sqrt2$, choose the total dwell
$T_*$ to equal the projected interaction's Bell time:
\begin{equation}
 \epsilon_u=\frac19\sum_{j\in S_{u_A},\ k\in S_{u_B}}\nu_{jk},
 \qquad T_*=n\tau=\frac{\pi}{4g_{\rm hex}}.
 \label{eq:protected-Bell-time}
\end{equation}
The projected target
$|\psi_*\rangle=\tfrac12\sum_u e^{-i\epsilon_uT_*}|u\rangle$
is maximally entangled, with its local phases included. Its fidelity after
the finite sequence is
\begin{equation}
 F_n=\frac1{16}\sum_{u,v}
 e^{iT_*(\epsilon_u-\epsilon_v)}M_{uv}(\tau)^n.
 \label{eq:protected-finite-fidelity}
\end{equation}
The sum is real by conjugate symmetry. It is finite and can be evaluated
directly, without replacing the dwell Hamiltonian by $H_L$.

\subsection{Residual localization and finite-frequency loss}

For a fixed total time $T=n\tau$, the accumulated local coherence is
$f(\tau)^n$. Its logarithm distinguishes dephasing that survives as
$\tau\to0$ from a loss proportional to the dwell length. Expanding
Eq.~\eqref{eq:protected-two-rate} near zero gives
\begin{equation}
 \log f(\tau)=-\lambda\tau-\frac{D_{\rm noise}}2\tau^2+O(\tau^3),
 \quad
 \lambda=\frac{4\gamma_1-\gamma_2}{3},
 \quad
 D_{\rm noise}=\frac49(\gamma_2-\gamma_1)^2.
 \label{eq:protected-noise-cumulants}
\end{equation}
Complete positivity implies $\lambda\ge0$. Quadratic diffusion,
$\Gamma(L)=\Lambda L^2$, has $\lambda=0$: more frequent correction removes
its logical dephasing. For Gaussian localization, let $\Lambda_0$ be
the event rate and $r_c$ the localization length. The full coherence-decay
kernel is
\begin{equation}
 \Gamma(L)=\Lambda_0[1-e^{-L^2/(4r_c^2)}],
 \label{eq:protected-Gaussian-kernel}
\end{equation}
Substitution into $\lambda=(4\gamma_1-\gamma_2)/3$ leaves the rate
\begin{equation}
 \lambda=\frac{\Lambda_0}{3}
 [3-4e^{-q}+e^{-4q}],\qquad q=\frac{r^2}{4r_c^2},
 \qquad
 \lambda=\frac{\Lambda_0r^4}{8r_c^4}+O(\Lambda_0r^6/r_c^6).
 \label{eq:protected-Gaussian-floor}
\end{equation}
At fixed $d,r_c,\Lambda_0$, shrinking this code reduces its gravitational
coupling as $r^6$ and its residual localization as $r^4$. Thus the
interaction-to-residual-noise ratio decreases for sufficiently small $r$.

For equal local residual rates, the continuously recovered state from
$|++\rangle$ is entangled at some time if and only if
\begin{equation}
 2g_{\rm hex}>\lambda.
 \label{eq:protected-entanglement-condition}
\end{equation}
To see this, remove the local Hamiltonian phases and apply a Hadamard to
each output. The state is an $X$ state whose potentially negative
partial-transpose eigenvalue is
\[
 \frac{1-e^{-2\lambda t}}4
 -\frac{e^{-\lambda t}}2|\sin(2g_{\rm hex}t)|.
\]
Negativity is therefore equivalent to
$|\sin(2g_{\rm hex}t)|>\sinh(\lambda t)$. The inequalities
$|\sin(2g_{\rm hex}t)|\le2g_{\rm hex}t$ and
$\sinh(\lambda t)\ge\lambda t$ prove necessity; expansion near zero
proves sufficiency. At $T_*$ the target fidelity is
\begin{equation}
 F_{\rm Zeno}=\frac{(1+e^{-\lambda T_*})^2}{4},
 \qquad F_{\rm Zeno}>\tfrac12
 \ \Longleftrightarrow\ 
 \lambda T_*<\log(1+\sqrt2).
 \label{eq:protected-Zeno-fidelity}
\end{equation}

Gravity introduces an additional finite-period loss because physical
states can leave the code during a dwell. We resolve the physical
frequencies into the same orthonormal error functions used by the recovery.
Let $h_{u,st}$ be the frequency component from logical sector $u$ into
error-function sector $(s,t)$, and let $\sigma_{uv}^2$ sum the squared
differences of these components for two logical sectors:
\begin{align}
 h_{u,st}&=\frac19\sum_{j\in S_{u_A},\ k\in S_{u_B}}
 f_s(\theta_j)f_t(\theta_k)\nu_{jk},
 \qquad s,t\in\{0,x,z\},\nonumber\\
 \sigma_{uv}^2&=\sum_{(s,t)\ne(0,0)}
 (h_{u,st}-h_{v,st})^2.
 \label{eq:protected-gravity-cumulant}
\end{align}
Here $h_{u,00}=\epsilon_u$ is the retained mean phase. The remaining
components determine how much the two logical sectors differ in their
excursions into error spaces. For identical local noise kernels, the
logarithm of the complete coherence multiplier is
\begin{equation}
 \log M_{uv}(\tau)
 =-[d_H(u,v)\lambda+i(\epsilon_u-\epsilon_v)]\tau
 -\frac{\tau^2}{2}
 [d_H(u,v)D_{\rm noise}+\sigma_{uv}^2]+O(\tau^3),
 \label{eq:protected-full-cumulant}
\end{equation}
where $d_H$ is the Hamming distance of the two logical strings. On each
parity triangle, the three-by-three matrix
$O_{sj}=f_s(\theta_j)/\sqrt3$ is orthogonal. Parseval's identity therefore
turns the gravitational second cumulant into the sum of squares
$\sigma_{uv}^2$. For the mixed term, the row identity
\[
 \sum_{j'\in S_{1-b}}W_{jj'}
 \Gamma(|\mathbf a_j-\mathbf a_{j'}|)
 =4\gamma_1-\gamma_2=3\lambda,
 \qquad j\in S_b,
\]
is independent of $j$. For equal parities it is zero. Consequently the
mixed noise--gravity second moment cancels against the corresponding
product of first moments. This proves Eq.~\eqref{eq:protected-full-cumulant}
without omitting a mixed term.

To control the truncation rather than only its formal order, we bound
every damping rate and frequency difference in the exact finite sum.
The following $Z_*$ bounds the modulus of each exponential's coefficient:
\[
 Z_*=2\max(\gamma_1,\gamma_2)
       +\max_{j,k}\nu_{jk}-\min_{j,k}\nu_{jk}.
\]
The absolute sum of the normalized weights in
Eq.~\eqref{eq:protected-exact-cycle} is at most $25/9$. For $\Re z\ge0$,
the integral Taylor remainder of $e^{-z\tau}$ after quadratic order has
magnitude at most $|z\tau|^3/6$. Hence
\begin{equation}
 |R_{uv}^{(3)}(\tau)|\le\frac{25}{54}(Z_*\tau)^3.
 \label{eq:protected-cycle-remainder}
\end{equation}
For parameter evaluation we use Eq.~\eqref{eq:protected-finite-fidelity};
the expansion identifies why finite recovery frequency loses signal.

\subsection{Finite control errors and elapsed time}

The exact $F_n$ assumes the stated recovery is implemented perfectly.
We compare each implemented cycle with that reference as a channel, so
that the error estimate includes inputs entangled with arbitrary memories.
Suppose each actual
joint cycle, including its decoder, re-encoding, reset and any retained
classical record, differs from the stated ideal cycle by at most
$\varepsilon$ in half-diamond norm. The bound is uniform over inputs,
ancillary systems and prior records. Suppose also that the initial
preparation has trace distance at most $\varepsilon_{\rm init}$ from
$|++\rangle\langle++|$. Channel telescoping and contraction of trace
distance give
\begin{equation}
 F_{\rm actual}\ge F_n-n\varepsilon-\varepsilon_{\rm init}.
 \label{eq:protected-control-fidelity}
\end{equation}
Leakage and flagged failures are included in the actual channel; no
outcome is discarded. If the error budgets are specified as the full
diamond norm and the full trace norm, respectively, their contribution
to Eq.~\eqref{eq:protected-control-fidelity} is half those budgets.

We can also bound $F_n$ without evaluating powers of the entire cycle
matrix. The first-order recovered evolution has residual dephasing rates
$\lambda_A,\lambda_B$; the second-order discrepancy accumulates in
proportion to $T_*\tau$. For each summand of
Eq.~\eqref{eq:protected-exact-cycle}, define its signed weight $c$,
total damping $\Gamma_{jj'kk'}$, and frequency difference $\delta\omega$
after removing the mean logical phase. We collect their second moments
in the nonnegative constant $B$:
\begin{align}
 c&=W_{jj'}W_{kk'}/81,\nonumber\\
 \Gamma_{jj'kk'}&=\Gamma_A(|\mathbf a_j-\mathbf a_{j'}|)
                 +\Gamma_B(|\mathbf b_k-\mathbf b_{k'}|),\nonumber\\
 \delta\omega&=\nu_{jk}-\nu_{j'k'}-(\epsilon_u-\epsilon_v),\nonumber\\
 \lambda_{uv}&=\lambda_A\mathbf1_{u_A\ne v_A}
                 +\lambda_B\mathbf1_{u_B\ne v_B},\nonumber\\
 B&=\frac1{32}\sum_{u\ne v}
 \left[\sum_{\text{terms}}|c|
       (\Gamma_{jj'kk'}^2+\delta\omega^2)+\lambda_{uv}^2\right].
 \label{eq:protected-control-B}
\end{align}
The sums over terms use exactly the indices of
Eq.~\eqref{eq:protected-exact-cycle}; $\lambda_{uv}$ is the residual
damping of logical coherence $uv$. This $B$ is a cycle-error coefficient
with units of inverse time squared. The fidelity is bounded by the
continuously recovered value minus finite-period, control and preparation
losses:
\begin{equation}
 F_{\rm actual}\ge
 \frac{(1+e^{-\lambda_AT_*})(1+e^{-\lambda_BT_*})}{4}
 -T_*\tau B-\frac{T_*}{\tau}\varepsilon-\varepsilon_{\rm init}.
 \label{eq:protected-closed-control}
\end{equation}
To prove it, remove the mean logical phase from $M_{uv}$ and compare its
first-order expansion with $e^{-\lambda_{uv}\tau}$. The scalar inequality
\[
 |e^{-z\tau}-1+z\tau|\le\tfrac12|z|^2\tau^2,
 \qquad \Re z\ge0,
\]
follows by the integral remainder. It bounds the one-cycle difference by
$\tau^2/2$ times the bracket in Eq.~\eqref{eq:protected-control-B}.
Both compared multipliers have magnitude at most one, so the difference
of their $n$th powers is at most $n$ times this bound. Summing the
off-diagonal terms in Eq.~\eqref{eq:protected-finite-fidelity} and using
Eq.~\eqref{eq:protected-control-fidelity} proves the result.

For equal quadratic noise, write $\gamma=\Gamma(r)$ and $D=d/r$.
At $D=3.6$, evaluating the finite sum separates a term quadratic in the
noise rate from a term quadratic in the gravitational frequency:
\begin{equation}
 B=\frac{56}{9}\gamma^2
   +0.00849138127713216\left(\frac{\mathcal G}{\hbar r}\right)^2.
 \label{eq:protected-B-calibration}
\end{equation}
The first term is recovery loss from random forces. The second charges
gravitational excursions into the syndrome spaces.
For a fixed per-cycle error budget, the loss
$T_*(\tau B+\varepsilon/\tau)$ is minimized at
$\tau=\sqrt{\varepsilon/B}$ before the integer constraint
$n=T_*/\tau$ is imposed. Thus increasing the recovery frequency indefinitely
is counterproductive when each operation has a nonzero error.
For a chosen finite $n$, Eq.~\eqref{eq:protected-closed-control} is the
applicable certificate. For equal residual rates, let
$\Delta=F_{\rm Zeno}-1/2-\varepsilon_{\rm init}>0$ be the fidelity margin
available for finite-period and control errors. Minimizing their sum
before imposing the integer-cycle constraint gives the sufficient
requirement
\begin{equation}
 \varepsilon<\frac{\Delta^2}{4BT_*^2}.
 \label{eq:protected-optimal-error-budget}
\end{equation}
One then evaluates the two nearest positive integers to
$T_*\sqrt{B/\varepsilon}$ in Eq.~\eqref{eq:protected-closed-control}.
Failure of this sufficient bound alone gives no impossibility result.

Pulse duration can be charged separately when the masses remain within
their balls throughout the pulse. Suppose the ideal local decoder and
reset take elapsed time $\delta$, and their calibrated half-diamond error
without the omitted drift is $\varepsilon_{\rm ctl}$. For the quadratic
generator in Eq.~\eqref{eq:protected-force-generator}, let
$\gamma=\Lambda r^2$ and $\omega=\mathcal G/(\hbar r)$. A sufficient
complete-pulse allowance is
\begin{equation}
 \varepsilon\le\varepsilon_{\rm ctl}
 +\delta\left[
 \frac{\omega}{2}\left(\frac1{D-2}-\frac1{D+2}\right)+4\gamma\right].
 \label{eq:protected-pulse-duration}
\end{equation}
The Hamiltonian commutator has diamond norm at most the spectral diameter
of $V/\hbar$, yielding the first term after division by two. The constraint
$\sum_iq_{Q,i}^2\le r^2I$ bounds one laboratory's diffusion generator in
diamond norm by $4\Lambda r^2$, using its completely positive and
anticommutator parts. The two laboratories therefore contribute at most
$4\gamma\delta$ in half-diamond norm. Duhamel's formula in the control
interaction picture proves Eq.~\eqref{eq:protected-pulse-duration}.
Any other omitted dynamics need their own allowance. With preparation
duration $T_{\rm prep}$ and readout duration $T_{\rm read}$, the total
elapsed time, including the $n$ recovery pulses, is
\begin{equation}
 T_{\rm wall}=T_{\rm prep}+n(\tau+\delta)+T_{\rm read}.
 \label{eq:protected-wall-time}
\end{equation}
These formulas specify what a decoder must achieve; a bounded
finite-dimensional recovery dilation alone does not establish the
corresponding motional operation on a massive object.

\subsection{A finite-cycle Bell-pair rate advantage under white-force noise}
\label{sec:protected-finite-cycle-rate}

We now compare Bell pairs produced per unit time when recovery has a finite
period and a nonzero error. The planar code gives a strictly larger rate
than every axial code in the following common class of reset protocols.
During each nominal dwell, Alice and Bob use fixed local encodings that
obey Eq.~\eqref{eq:protected-KL}. A local recovery returns each laboratory
to its code and discards its quantum syndrome. All retained logical
systems, memories and classical records are included in the channel.
Boundary LOCC may depend on the records and may change the protected
codes. Initial entanglement is absent, and every cycle used to generate
an entangled seed is counted. Local memory processing and distillation
have the same ideal time convention on both sides of the comparison.

Write the nominal channel for a dwell of duration $\tau>0$ as
\begin{equation}
 \Phi_\tau=\mathcal R e^{\tau\mathcal L}\mathcal E,
 \qquad \mathcal R\mathcal E=\mathrm{id},
 \label{eq:finite-cycle-nominal}
\end{equation}
where $\mathcal E$ is a product of local isometric encodings and
$\mathcal R$ is a product of local recovery channels. The identity acts
on the complete retained system. Registers may be embedded in a fixed
ambient space so that a cycle and its first-order generator have the
same domain and codomain. The generator $\mathcal L$ consists of the
Newtonian Hamiltonian and local terms. The axial converse will use the trace norm after partial transpose,
which bounds overlap with a maximally entangled target. We therefore
control a cycle in both the ordinary norm and the norm after transposing
Bob's input and output. For a map $\mathcal M$, let
$\mathcal M^\theta=\Theta_B\mathcal M\Theta_B$, where $\Theta_B$ transposes
all Bob's registers in fixed bases. We assume uniformly over the allowed
codes, memories and conditional branches that
\begin{equation}
 \|\mathcal L\|_\diamond,\ \|\mathcal L^\theta\|_\diamond\le M<\infty.
 \label{eq:finite-cycle-generator-budget}
\end{equation}
The common finite constant $M$ bounds both generator norms independently
of the code and retained memory. The implemented complete cycle
$\widetilde\Phi_\tau$ obeys both error bounds
\begin{equation}
 \frac12\|\widetilde\Phi_\tau-\Phi_\tau\|_\diamond\le\epsilon,
 \qquad
 \frac12\|(\widetilde\Phi_\tau-\Phi_\tau)^\theta\|_\diamond
       \le\epsilon e^{M\tau}.
 \label{eq:finite-cycle-error-budget}
\end{equation}
For example, product local recovery faults whose local half-diamond errors
sum to at most $\epsilon$ satisfy these bounds. Finite recovery pulses,
gravity during the pulses and any departure from the nominal confinement
must be included in this complete-cycle comparison.

The axial estimate needed below is the product-interaction bound already
proved in Theorem~\ref{thm:protected-axial}, with the entropy-rate constant
$\beta$ removed. Define the frequency
\begin{equation}
 J_{\rm ax}=\frac{\mathcal G}{\hbar d}F(r/d)
 =\frac{\mathcal G}{\hbar r}
   \int_0^\infty e^{-(d/r)v}
       \left[2\sum_{k\ge3}I_k(v)\right]^2\,dv,
 \label{eq:finite-cycle-Jax}
\end{equation}
where $I_k$ are modified Bessel functions. In particular, the ratio
$g/J_{\rm ax}>2.5104$ at $d/r=3.6$ follows from
Eq.~\eqref{eq:protected-hexagon-gap}, with $g=g_{\rm hex}$.

The theorem compares a converse for every axial protocol in this class
with a distillable rate from independent planar blocks. In the axial
bound, $q$ is an upper bound on the one-cycle amplification of partial-
transpose norm. For the planar construction, $p$ bounds Bell infidelity
and $Y(p)$ is the corresponding hashing yield in Bell pairs per block.
We write $h_2(p)=-p\log_2p-(1-p)\log_2(1-p)$.

\begin{theorem}[Finite-cycle rate separation]
\label{thm:protected-finite-cycle-rate}
Consider reset protocols satisfying
Eqs.~\eqref{eq:finite-cycle-nominal}--\eqref{eq:finite-cycle-error-budget},
with a fixed dwell $\tau$ and a hard budget of $N$ cycles. Every adaptive
axial protocol that aims to produce $m$ Bell pairs has target fidelity
\begin{equation}
 F_{2^m}\le 2^{-m}q^N,
 \qquad
 q=e^{2J_{\rm ax}\tau}+(M^2\tau^2+2\epsilon)e^{M\tau}.
 \label{eq:finite-cycle-strong-converse}
\end{equation}
Its asymptotic rate per unit dwell time therefore satisfies
$R_{\rm ax}\le\log_2q/\tau$, with exponentially vanishing fidelity above
this rate.

Suppose the planar code is subject to white-force diffusion
$\Gamma(L)=aL^2$, with $0<a<\infty$, and its nominal local recovery exactly
corrects $\{I,x,y,z\}$. Let independent, identically prepared blocks use
$T=\pi/(4g)=n\tau$ of dwell time. If initialization has half-trace error at
most $\epsilon_{\rm init}$, define
\begin{equation}
 \eta=T\left(\frac{M^2\tau}{2}+\frac{\epsilon}{\tau}\right),
 \qquad p=\epsilon_{\rm init}+\eta\le\frac34,
 \qquad Y(p)=1-h_2(p)-p\log_2 3.
 \label{eq:finite-cycle-planar-error}
\end{equation}
With $Y(p)>0$, twirling and one-way hashing achieve
$R_{\rm pl}\ge Y(p)/T$. If each planar cycle has additional charged time
at most $\chi\tau$, the charged-time rate instead satisfies
$R_{\rm pl}\ge Y(p)/[T(1+\chi)]$; the upper bound
$R_{\rm ax}\le\log_2q/\tau$ remains valid for axial charged time.

For $g/J_{\rm ax}>2.5104$, $M\tau\le0.01$, $\eta\le0.004$ and
$\epsilon_{\rm init}\le0.001$, these bounds give
\begin{equation}
 TR_{\rm pl}\ge\frac{0.946660}{1+\chi},\qquad
 TR_{\rm ax}<0.914375.
 \label{eq:finite-cycle-numerical-gap}
\end{equation}
Thus the planar rate is strictly larger, including for a positive
recovery-time overhead $\chi\le0.01$.
\end{theorem}

\begin{proof}
We first identify the part of the axial first-order generator that can
increase entanglement. The Chebyshev expansion
$e^{v t}=I_0(v)+2\sum_{k\ge1}I_k(v)T_k(t)$ has nonnegative coefficients
for $v\ge0$. Removing $T_0,T_1,T_2$ leaves a tail of operator norm at most
$E_3(v)=2\sum_{k\ge3}I_k(v)$. The removed terms compress to scalars on
every protected code. Substitution into the axial Laplace representation
of $1/(d+x_B-x_A)$ therefore writes the nonlocal compressed Hamiltonian as
$H_{\rm nl}/\hbar=\sum_\alpha A_\alpha\otimes B_\alpha$, with Hermitian
factors and $\sum_\alpha\|A_\alpha\|\|B_\alpha\|\le J_{\rm ax}$.
The sum can equivalently be a norm-convergent integral. Expanding the
Bessel functions gives
$E_3(v)=\sum_{n\ge3}e_n v^n/n!$ with exactly the coefficients in
Eq.~\eqref{eq:protected-ek}. Termwise integration proves both expressions
for $J_{\rm ax}$ in Eq.~\eqref{eq:finite-cycle-Jax}.

We next connect this compressed interaction bound to a finite cycle.
The first derivative of the recovered cycle at zero dwell is
$\mathcal G_0=\mathcal R\mathcal L\mathcal E$.
It is a bounded GKSL generator: the CPTP curve $\Phi_\tau$ starts at the
identity, so its derivative is conditionally completely positive and
trace preserving. A channel that inverts an isometric encoding erases
coherences between that code and its orthogonal complement. To verify
this, write $R_kW=c_kI$ for its Kraus operators; trace preservation gives
$\sum_k c_k^*R_kv=0$ for $v\perp\operatorname{ran}W$. Consequently the
Hamiltonian part of $\mathcal G_0$ is the commutator with $W^\dagger HW$;
the recovered local generators remain local.

Conjugating a local channel by full transpose gives another channel.
Thus the local part of $\mathcal G_0^\theta$ generates a contractive
semigroup. For a product Hamiltonian, the transposed commutator is the
difference of two left-right multiplication maps, each with diamond norm
at most $\|A_\alpha\|\|B_\alpha\|$. The bounded perturbation expansion
then yields
\[
 \|e^{\tau\mathcal G_0^\theta}\|_\diamond
 \le e^{2J_{\rm ax}\tau}.
\]
Because the local maps $\mathcal R^\theta$ and $\mathcal E^\theta$ have
diamond norm one, $\|\mathcal G_0^\theta\|_\diamond\le M$.
The constant and linear terms of $\Phi_\tau^\theta$ and
$e^{\tau\mathcal G_0^\theta}$ agree. Each Taylor remainder is at most
$M^2\tau^2e^{M\tau}/2$, so
\[
 \|\Phi_\tau^\theta-e^{\tau\mathcal G_0^\theta}\|_\diamond
 \le M^2\tau^2e^{M\tau}.
\]
Adding Eq.~\eqref{eq:finite-cycle-error-budget} proves
$\|\widetilde\Phi_\tau^\theta\|_\diamond\le q$.

The one-cycle estimate composes across an adaptive experiment because
intermediate LOCC, including its classical records, is contractive after
partial transpose. Starting with separable laboratories therefore gives
$\|\rho_N^\theta\|_1\le q^N$ even with adaptive choices of codes and
arbitrary retained memories. The partial transpose of a maximally
entangled state of dimension $D$ is $1/D$ times the swap, so
$F_D\le\|\rho_N^\theta\|_1/D$.
Taking $D=2^m$ proves Eq.~\eqref{eq:finite-cycle-strong-converse}.
The cycle count bounds every branch; an expected stopping-time constraint
alone is a different assumption.

For the planar code, exact correction makes the first-order recovered
white-force generator vanish on the logical quantum systems. The logical
Hamiltonian is $\hbar g Z_AZ_B$, up to local terms, so evolution from a
product of logical $|+\rangle$ states for $T=\pi/(4g)$ gives a Bell pair
up to local unitaries. In ordinary diamond norm both
$e^{s\mathcal L}$ and $e^{s\mathcal G_0}$ are contractions.
Their integral Taylor remainders give
\[
 \frac12\|\Phi_\tau-e^{\tau\mathcal G_0}\|_\diamond
 \le\frac{M^2\tau^2}{2}.
\]
Telescoping $n$ cycles and adding the initialization error show that the
Bell overlap is at least $1-p$. Local twirling gives a Bell-diagonal state
of entropy at most $h_2(p)+p\log_2 3$. Independent repetition, including
reset of the fault process and environment, permits the one-way hashing
rate $Y(p)$ per block\cite{BennettDiVincenzoSmolinWootters1996}.
Dividing by the block duration proves the lower rate bounds.

Finally, $\log(1+x)\le x$ and $J_{\rm ax}\ge0$ imply
\[
 TR_{\rm ax}\le
 \frac{\pi}{2(g/J_{\rm ax})\ln2}
 +\frac{2\eta e^{M\tau}}{\ln2}.
\]
At the stated parameter bounds, $Y(0.005)=0.946660495162600\ldots$ and
the last expression is smaller than $0.914374281718837\ldots$.
The difference exceeds $0.03228$ for $\chi=0$ and $0.02291$ for
$\chi=0.01$.
\end{proof}

Every finite $M$ permits strictly positive control tolerances. Choose
$\tau=T/n$ with
\[
 0<\tau\le\min\left\{\frac{0.004}{TM^2},\frac{0.01}{M}\right\},
 \qquad
 0<\epsilon\le\frac{0.002\tau}{T}.
\]
Then $\eta\le0.004$. These are sufficient specifications for the matched
reset architecture. In a fixed-position mode model with no unbounded
kinetic term, one may take
$M=2\mathcal G/[\hbar(d-2r)]+8ar^2$ for both geometries:
the Laplace-angular product representation bounds the Hamiltonian
commutator in ordinary and transposed diamond norm, and
$\sum_i x_i^2\le r^2 I$ bounds the local white-force generators.
Finite-energy packets require their own dynamical and recovery
justification. Likewise, the Gaussian-localization example with
$F\ge0.8170$ does not supply the hypotheses or fidelity required by this
rate theorem.

Both norms in Eq.~\eqref{eq:finite-cycle-error-budget} matter when memory
dimension is unrestricted. A global fault that outputs a flagged
$D$-dimensional maximally entangled state with probability $\epsilon$
can have ordinary half-diamond error $\epsilon$ while its output partial
transpose norm is $1-\epsilon+\epsilon D$. Local recovery faults avoid
this dimension dependence because full transpose preserves the diamond
norm of a local map. This example explains the two error specifications;
it does not provide an uncounted source of physical entanglement.
\subsection{A fixed-energy encoded construction under ideal continuous recovery}
\label{sec:fixed-energy-gaussian-recovery}

The residual Gaussian localization of the hexagon raises a different
question: can another code approach a perfect Bell pair at the same
geometry and finite noise strength? We construct such a family for the
full Newtonian interaction on two disjoint balls of radius $r$, with
centre separation $d>2r$. Higher angular order suppresses uncorrectable
noise, while a decreasing occupation of that angular component keeps
the mean kinetic energy of every encoded state fixed.
Recovery is ideal and local, with unrestricted noiseless auxiliary systems
and syndrome reset. This is an encoded-energy statement; it does not bound
momentum bandwidth, recovery work or the energy of every noisy trajectory.

Let the local kinetic operators be the Dirichlet Laplacians
$K_\alpha=-\hbar^2\Delta/(2m_\alpha)$ on the balls. This explicit confinement
keeps every finite dwell inside the supported domain. Each local auxiliary
qutrit is noiseless. For laboratory $\alpha=A,B$, Gaussian localization
is a Poisson process of rate $\Lambda_\alpha$: one event applies the
channel $\mathcal J_\alpha$ with localization length $\ell_\alpha$.
On the motional coordinates, let
\begin{equation}
 \mathcal L_\alpha=\Lambda_\alpha(\mathcal J_\alpha-\mathrm{id}),
 \qquad
 (\mathcal J_\alpha\rho)(\mathbf x,\mathbf x')
 =e^{-|\mathbf x-\mathbf x'|^2/(4\ell_\alpha^2)}
   \rho(\mathbf x,\mathbf x'),
 \label{eq:fixed-energy-gaussian-model}
\end{equation}
where $0\le\Lambda_\alpha<\infty$, $0<\ell_\alpha<\infty$, and the two
noise processes are independent. Gravity is their only nonlocal coupling.

Choose cylindrical coordinates $(s,z,\theta)$ with both axes parallel to
the line joining the centres and with the same angular orientation.
Here $s$ is the transverse radius, $z$ the axial coordinate and $\theta$
the azimuth. We keep the radial and axial profile fixed as the angular
order increases. Choose a smooth $\chi(s,z)$, normalized in $s\,ds\,dz$,
with support strictly inside
\begin{equation}
 0<s_-\le s\le s_+,\qquad |z|\le h,
 \qquad s_+^2+h^2<r^2.
 \label{eq:fixed-energy-profile}
\end{equation}
The positive inner radius $s_-$ keeps the angular kinetic-energy
integral finite. We use the same profile in both laboratories and separate
its radial--axial energy $E_{\chi,\alpha}$ from the inverse-square radial
moment $M_{-2}$ that multiplies the angular energy:
\begin{equation}
 E_{\chi,\alpha}=
 \frac{\hbar^2}{2m_\alpha}\int
 (|\partial_s\chi|^2+|\partial_z\chi|^2)s\,ds\,dz,
 \qquad
 M_{-2}=\int s^{-2}|\chi|^2s\,ds\,dz.
\end{equation}

\begin{proposition}[Fixed encoded mean energy and arbitrarily accurate Bell generation]
\label{prop:fixed-energy-gaussian-recovery}
Fix $0<\eta_A,\eta_B\le1$. Under the model above, local ideal continuous
recovery permits Bell-state fidelity arbitrarily close to one at the fixed
geometry $(r,d)$, while every encoded state of mass $\alpha$ has mean kinetic
energy
\begin{equation}
 E_{\chi,\alpha}+
 \frac{\eta_\alpha\hbar^2}{2m_\alpha}M_{-2}.
 \label{eq:fixed-energy-encoded-energy}
\end{equation}
For every finite requested accuracy, a finite code order and a finite
interaction time suffice. At each such order and time, a finite number of
ideal recovery operations approximates the continuous-recovery output to
arbitrary additional accuracy. There is no uniform bound on that number
or on the resources required to implement the recovery.
\end{proposition}

\begin{proof}
For angular order $n\ge2$, an occupied harmonic costs energy
proportional to $n^2$. We therefore choose its probability to be
$w_\alpha=\eta_\alpha/n^2$, where $0<\eta_\alpha\le1$ fixes the angular
energy allowance. The auxiliary levels $|0\rangle_a,|1\rangle_a$ retain
the logical label in the zero-angular-order part, while the shared level
$|2\rangle_a$ permits a gravitational transition between the logical states.
Encode the logical basis as
\begin{equation}
 \Psi_{j,\alpha}(s,z,\theta)=
 \frac{\chi(s,z)}{\sqrt{2\pi}}
 \left[\sqrt{1-w_\alpha}\,|j\rangle_a+
 \sqrt{w_\alpha}\,e^{i(2j-1)n\theta}|2\rangle_a\right],
 \qquad j=0,1.
 \label{eq:fixed-energy-qutrit-code}
\end{equation}
Let $P_\alpha$ project onto their span and put $P=P_A\otimes P_B$.
The auxiliary labels make these states orthogonal. Direct integration of
the Dirichlet kinetic quadratic form gives
\begin{equation}
 P_\alpha K_\alpha P_\alpha=
 \left(E_{\chi,\alpha}+
 \frac{w_\alpha n^2\hbar^2}{2m_\alpha}M_{-2}\right)P_\alpha.
\end{equation}
The off-diagonal matrix element vanishes by angular orthogonality, so this
energy is the same for every logical state, including a logical subsystem
entangled with other systems. Substituting $w_\alpha n^2=\eta_\alpha$ proves
Eq.~\eqref{eq:fixed-energy-encoded-energy}.

\paragraph{An explicit recovery.}
To recover a low angular-order kick, we must separate its angular
transfer from the logical label. Suppress the laboratory index and write
$|\ell,j\rangle$ for angular harmonic $e^{i\ell\theta}/\sqrt{2\pi}$
with auxiliary level $j$. A multiplication error
$f_\ell(s,z)e^{i\ell\theta}$ gives the radial function $f_\ell\chi$ and
the angular--auxiliary vector
\begin{equation}
 v_{j,\ell}=\sqrt{1-w}\,|\ell,j\rangle+
             \sqrt w\,|\ell+(2j-1)n,2\rangle.
\end{equation}
For $|\ell|,|\ell'|\le n-1$,
\begin{equation}
 \langle v_{j,\ell}|v_{k,\ell'}\rangle
 =\delta_{jk}\delta_{\ell\ell'}.
 \label{eq:fixed-energy-error-orthogonality}
\end{equation}
Indeed, a cross-logical overlap of the common auxiliary level would require
$\ell-\ell'=\pm2n$. Thus an isometry on this correctable subspace maps
\begin{equation}
 |f\rangle_{\rm radial}|v_{j,\ell}\rangle
 \longmapsto |j\rangle_L|f,\ell\rangle_{\rm syndrome}.
\end{equation}
Trace out the syndrome, prepare the fixed profile and re-encode. On the
orthogonal complement, a trace-and-prepare channel completes this to a
normal deterministic recovery. This explicitly resets the radial
continuum as well as the angular syndrome. It uses unrestricted local
operations; it is not a finite-control implementation. In particular it
corrects $I,x,y,z$, since these have angular orders zero and one.

\paragraph{The entire residual noise channel.}
The recovery corrects every transfer of order below $n$. We now bound
the total probability of all remaining transfers, which controls errors
on every logical input. With the laboratory index suppressed, $\ell$
denotes its localization length. The Gaussian jump channel in
Eq.~\eqref{eq:fixed-energy-gaussian-model} is
the mixture of $e^{i\mathbf q\cdot\mathbf x}$ with density
$\ell^3\pi^{-3/2}e^{-\ell^2|\mathbf q|^2}$. Expanding a transverse kick
into angular harmonics and averaging its azimuth eliminates cross terms
between distinct angular transfers. The transfer-$\ell'$ Kraus operators
are radial multipliers times $e^{i\ell'\theta}$. At radius $s$, their
total probability is
\begin{equation}
 p_{\ell'}(s)=e^{-a(s)}I_{|\ell'|}(a(s)),
 \qquad a(s)=s^2/(2\ell^2),
\end{equation}
where $I_j$ is the modified Bessel function. Equivalently, $\ell'$ has the
distribution of the difference of two independent Poisson variables of
mean $a(s)/2$. Let $q_{\alpha,n}$ denote the actual probability of
$|\ell'|\ge n$, averaged over $|\chi(s,z)|^2s\,ds\,dz$. Since the absolute
difference is at most the sum of those Poisson variables,
\begin{equation}
 q_{\alpha,n}\le \frac{a_\alpha^n}{n!},
 \qquad a_\alpha=\frac{s_+^2}{2\ell_\alpha^2}.
 \label{eq:fixed-energy-factorial-tail}
\end{equation}
Here the last bound is the Poisson factorial-moment inequality.

The probability of each angular-transfer sector is independent of the
logical input and of $w_\alpha$: its Kraus norm depends only on $(s,z)$,
and the compressed radial multiplication operator is scalar. All low
transfers are corrected exactly. Consequently, on logical inputs, the
recovered jump channel is exactly
\begin{equation}
 (1-q_{\alpha,n})\mathrm{id}+q_{\alpha,n}\Phi_{\alpha,n}
 \label{eq:fixed-energy-residual-channel}
\end{equation}
for a channel $\Phi_{\alpha,n}$. This controls the full residual channel,
including logical flips and leakage handled by the recovery, rather than
one selected coherence.

\paragraph{The retained Newtonian coupling.}
Write $\mathcal G=Gm_Am_B$. The two basis states have the same position
density; their transition density is
$w_\alpha|\chi|^2e^{2in\theta}/(2\pi)$. With the aligned cylindrical axes,
the potential depends on the angles only through
$\phi=\theta_A-\theta_B$. Simultaneous rotational invariance therefore
eliminates single logical flips and the $|00\rangle\langle11|$ term. It
gives exactly, with $V_0$ the common diagonal potential energy,
\begin{equation}
 PVP=V_0I-\hbar w_Aw_Bg_{2n}
 (|01\rangle\langle10|+|10\rangle\langle01|).
 \label{eq:fixed-energy-exchange}
\end{equation}
For clarity, writing $\chi_\alpha=\chi(s_\alpha,z_\alpha)$, $g_m$ uses
the complex Fourier coefficient convention:
\begin{equation}
 \begin{split}
 g_m={}&\frac{\mathcal G}{\hbar}
 \int |\chi_A|^2|\chi_B|^2
 \left[\frac1{2\pi}\int_0^{2\pi}
 \frac{e^{im\phi}\,d\phi}
 {\sqrt{A-2s_As_B\cos\phi}}\right]
 s_A\,ds_A\,dz_A\,s_B\,ds_B\,dz_B,\\
 A={}&(d+z_B-z_A)^2+s_A^2+s_B^2.
 \end{split}
 \label{eq:fixed-energy-Fourier-convention}
\end{equation}
The bracket is real and positive for $m\ge1$. To compare the surviving
signal with the factorial noise tail, we need a positive lower bound on
this Fourier coefficient that is uniform over the fixed profiles. Bound
the denominator scale by $Q$, the radial product ratio from below by $b$,
and denote the reciprocal-square-root series coefficients by $c_k$:
\begin{equation}
 Q=(d+2h)^2+2s_+^2,\qquad b=s_-^2/Q,
 \qquad c_k=\binom{2k}{k}/4^k.
\end{equation}
The binomial expansion of the reciprocal square root has nonnegative
contributions to Fourier order $m$. Its order-$m$ term contributes
$c_m A^{-1/2}(s_As_B/A)^m$: the $2^m$ from the potential expansion cancels
the $2^{-m}$ coefficient of $e^{im\phi}$ in $\cos^m\phi$. Since $A\le Q$
and $s_As_B\ge s_-^2$, normalization of the profiles gives
\begin{equation}
 g_{2n}\ge \frac{\mathcal G}{\hbar\sqrt Q}c_{2n}b^{2n}
 \ge \frac{\mathcal G}{\hbar\sqrt Q(4n+1)}b^{2n}>0.
 \label{eq:fixed-energy-geometric-signal}
\end{equation}
The last inequality follows because the central binomial coefficient is
the largest of the $4n+1$ coefficients whose sum is $2^{4n}$.

\paragraph{Continuous recovery with the full kinetic Hamiltonian.}
Let $\mathfrak E_n$ be the two-laboratory code isometry channel and let
$\mathfrak D_n$ be the product of the two decoding/reset channels above,
with $\mathfrak D_n\mathfrak E_n=\mathrm{id}$ on the logical qubits. Let
$\mathcal L$ generate the physical evolution with
$H=K_A+K_B+V$ and the two noise generators in
Eq.~\eqref{eq:fixed-energy-gaussian-model}. The potential is bounded on
the disjoint balls, and the Poisson noise generators are bounded on trace
class. Hence this is a strongly continuous physical semigroup with the
Dirichlet kinetic Hamiltonian retained.

We must also show that repeated physical recovery approaches the
claimed logical evolution while retaining kinetic dynamics. For each
fixed $n$, the smooth compactly supported code vectors lie in
the Hamiltonian domain. Their matrix units therefore lie in the
trace-class generator domain. On the finite-dimensional logical input,
\begin{equation}
 M_{\tau,n}:=\mathfrak D_n e^{\tau\mathcal L}\mathfrak E_n
 =\mathrm{id}+\tau\mathcal G_n+o(\tau)
 \label{eq:fixed-energy-cycle-derivative}
\end{equation}
in diamond norm. To identify the Hamiltonian part, note that a decoder
which acts identically on the code kills cross operators between the
code and its orthogonal complement. A Stinespring isometry sends a code
vector to its logical vector tensored with one fixed syndrome state;
orthogonality forces every complementary vector to have zero component
in this same syndrome sector. Thus the Hamiltonian derivative after
decoding is the commutator with the compressed Hamiltonian. The local
kinetic compressions are scalar by Eq.~\eqref{eq:fixed-energy-encoded-energy}.
Combining Eqs.~\eqref{eq:fixed-energy-residual-channel} and
\eqref{eq:fixed-energy-exchange} gives
\begin{equation}
 \mathcal G_n(\rho)=
 -\frac{i}{\hbar}[PVP,\rho]
 +\sum_{\alpha=A,B}\Lambda_\alpha q_{\alpha,n}
   (\Phi_{\alpha,n}\otimes\mathrm{id}_{\bar\alpha}-\mathrm{id})(\rho),
 \label{eq:fixed-energy-logical-generator}
\end{equation}
where $PVP$ denotes its logical matrix. The finite-dimensional product
formula now implies
$M_{T/N,n}^{\circ N}\to e^{T\mathcal G_n}$ in diamond norm for every fixed
$n$ and finite $T$. This proves the finite-cycle approximation assertion
without replacing the kinetic Hamiltonian by a static phase model.

\paragraph{Bell fidelity.}
We can now compare the time needed to generate entanglement with the
probability of a residual jump during that time. From the separable
logical state $|01\rangle$, the noiseless exchange in
Eq.~\eqref{eq:fixed-energy-exchange} reaches a Bell state at the following
$T_n$. The constant $T_*$ in this subsection is the geometry-dependent
prefactor of its upper bound:
\begin{equation}
 T_n=\frac{\pi}{4w_Aw_Bg_{2n}}
 \le T_*n^4(4n+1)b^{-2n},\qquad
 T_* =\frac{\pi\hbar\sqrt Q}
 {4\mathcal G\eta_A\eta_B}.
\end{equation}
The residual terms in Eq.~\eqref{eq:fixed-energy-logical-generator} have
state-independent Poisson rates. The no-residual-jump trajectory has
the ideal Bell output, so its probability bounds the fidelity from below.
Let $F_n$ be the overlap with that Bell target in the continuous-recovery
model. With $\Lambda=\Lambda_A+\Lambda_B$ and $a=\max(a_A,a_B)$,
\begin{equation}
 1-F_n\le
 \sum_\alpha\Lambda_\alpha q_{\alpha,n}T_n
 \le\Lambda T_*n^4(4n+1)
       \frac{(a/b^2)^n}{n!}\longrightarrow0.
 \label{eq:fixed-energy-Bell-error}
\end{equation}
For any requested error, first choose a finite $n$ making this bound
small, and then choose a finite number of cycles making the approximation
in Eq.~\eqref{eq:fixed-energy-cycle-derivative} as accurate as required.
All interaction time is included in $T_n$, and no entangled seed is used.
\end{proof}

The encoded-state energy is fixed immediately after each recovery.
During a finite dwell, localization kicks can heat the masses and gravity
can exchange kinetic and potential energy. Bounding these two changes
extends the energy statement to every time within a dwell, after averaging
over the unobserved noise.

\begin{proposition}[A uniform mean kinetic-energy bound during finite dwells]
\label{prop:fixed-energy-active-mean}
Let every dwell have length at most a fixed $\tau_{\max}<\infty$, and let
each ideal recovery re-encode into Eq.~\eqref{eq:fixed-energy-qutrit-code}.
Write $E_{\rm code}=\sum_\alpha[E_{\chi,\alpha}+
\eta_\alpha\hbar^2M_{-2}/(2m_\alpha)]$. The mean localization heating rate
$h_{\rm heat}$ and the range $\operatorname{osc}(V)$ of the pair potential
are
\begin{equation}
 h_{\rm heat}=\sum_{\alpha=A,B}
 \frac{3\Lambda_\alpha\hbar^2}{4m_\alpha\ell_\alpha^2},\qquad
 \operatorname{osc}(V)=\mathcal G\left(\frac1{d-2r}-\frac1{d+2r}\right).
\end{equation}
During every active dwell, the total kinetic energy, averaged over the
unobserved localization noise and all previous outcomes, obeys
\begin{equation}
 \mathbb E\langle K_A+K_B\rangle
 \le E_{\rm code}+\operatorname{osc}(V)+h_{\rm heat}\tau_{\max}.
 \label{eq:fixed-energy-active-mean}
\end{equation}
The bound is independent of the code order and number of cycles. Thus the
finite-cycle approximations in Proposition~\ref{prop:fixed-energy-gaussian-recovery}
can also be chosen with a uniform unconditional active-dwell mean-energy
bound. No bound on recovery work or individual noise-conditioned
trajectories is asserted.
\end{proposition}

\begin{proof}
The multiplication unitary $e^{i\mathbf q\cdot\mathbf x_\alpha}$ preserves
the Dirichlet quadratic-form domain and commutes with $V$. Its change of
the kinetic quadratic form is
\begin{equation}
 \frac{\hbar}{m_\alpha}\mathbf q\cdot\langle\mathbf p_\alpha\rangle
 +\frac{\hbar^2|\mathbf q|^2}{2m_\alpha}.
\end{equation}
The first term averages to zero, and the Gaussian kick distribution has
$\mathbb E|\mathbf q|^2=3/(2\ell_\alpha^2)$. Between jumps the mean of
$H=K_A+K_B+V$ is conserved. The Poisson unraveling therefore gives
$\mathbb E\langle H(t)\rangle=\mathbb E\langle H(0)\rangle+h_{\rm heat}t$.
This argument also establishes finite mean energy by summing the finite
mean number of jumps and their finite second moments. Each dwell starts
with kinetic energy $E_{\rm code}$, while the change in mean potential
energy is bounded by $\operatorname{osc}(V)$. This proves
Eq.~\eqref{eq:fixed-energy-active-mean}. For any finite target interaction
time, the continuous-recovery approximation may use arbitrarily many
cycles, so imposing $\tau\le\tau_{\max}$ does not restrict that limit.
\end{proof}

\paragraph{Scope of the resources.}
The code energy in Eq.~\eqref{eq:fixed-energy-encoded-energy} holds after
each ideal re-encoding and throughout the ideal Zeno evolution. It is
not the same as a common kinetic-energy bound at every time on every
noise-conditioned history. Equation~\eqref{eq:fixed-energy-active-mean}
adds a uniform unconditional mean bound during finite dwells; its
allowance includes the potential variation and noise heating, and need
not equal the smaller encoded-state allowance. The recovery explicitly disposes of a radial
syndrome with an infinite-dimensional state space; approximating that
operation at finite precision requires its own resources. Smooth compact
profiles and high angular orders also do not impose a momentum cutoff.
The occupied angular harmonic is $n$, its probability is proportional to
$n^{-2}$, and the kinetic-energy second moment grows as $n^2$. There is
therefore no uniform nonrelativistic bandwidth guarantee as $n\to\infty$.
The result does not provide a fixed pulse-error allowance, bounded
recovery work or a finite-rate comparison with other protected codes.

For the ideal resource convention, a finite choice with Bell infidelity
$\delta<0.1$ also gives a positive achievable Bell-pair rate. Independent
repetition, bilateral twirling and one-way hashing give\cite{BennettDiVincenzoSmolinWootters1996}
\begin{equation}
 C_{\rm ideal}\ge
 \frac{1-h_2(\delta)-\delta\log_2 3}{T_n}>0,
\end{equation}
where $h_2$ is the binary entropy, and local processing and noiseless
storage are free. This is an
achievable rate under the stated ideal recovery assumptions, not the
optimal protected capacity or a practical-rate estimate.

\subsection{An operational tolerance for imperfect position-error correction}
\label{sec:approximate-position-correction}

Exact correction removes the low position moments from the logical
interaction. To quantify an imperfect cancellation, we use a specified
calibration channel. Its strength is fixed; its recovery infidelity is
neither the physical force-noise strength nor the error of a complete
experimental recovery cycle.

Write $X_j=x_j/r$ on a mass supported in its radius-$r$ ball, with local
ancillary identities understood. We construct a calibration channel
whose exact Knill--Laflamme conditions are precisely the required first-
and second-moment conditions. The identity and position errors share an
output flag $|s\rangle$; a second, orthogonal flag $|f\rangle$ supplies
the remaining probability needed for trace preservation. Its Kraus
operators are
{\small
\begin{equation}
 K_0=|s\rangle\otimes I/\sqrt2,\qquad
 K_j=|s\rangle\otimes X_j/\sqrt2\quad(j=x,y,z),\qquad
 K_f=|f\rangle\otimes\sqrt{(I-\sum_jX_j^2)/2},
 \quad\langle s|f\rangle=0.
 \label{eq:position-calibration-channel}
\end{equation}
}
The support constraint makes the square root positive, and the Kraus
operators are complete. The flag makes $K_f^\dagger K_j=0$ for
$j=0,x,y,z$. Exact correctability is therefore equivalent to the
Knill--Laflamme conditions for $\{I,x,y,z\}$: the remaining failure
product is a linear combination of the successful quadratic products.

Let $\mathcal N$ denote this calibration channel and, for encoding
isometry $W$, let $\mathcal W(\rho)=W\rho W^\dagger$ and
$\mathcal N_P=\mathcal N\circ\mathcal W$. Its complementary channel
$\widehat{\mathcal N_P}$ outputs the environment register indexed by the
five Kraus operators. Exact recovery is possible when this environment
learns nothing about the logical input. We quantify approximate recovery
by its distance from a constant environmental output:
\begin{equation}
 \delta_P=\frac12\inf_{\sigma}
 \|\widehat{\mathcal N_P}-[\rho\mapsto\sigma\operatorname{Tr}\rho]
 \|_\diamond,
 \label{eq:position-environment-defect}
\end{equation}
where $\sigma$ is a normalized environmental state. References are
included in this norm. For finite code spaces the infimum is attained;
the bounds below are uniform in their dimensions and pass to the stated
operational limits.

Use $\delta_A,\delta_B$ for the defects of Alice's and Bob's codes.
Their residual first moments can restore a dipole interaction; residual
second moments also couple to higher moments. The following bound keeps
these contributions separate. In this subsection $D=d-2r$ is the minimum
physical clearance between the support balls.

\begin{proposition}[Capacity under approximate position-error correction]
\label{prop:approximate-position-capacity}
Let $\mathcal G=Gm_Am_B$, $D=d-2r>0$, and let $H_L$ be the Newtonian
Hamiltonian compressed to two fixed codes. Its ideal coherent capacity
satisfies
\begin{equation}
 C_E(H_L)\le\frac{\beta\mathcal G}{\hbar}
 \left[
 \frac{16\delta_A\delta_Br^2}{D^3}
 +\frac{8\sqrt2(\delta_A+\delta_B)r^4}{D^5}
 +\frac{20r^6}{D^7}\right].
 \label{eq:approximate-position-capacity}
\end{equation}
If a calibration recovery $\mathcal R$ has worst-case entanglement
infidelity
\begin{equation}
 \varepsilon_F=1-\inf_{\Psi}
 \langle\Psi|[(\mathcal R\mathcal N_P)\otimes\mathrm{id}](\Psi)
 |\Psi\rangle,
 \label{eq:position-recovery-infidelity}
\end{equation}
then $\delta_P\le\sqrt{\varepsilon_F}$. The infimum is over pure states
of the logical input and a reference. The capacity in
Eq.~\eqref{eq:approximate-position-capacity} concerns the coherent
compressed Hamiltonian, not its capacity in the presence of residual
logical noise.
\end{proposition}

\begin{proof}
Choose a minimizing $\sigma$ in Eq.~\eqref{eq:position-environment-defect}
and write $\mathcal D=\widehat{\mathcal N_P}-\sigma\operatorname{Tr}$.
Complete norm duality gives
$\|\mathcal D^*\|_{\rm cb}=\|\mathcal D\|_\diamond=2\delta_P$, where
$\|\cdot\|_{\rm cb}$ is the completely bounded operator norm. We use
this identity to bound compressed position moments by environmental
observables. The off-diagonal Kraus blocks contain first moments and the
position--position blocks contain second moments. In the environment let
$|t_\theta\rangle=\cos\theta|x\rangle+\sin\theta|y\rangle$,
which is orthogonal to the Kraus labels $|0\rangle$ and $|z\rangle$.
The following four self-adjoint observables select the real and imaginary
parts of the first and second harmonic moments in direction $\theta$:
\begin{align*}
 H_{1,R}&=|0\rangle\langle z|+|z\rangle\langle0|,&
 H_{1,I}&=|0\rangle\langle t_\theta|
                   +|t_\theta\rangle\langle0|,\\
 H_{2,R}&=2(|z\rangle\langle z|
                   -|t_\theta\rangle\langle t_\theta|),&
 H_{2,I}&=2(|z\rangle\langle t_\theta|
                   +|t_\theta\rangle\langle z|).
\end{align*}
To retain a dimension-independent bound, we bound each real--imaginary
pair as one operator column. Its norm is the square root of the norm
of the sum of the two component squares, giving
\[
 \left\|\binom{H_{1,R}}{H_{1,I}}\right\|=\sqrt2,
 \qquad
 \left\|\binom{H_{2,R}}{H_{2,I}}\right\|=2\sqrt2.
\]
Indeed, their sums of squares are respectively
$2|0\rangle\langle0|+|z\rangle\langle z|
 +|t_\theta\rangle\langle t_\theta|$
and $8(|z\rangle\langle z|+|t_\theta\rangle\langle t_\theta|)$.

Put $q_\theta=x_z+i(x_x\cos\theta+x_y\sin\theta)$.
The complementary $0j$ and $ij$ blocks contain
$W^\dagger X_jW/2$ and $W^\dagger X_iX_jW/2$.
Thus applying $\mathcal D^*$ to the displayed columns gives the real
and imaginary parts of $W^\dagger(q_\theta/r)^nW$, for $n=1,2$,
after subtracting their appropriate scalar values. Complete norm duality,
including its rectangular-column amplification, consequently gives
\begin{equation}
 a_1\le2\sqrt2\,\delta_Pr,\qquad
 a_2\le4\sqrt2\,\delta_Pr^2,
 \label{eq:position-low-moment-columns}
\end{equation}
where $a_n$ is the norm of the column of centered real and imaginary
parts. These are complete-norm estimates, not entrywise moment estimates.
For $n\ge3$ no subtraction is needed: compression of
$|q_\theta|\le r$ gives $a_n\le r^n$.

The Newtonian Laplace representation contains compressed exponentials
of $q_\theta$. Subtract their scalar terms through order two, which affect
only local Hamiltonians in the separated product. With $s\ge0$ the Laplace
variable and $x=sr$, the moment bounds give the real--imaginary column
norm at most
\[
 2\sqrt2\,\delta_P(x+x^2)+\sum_{n\ge3}\frac{x^n}{n!}
 \le e^x\left(2\sqrt2\,\delta_Px+\frac{x^3}{6}\right).
\]
Subtracting a scalar factor on either side changes the separated
Newtonian interaction only by local terms. Apply the real sum-of-squares
entropy-rate bound to both angular components together, then integrate
the two column bounds against $e^{-ds}$. The three terms use
$\int_0^\infty s^ne^{-Ds}ds=n!/D^{n+1}$ for $n=2,4,6$.
Their coefficients are $16$, $8\sqrt2$ and $20$, proving
Eq.~\eqref{eq:approximate-position-capacity} without a dimension factor.

For the last assertion, use the root-fidelity convention in the
complementary-channel recovery theorem of B\'eny and Oreshkov
\cite{BenyOreshkov2010}. A recovery with the displayed squared
entanglement infidelity gives worst-case root fidelity at least
$\sqrt{1-\varepsilon_F}$. Their theorem gives one constant environmental
channel with at least the same worst-case fidelity. The state
fidelity--trace-distance inequality, uniformly including references,
then gives $\delta_P\le\sqrt{\varepsilon_F}$. For each finite code,
$\mathcal N_P$ has output support of dimension at most five times the
code dimension, so the finite-dimensional theorem applies even when
the physical wavefunctions have continuous support.
\end{proof}

\subsection{Finite-energy codes attaining the correction-precision scaling}
\label{sec:approximate-position-attainment}

We now construct codes whose retained coherent interaction has both
the $d^{-7}$ exact-correction term and a dipole term proportional to the
allowed calibration infidelity. A small imbalance in reflected packet
weights changes the first moment while preserving scalar second moments.
Let $\ell=r/2$, and choose one normalized radial smooth packet with
support radius $w<r/8$. Translate it to the four disjoint centers
$\ell,-\ell/2,-\ell,\ell/2$ on the axis. Write $|z\rangle$ for the
packet centered at $z\hat z$, and let $v$ be its variance in each
coordinate. Define
\begin{equation}
 |0_t\rangle=\sqrt{\frac{1+2t}{3}}|\ell\rangle
             +\sqrt{\frac{2-2t}{3}}|-\ell/2\rangle,
 \qquad |1_t\rangle=\mathsf{Reflection}|0_t\rangle,
 \quad 0\le t\le\tfrac12.
 \label{eq:approximate-four-packets}
\end{equation}
Their supports lie inside the balls. Packet width and kinetic energy can
remain fixed as $t$ and $d$ vary. On the encoded qubit,
\[
 W^\dagger X_zW=\mu Z,\qquad
 W^\dagger X_z^2W=bI,\qquad
 \mu=t\ell/r,\qquad b=\frac{(1+t)\ell^2/2+v}{r^2}.
\]
The transverse first moments and mixed second moments vanish, while
$W^\dagger X_x^2W=W^\dagger X_y^2W=(v/r^2)I$.
All off-diagonal logical matrix elements vanish by disjoint support.

Only the complementary $0z$ and $z0$ blocks depend on the logical input.
With $H_{0z}=|0\rangle\langle z|+|z\rangle\langle0|$, the difference
from the midpoint constant channel is
\[
 \mathcal D(\rho)=\frac\mu2\operatorname{Tr}(Z\rho)H_{0z}.
\]
Its diamond norm is $\mu$: the functional has complete norm one and
$\|H_{0z}\|_1=2$. Conversely, the two logical basis states give
environment outputs separated by trace norm $2\mu$, so no constant
channel has diamond distance smaller than $\mu$ from both. Hence
\begin{equation}
 \delta_P=\mu/2.
 \label{eq:approximate-four-packet-delta}
\end{equation}

The environmental defect measures distinguishability, but we also need
a recovery whose infidelity has the claimed order. Split the $z$ error
into its component inside the code and an orthogonal error isometry $E_z$:
\[
 X_zW=\mu WZ+\sqrt{b-\mu^2}\,E_z,
 \qquad W^\dagger E_z=0,\quad E_z^\dagger E_z=I.
\]
The normalized $X_xW$, $X_yW$ and $E_z$ are mutually orthogonal error
isometries, all orthogonal to $W$. The flagged failure map has scalar
squared norm and an orthogonal output flag, so it is another correctable
isometry. Project onto these syndrome ranges and map each isometry back
to the code; complete the map arbitrarily on their orthogonal complement.
This is a trace-preserving recovery that fixes the code. It gives exactly
\begin{equation}
 \mathcal R\mathcal N_P
 =(1-\mu^2/2)\mathrm{id}+(\mu^2/2)Z(\cdot)Z,
 \qquad \varepsilon_F=\mu^2/2=2\delta_P^2.
 \label{eq:approximate-four-packet-recovery}
\end{equation}
The minimum fidelity is reached by a logical state with zero $Z$
expectation, including a maximally entangled input. This recovery is
certified explicitly; its global optimality is not needed or asserted.

The compressed Newtonian Hamiltonian is diagonal. Radial averaging of
the nonsingular Newtonian kernel, first over one packet and then the
other, gives exactly the potential between their centers. Let $m_j$ be
the half-difference of the two logical distributions' centre moments.
These moment coefficients, indexed by degree $j$, are distinct from the
physical masses $m_A,m_B$.
Even $m_j$ vanish, and for odd $j$,
\[
 m_j=\frac{\ell^j}{3}
       \left[(1+2t)-\frac{2-2t}{2^j}\right]\ge0,
 \qquad m_1=t\ell,\quad m_3=(1+3t)\ell^3/4.
\]
All surviving odd moment coefficients are nonnegative, so the terms of
the convergent centre expansion add with the same sign. Keeping just the
dipole--dipole and cubic--cubic terms gives the positive Ising energy
\[
 J_t=\mathcal G\sum_{\substack{j,k\ge1\\j,k\text{ odd}}}\binom{j+k}{j}
                  \frac{m_jm_k}{d^{j+k+1}}
 \ge\frac{2\mathcal Gm_1^2}{d^3}
      +\frac{20\mathcal Gm_3^2}{d^7}.
\]
Choose $t=2\sqrt{2\varepsilon}$ for $0\le\varepsilon\le1/32$.
Equation~\eqref{eq:approximate-four-packet-recovery} has infidelity
exactly $\varepsilon$, while
\begin{equation}
 J_t\ge4\mathcal G\varepsilon\frac{r^2}{d^3}
       +\frac{5\mathcal G}{256}\frac{r^6}{d^7}.
 \label{eq:approximate-position-lower}
\end{equation}

Let $C_{\rm comp,opt}(\varepsilon;r,d)$ denote the supremum of coherent
compressed-Hamiltonian capacity over codes for which each local
calibration admits recovery infidelity at most $\varepsilon$.
For $d\ge4r$ and $0\le\varepsilon\le1/32$, the upper bound and the
four-packet construction yield
\begin{equation}
 C_{\rm comp,opt}(\varepsilon;r,d)
 \asymp\frac{\mathcal G}{\hbar}
       \left(\frac{r^6}{d^7}
                   +\varepsilon\frac{r^2}{d^3}\right).
 \label{eq:approximate-position-crossover}
\end{equation}
For example, a common lower constant is $5\beta/256$ and a common
upper constant is $\beta(2560+256\sqrt2)$, independent of dimensions,
separation and tolerance. For the upper bound, use $D\ge d/2$,
$\delta_A,\delta_B\le\sqrt\varepsilon$ and the
arithmetic--geometric mean inequality on the mixed term. The lower
bound uses the established Ising capacity $\beta J_t/\hbar$, with
seed preparation charged.

Thus a uniform guarantee of the exact-correction $d^{-7}$ order requires
calibration tolerance $\varepsilon=O((r/d)^4)$; in the complementary
norm its corresponding worst-case order is $\delta=O((r/d)^2)$.
Necessity here concerns the class of all admitted codes, not every
individual code or every chosen recovery. For the physical generator
$\gamma\sum_j\mathcal D[X_j]$, the same syndrome recovery leaves the
logical generator $\gamma\mu^2\mathcal D[Z]$ in the infinitely frequent
recovery limit. The restored dipole therefore brings residual noise as
well as coherent signal. Equation~\eqref{eq:approximate-position-crossover}
is not an achievable noisy-throughput statement or a complete-cycle
control-error budget.

\subsection{Information lost through an exposed radiative reset}
\label{sec:exposed-reset}

A recovery must remove its syndrome without revealing the logical state.
We can quantify the lost information by asking how much logical
coherence any deterministic recovery can restore after a photon escapes.
The following calculation answers this question for one reset model. During an unobserved isotropic emission of one photon of wave
number $k$, suppose the logical information remains in the alternating
three-vertex states of a radius-$\rho$ hexagon. The point-mode photon
channel multiplies a spatial coherence between vertices $\mathbf q_i$
and $\mathbf q_j$ by $\operatorname{sinc}(k|\mathbf q_i-\mathbf q_j|)$,
where $\operatorname{sinc}z=\sin z/z$ and its value at zero is one.
This is a specified emission channel, not a model of every possible
syndrome reset.

The photon preserves the populations on the two disjoint vertex sets
but suppresses their cross coherence. The trace norm of the resulting
cross-coherence matrix will determine the best recovery. We use
entanglement fidelity with a maximally entangled logical reference, so
perfect recovery has fidelity one.

\begin{proposition}[Optimal recovery after an exposed photon]
\label{prop:exposed-photon}
Let $C$ be the $3\times3$ matrix of photon coherence factors between the
even and odd vertices. Optimizing over all deterministic local recovery
channels, the entanglement fidelity of the recovered logical qubit is
\begin{equation}
 F_{e,\mathrm{opt}}=\frac12+\frac{\|C\|_1}{6}
 =\frac{1+\eta_{\mathrm{ph}}}{2},\qquad
 \eta_{\mathrm{ph}}=\frac{|4\operatorname{sinc}z-
                       \operatorname{sinc}(2z)|}{3},\quad z=k\rho.
 \label{eq:exposed-photon-fidelity}
\end{equation}
The smallest half-diamond distance from the ideal logical identity is
$(1-\eta_{\mathrm{ph}})/2$. For $z>0$,
$\eta_{\mathrm{ph}}\le\min\{1,3/(2z)\}$.
\end{proposition}

\begin{proof}
Let $\mathcal H_0,\mathcal H_1$ be the orthogonal even- and odd-vertex
spaces. The channel sends the logical off-diagonal operator to $C/3$
from $\mathcal H_1$ to $\mathcal H_0$, and the two logical populations
to normalized states on their respective spaces. For any recovery,
each population contribution to entanglement fidelity is at most one.
The adjoint recovery is unital and completely positive, so the relevant
block of its image of $|1\rangle\langle0|$ has operator norm at most
one. The coherence contribution is therefore at most $\|C\|_1/3$.
These two bounds give $F_e\le1/2+\|C\|_1/6$.

To attain the coherence bound, align each left and right singular
vector of $C$ with the corresponding logical basis vector. For a
singular-value decomposition
$C=\sum_j s_j|u_j\rangle\langle v_j|$, take recovery Kraus operators
$R_j=|0\rangle\langle u_j|+|1\rangle\langle v_j|$.
They are complete on the six modes, preserve both populations, and
retain coherence $\sum_j s_j/3$. Thus the bound is attained by a
logical dephasing channel. Testing its Choi state against the ideal
Bell projector proves the corresponding lower bound on half-diamond
distance; the attaining dephasing channel has exactly that distance.

Order the odd vertices opposite the even vertices. With
$a=\operatorname{sinc}z$ and $b=\operatorname{sinc}(2z)=a\cos z$,
we have $C=aJ+(b-a)I$, where $J$ is the all-ones matrix.
Its eigenvalues are $b+2a,b-a,b-a$, giving
$\|C\|_1=|a|(4-\cos z)=|4\operatorname{sinc}z-
\operatorname{sinc}(2z)|$. Finally
$|4\sin z-\tfrac12\sin2z|\le9/2$ proves the stated bound.
\end{proof}

A reset may emit only occasionally, without revealing which cycles
emitted. If the photon occurs with input-independent probability
$p_{\mathrm{ph}}$ and otherwise the channel is the identity, the same
recovery must act on both possibilities. Linearity of entanglement
fidelity and the bound for one emitted photon give, for every such recovery,
\begin{equation}
 \frac12\|\mathcal R[(1-p_{\mathrm{ph}})\mathrm{id}
          +p_{\mathrm{ph}}\mathcal D_{\mathrm{ph}}]\mathcal E
          -\mathrm{id}_L\|_\diamond
 \ge\frac{p_{\mathrm{ph}}(1-\eta_{\mathrm{ph}})}2.
 \label{eq:exposed-photon-budget}
\end{equation}
Here $\mathcal E$ encodes the logical qubit. An exact common-recovery
formula follows by replacing $C$ with
$C_p=(1-p_{\mathrm{ph}})J+p_{\mathrm{ph}}C$ in
Eq.~\eqref{eq:exposed-photon-fidelity}. Thus the bound does not assume
that the recovery knows whether a photon was emitted.
For $\rho=125\,\mu\mathrm m$ and wavelength $800\,\mathrm{nm}$,
$\eta_{\mathrm{ph}}=0.00135812218\ldots$ and
$F_{e,\mathrm{opt}}=0.50067906109\ldots$.
An isolated reset error at most $10^{-6}$ relative to logical identity
therefore requires $p_{\mathrm{ph}}\le2.00272\times10^{-6}$.

This last comparison is to an ideal reset. It must not be equated
without further calculation to the complete-cycle error $\varepsilon$
relative to a noisy reference cycle. For example, a preceding reference
dephasing with coherence factor $s$ multiplies the additional channel
distance by $s$. Nor does a single-cycle lower bound alone imply an
accumulated decay rate. An exponential decay requires a specified
independent, input-independent event process and repeated re-encoding
into the same exposed logical basis.

There is a controlled finite-width extension. Suppose the six normalized
packets have disjoint supports and mean-square displacements from their
centres at most $\sigma^2$. Compare the true recoil unitary
$e^{ik\mathbf n\cdot\mathbf X}$ with its branch-centre value.
Averaging $|e^{iu}-1|^2\le u^2$ over isotropic directions gives a
Stinespring-isometry distance at most $k\sigma/\sqrt3$, uniformly over
logical inputs and references. The channel half-diamond distance, and
hence the change in optimized entanglement fidelity, is at most this
quantity. The point-mode bound remains informative when this allowance
is small. Transferring the logical information to an unaffected local
system before emission changes the encoding and avoids the premise of
this obstruction; realizing that transfer and reset remains a control
problem.

\section{Further geometry, energy and confinement results}
\label{sec:extensions}
\label{sec:si-extensions}

The main capacity and mean-time results leave several distinct resource
questions. A bound on every history's duration permits a different
statistical converse; a common kinetic-energy budget prevents arbitrarily
sharp endpoint localization; finite barriers require control of leakage;
and near-contact geometry can favour more than two spatial levels. We
establish these extensions with their respective assumptions. The
collinear decomposition comes first because it supplies the exact
interaction strength used by both time and energy bounds.

\subsection{Confined Newtonian interaction and confidence at bounded time}
\label{sec:newtonian-confidence}

We first impose a hard bound on the interaction duration of every
history, including histories selected by feedback. For collinear supports,
the full Newtonian potential gives an exponential bound on the power of
a test with prescribed size. The position domains fix a finite nonlocal
interaction strength after local gravitational phases have been removed.
We evaluate that strength exactly and use it to bound the final state's
departure from the separable set. The statistical conversion is an instance
of the established max-relative-entanglement method for bipartite
interactions\cite{DasBaumlWilde2020}; product-decomposition
growth estimates also underlie Hamiltonian circuit-cutting
bounds\cite{HarrowLowe2025}.

Let $\mathcal G=Gm_A m_B$, $|x_A|\le\ell_A$, $|x_B|\le\ell_B$, and
$d_{\min}=d-\ell_A-\ell_B>0$, throughout all interaction and local-control
steps. The only quantum interaction between the laboratories is
$V=-\mathcal G/(d+x_B-x_A)$. We allow initially separable local memories,
local instruments and classical feedback, with all flags retained. A final
effect $0\le M\le I$ has size at most $\alpha$ on every separable state of
the complete laboratories and power $p>\alpha$ on the actual output.
The effect $M$ includes the trusted verification measurement, evaluated on the laboratories before readout. Reading only an already produced acceptance flag would have all-separable size one.
All logarithms below are natural unless a base is displayed.

\subsection{The nonlocal strength of the full potential}

A local phase cannot entangle the laboratories, so we remove it before
bounding the interaction. In the Laplace representation of the reciprocal
separation, the two position factors are $e^{sx_A}$ and $e^{-sx_B}$,
with $s\ge0$ of units inverse length. Subtracting each factor's spectral
midpoint minimizes its operator norm. Define
\[
 A_s=e^{sx_A}-\cosh(s\ell_A)I,\qquad
 B_s=e^{-sx_B}-\cosh(s\ell_B)I.
\]
Their norms are $\sinh(s\ell_A)$ and $\sinh(s\ell_B)$. Expanding the product
gives $V=V_A\otimes I+I\otimes V_B+V_{\rm int}$, up to a scalar, with
\begin{equation}
 V_{\rm int}=-\mathcal G\int_0^\infty e^{-sd}A_s\otimes B_s\,ds.
 \label{eq:newton-centered-integral}
\end{equation}
Every integral is bounded because $d_{\min}>0$. After centering, the
integral of the product norms bounds the strength that can generate
entanglement. Dividing by $\hbar$ defines its frequency $\kappa$:
\begin{align}
 \kappa&=\frac{\mathcal G}{\hbar}\int_0^\infty
 e^{-sd}\sinh(s\ell_A)\sinh(s\ell_B)\,ds\nonumber\\
 &=\frac{\mathcal G}{4\hbar}\left[
 \frac1{d-\ell_A-\ell_B}+\frac1{d+\ell_A+\ell_B}
 -\frac1{d-\ell_A+\ell_B}-\frac1{d+\ell_A-\ell_B}\right].
 \label{eq:newton-kappa}
\end{align}
For small allowed displacements,
$\kappa=2\mathcal G\ell_A\ell_B/(\hbar d^3)
+O[\mathcal G\ell_A\ell_B(\ell_A^2+\ell_B^2)/(\hbar d^5)]$.
We use the full expression in all bounds below.

We can certify that no alternative subtraction of local terms lowers
this product-norm sum. Evaluate the potential at the four support corners
with alternating signs: every local function cancels, leaving a lower
bound on any sum of product norms. Thus Eq.~\eqref{eq:newton-kappa} is
also the minimum product-norm sum of $V/\hbar$ modulo local operators.
Indeed, consider any representation by
local summands and product terms $\sum_j c_jX_j\otimes Y_j$, including
limits in the projective operator norm. The alternating expectation at the
four support corners cancels the local summands and has magnitude at most
$4\sum_j|c_j|\|X_j\|\|Y_j\|$. For $V/\hbar$ its value is $4\kappa$.
Smooth normalized packets approaching the corners justify these
expectations without position eigenstates. Equation~\eqref{eq:newton-centered-integral}
attains the matching product-norm sum. If either $\ell_A$ or $\ell_B$ is
zero, $\kappa=0$ and the potential is local; the divisions by $\kappa$
below concern positive widths.

\subsection{A hard time budget, including adaptive histories}

We follow subnormalized pure branches so that local filtering cannot hide
its failure probability. For a bipartite vector with coefficient operator
$C$, the nuclear norm $\|C\|_1$ is its sum of Schmidt amplitudes. This norm
has a dimension-independent interaction bound and a useful averaged
monotonicity under local instruments. Its square also equals the trace
of an explicit separable positive operator that dominates the branch
state. This converts its growth bound into a bound on any final test,
without assuming that the output is close to a Bell state.

\begin{theorem}[Confined Newtonian bound at a hard time budget]
\label{thm:newton-hard-time}
Under the supported interaction and separable-input assumptions above,
suppose every history has total interaction duration at most $T_{\max}$.
For any final test of size $0<\alpha<p\le1$,
\begin{equation}
 p\le\alpha e^{2\kappa T_{\max}},\qquad
 T_{\max}\ge\frac{\ln(p/\alpha)}{2\kappa}.
 \label{eq:newton-hard-time}
\end{equation}
The statement holds for finite protocols and their normal trace-norm
limits with the same pathwise time bound and support constraint.
\end{theorem}

\begin{proof}
Remove local Hamiltonians in a local interaction picture. Their unitaries
preserve the norms of the factors in
Eq.~\eqref{eq:newton-centered-integral}. The nuclear-norm ideal property,
$\|ACB^{\mathsf T}\|_1\le\|A\|\|B\|\|C\|_1$, applied to the integral
evolution equation gives
\[
 \|C(t+\delta)\|_1\le e^{\kappa\delta}\|C(t)\|_1.
\]
This argument applies on the nuclear class and does not require
differentiable Schmidt coefficients.

We must also control the branches created by intermediate measurements.
For a local Kraus family $\{K_i\}$, write
$C=\sum_j s_j|u_j\rangle\langle\bar v_j|$, where $s_j\ge0$. The triangle
inequality and Kraus completeness give
\begin{align*}
 \sum_i\|K_iC\|_1^2
 &\le\sum_{j,k}s_js_k\sum_i\|K_i u_j\|\|K_i u_k\|\\
 &\le\sum_{j,k}s_js_k=\|C\|_1^2.
\end{align*}
The second inequality is Cauchy--Schwarz over $i$. The same proof applies
to operations in Bob's laboratory. Thus local instruments cannot increase
the sum over outcomes, even though a normalized selected outcome can have
more entanglement.

Let $C_h$ describe a subnormalized branch and $t_h$ its accumulated
interaction duration. The discounted sum
$\sum_h e^{-2\kappa t_h}\|C_h\|_1^2$ does not increase under either
interaction or a local instrument. It starts at one for a pure-product
decomposition of a separable input. The pathwise bound therefore implies
$\sum_h\|C_h\|_1^2\le e^{2\kappa T_{\max}}$.

To turn the branch-norm estimate into a testing bound, we construct the
separable dominator explicitly. For
$|\psi_h\rangle=\sum_j s_{hj}|u_{hj}v_{hj}\rangle$, put
$S_h=\sum_j s_{hj}$. Weighted Cauchy--Schwarz shows that the separable
positive operator
\[
 T_h=S_h\sum_j s_{hj}|u_{hj}v_{hj}\rangle\langle u_{hj}v_{hj}|
 \ \ge\ |\psi_h\rangle\langle\psi_h|
\]
has trace $S_h^2$. Summing over all histories, including their flags, gives
a separable positive dominator of the output with trace at most
$e^{2\kappa T_{\max}}$. The size bound on $M$ gives
Eq.~\eqref{eq:newton-hard-time}. General separable inputs follow by finite
separable-mixture approximation. The final probability passes to
trace-norm protocol limits because $M$ is bounded.
\end{proof}

The dominator also gives $1+R_g(\rho)\le e^{2\kappa T_{\max}}$ whenever
the displayed branch construction is used, where $R_g$ is generalized
entanglement robustness. Equation~\eqref{eq:newton-hard-time} requires a
bound on every history. An expectation cannot replace $T_{\max}$ in its
exponential: a protocol may rarely run a long entangling sequence and
otherwise stop immediately. The entropy argument in
Section~\ref{sec:si-evidence} gives the bound whose cost is linear in
expected time. For parallel interactions,
the resource in this proof is the history-dependent sum
$\int\sum_j\kappa_j(t)\,dt$; unlimited parallel pairs do not obey the same
wall-clock bound as one pair.

\subsection{When the endpoint construction applies to other potentials}
\label{sec:newton-fixed-sign-kernels}

The endpoint construction works because all the product interactions
favour the same two local positions. We can identify this property directly
from the potential. Let $W(x,y)$ be a real $C^2$ interaction energy on
$[x_0,x_1]\times[y_0,y_1]$, with both intervals of positive length. We keep
the ideal local-control and supported-protocol convention of
Proposition~\ref{prop:newton-capacity}. In particular, local memories are
free and packet energies may grow along an approximating sequence.
The mixed derivative $W_{xy}$ removes all terms that depend on only one
position. We integrate its absolute value to bound the total strength
of the remaining product interactions:
\begin{equation}
 \kappa_W^{\rm abs}=\frac1{4\hbar}
 \int_{x_0}^{x_1}\!du\int_{y_0}^{y_1}\!dv\,
 |\partial_x\partial_yW(u,v)|.
 \label{eq:fixed-sign-strength}
\end{equation}

\begin{proposition}[Capacity for a mixed derivative of fixed sign]
\label{prop:fixed-sign-capacity}
The ideal entanglement capacity of $W$ is at most
$\beta\kappa_W^{\rm abs}$. If $\partial_x\partial_yW$ is everywhere
nonnegative or everywhere nonpositive, then
\begin{equation}
 C_W=\beta\kappa_W,\qquad
 \kappa_W=\frac{|W(x_1,y_1)+W(x_0,y_0)
                         -W(x_1,y_0)-W(x_0,y_1)|}{4\hbar}
 =\kappa_W^{\rm abs}.
 \label{eq:fixed-sign-capacity}
\end{equation}
For $\kappa_W>0$, the interaction is reversibly equivalent to
$\hbar\kappa_W Z\otimes Z$ in the closure of finite ideal-control
protocols. If $\kappa_W=0$ under the sign condition, $W$ is local.
\end{proposition}

\begin{proof}
We first separate the local terms by integrating the mixed derivative.
For the position operators $X,Y$, let
$P_u=\mathbf1_{X\ge u}$ and $Q_v=\mathbf1_{Y\ge v}$ be their spectral
step projections. Integrating these steps reconstructs the potential
from its mixed derivative; centering each projection removes only local
terms. The fundamental theorem of calculus gives, modulo local operators,
\[
 W(X,Y)=\int_{x_0}^{x_1}\!du\int_{y_0}^{y_1}\!dv\,
 W_{xy}(u,v)(P_u-I/2)\otimes(Q_v-I/2).
\]
Each centred projection has spectral diameter at most one and norm at
most $1/2$. Thus the established product-Hamiltonian simulation and
entropy-rate bounds\cite{ChildsLeungVidal2004} give the total Ising
strength $\kappa_W^{\rm abs}$ and rate $\beta\kappa_W^{\rm abs}$.
For continuous positions, approximate $X,Y$ by finite position grids.
The error is bounded by
\[
 \|W(X,Y)-W(X_\delta,Y_\delta)\|
 \le \|W_x\|_\infty\|X-X_\delta\|
       +\|W_y\|_\infty\|Y-Y_\delta\|.
\]
The finite-grid simulation therefore converges on each finite protocol,
by the same channel-error argument as above.

When $W_{xy}$ has one sign, its integral is the corner difference in
Eq.~\eqref{eq:fixed-sign-capacity}. On the two endpoint states, every
centred projection restricts to the same $Z/2$, up to the common choice
of basis order. Hence all terms add with the same sign and realize an
Ising interaction of strength $\kappa_W$. Smooth stationary packets
approaching these endpoints give the reverse simulation: the displayed
derivative bound controls the potential variation over each branch
rectangle, and Duhamel's inequality controls the full evolution with
the local kinetic terms retained. Each packet has finite energy.
Choosing its width separately for each finite block proves equality
and reversible simulation. If the corner difference vanishes under
the sign condition, continuity implies $W_{xy}=0$ throughout, so
$W(x,y)=W(x,y_0)+W(x_0,y)-W(x_0,y_0)$ is local.
\end{proof}

We can now see why an axial result need not extend to two displaced
parallel lines. For transverse separation $b\ge0$, their interaction is
\[
 W_b(x,y)=-\frac{\mathcal G}{\sqrt{(d+y-x)^2+b^2}},\qquad
 (W_b)_{xy}=\frac{\mathcal G[2(d+y-x)^2-b^2]}
 {[(d+y-x)^2+b^2]^{5/2}}.
\]
Take $|x|\le\ell_A$, $|y|\le\ell_B$ and
$z_{\min}=d-\ell_A-\ell_B>0$, $z_{\max}=d+\ell_A+\ell_B$.
The sign of the mixed derivative is fixed whenever its numerator keeps
one sign over the entire allowed range. The corner formula is therefore
exact in either regime
\begin{equation}
 b^2\le2z_{\min}^2\quad\hbox{or}\quad b^2\ge2z_{\max}^2.
 \label{eq:fixed-sign-parallel}
\end{equation}
Between these regimes, opposite signs can cancel in the outer corner
difference while smaller spatial codes still interact. In units with
$\mathcal G=d=1$, take $\ell_A=\ell_B=0.1$. The outer difference is
positive at $b=0$ and behaves as $-0.04/b^3+O(b^{-5})$ for large $b$,
so it has a zero. One zero is $b_*=1.403173365214517\ldots$.
At this value, the rectangle
$x\in[0.08,0.1]$, $y\in[-0.1,-0.08]$ has corner difference
\begin{equation}
 W_{b_*}(0.1,-0.08)+W_{b_*}(0.08,-0.1)
 -W_{b_*}(0.1,-0.1)-W_{b_*}(0.08,-0.08)
 =-2.2025884191\ldots\times10^{-5}.
 \label{eq:parallel-zero-counterexample}
\end{equation}
Thus this smaller code has positive capacity even though the outer
corner expression is zero. The conclusion also follows without the
numerical root: $(W_b)_{xy}$ cannot vanish throughout a rectangle of
positive area, so some smaller corner difference is nonzero at any
such cancellation. The fixed-sign condition is sufficient for the
endpoint formula; it cannot simply be omitted.

\subsection{A strict capacity gap at fixed kinetic energy}
\label{sec:newton-fixed-energy}

Approaching the endpoint capacity requires increasingly narrow position
states. We now show that a fixed kinetic-energy budget prevents any
protocol from reaching that capacity. The reason is simple: a wavefunction
that vanishes at the confinement boundary must spend kinetic energy to
place its probability near that boundary. A finite budget therefore
leaves a definite fraction of the mass's position distribution in the
interior, where the product factors in the Newtonian potential are smaller.

In this subsection the masses are equal, $m_A=m_B=m$, their collinear
supports are $[-\ell,\ell]$, and $d>2\ell$. Thus
$\mathcal G=Gm^2$ and $\kappa=\kappa(\ell,\ell)$ in
Eq.~\eqref{eq:newton-kappa}. Let $K_j$ be the Dirichlet kinetic-energy
quadratic form $-\hbar^2\partial_{x_j}^2/(2m)$. Vector-valued
wavefunctions belong to $H_0^1([-\ell,\ell])$, with arbitrary local
memories included in their coefficient space. Let $\rho_{j,h}(t)$ be the
conditional state of mass $j$ in recorded history $h$. We impose the bound
\begin{equation}
 \operatorname{Tr}(\rho_{j,h}(t)K_j)\le\mathcal E,
 \qquad j=A,B,
 \label{eq:newton-active-energy}
\end{equation}
in every normalized recorded history $h$ throughout each interaction.
Here $\mathcal E$ is one fixed budget for the entire protocol, including
its gravitational seed preparation. Local operations between interactions
remain free, and must return an admissible state before interaction resumes.
The Dirichlet ground-state bound requires
$\mathcal E\ge\pi^2\hbar^2/(8m\ell^2)$. Define
$\delta_{\mathcal E}=\hbar/\sqrt{2m\mathcal E}$, so
$0<\delta_{\mathcal E}\le2\ell/\pi<\ell$. We will show that at least half
the position probability lies outside boundary strips of this width.
That forced interior probability reduces the second moments of the
centered interaction factors and gives the positive frequency gap below.

\begin{proposition}[Finite-energy capacity and confidence bounds]
\label{prop:newton-fixed-energy}
For finite initially separable protocols obeying
Eq.~\eqref{eq:newton-active-energy}, put
\begin{align}
 \Delta\kappa_{\mathcal E}
 &=\frac{\mathcal G}{2\hbar}\left[
 \frac1{d-\delta_{\mathcal E}}-\frac1d
 +\frac1{d+2\ell-\delta_{\mathcal E}}
 -\frac1{d+2\ell-2\delta_{\mathcal E}}\right],
 \label{eq:newton-energy-gap}\\
 \Gamma_{\mathcal E}&=\beta(\kappa-\Delta\kappa_{\mathcal E}).
 \nonumber
\end{align}
Then $0<\Delta\kappa_{\mathcal E}<\kappa$, and the asymptotic
entanglement capacity under this energy constraint satisfies
\begin{equation}
 C_{\rm kin}(\mathcal E)\le\Gamma_{\mathcal E}<\beta\kappa.
 \label{eq:newton-fixed-energy-rate}
\end{equation}
Any final test of all-separable size $0<\alpha<p\le1$ obeys
\begin{equation}
 d_{\rm bin}(p\|\alpha)
 \le\Gamma_{\mathcal E}\ln2\,\mathbb E T_{\rm int}.
 \label{eq:newton-fixed-energy-confidence}
\end{equation}
The bounds extend to normal protocol limits admitting energy-respecting,
bounded-duration approximants with convergent output probabilities and
expected interaction time.
\end{proposition}

\begin{proof}
We first bound the probability near the two support endpoints. For a
normalized vector-valued wavefunction, the zero boundary trace and
Cauchy--Schwarz give
\[
 \|\psi(x)\|^2\le(\ell-x)\int_x^\ell\|\psi'(u)\|^2du,
 \qquad \ell-\delta_{\mathcal E}\le x\le\ell.
\]
The vanishing boundary value expresses the wavefunction inside the
strip as an integral of its derivative. Integrating the resulting
probability bound and adding the disjoint left endpoint strip gives
\begin{equation}
 \Pr\{|x|>\ell-\delta_{\mathcal E}\}
 \le\frac{\delta_{\mathcal E}^2}{2}
       \int_{-\ell}^{\ell}\|\psi'(x)\|^2dx
 \le\frac{m\mathcal E\delta_{\mathcal E}^2}{\hbar^2}
 =\frac12.
 \label{eq:newton-boundary-mass}
\end{equation}
Averaging this quadratic-form inequality proves it for mixed
conditional states, including states entangled with ancillary systems.
Hence at least half the probability lies in
$[-\ell+\delta_{\mathcal E},\ell-\delta_{\mathcal E}]$ on every recorded
history.

For $s>0$, the factors $A_s,B_s$ in
Eq.~\eqref{eq:newton-centered-integral} have absolute value at most
$S_s=\sinh(s\ell)$. On the interior interval their absolute value is
at most $L_s=\cosh(s\ell)-e^{-s(\ell-\delta_{\mathcal E})}$.
Indeed, comparison of the two endpoint values leaves the nonnegative
difference $2[\cosh(s\ell)-\cosh(s(\ell-\delta_{\mathcal E}))]$.
Since $0<L_s<S_s$, the probability bound implies
\begin{equation}
 \operatorname{Tr}(\rho_{A,h}A_s^2),\quad
 \operatorname{Tr}(\rho_{B,h}B_s^2)
 \le Q_s:=\frac{S_s^2+L_s^2}{2}<S_s^2.
 \label{eq:newton-interior-moment}
\end{equation}

An operator-norm bound alone would miss this restriction on the occupied
state. We instead bound the entropy derivative by the actual second
moments and denote the Schmidt entropy in bits by $S_2$. For a
pure state with Schmidt probabilities $\lambda_i$ and a product
Hamiltonian $A\otimes B$, the entropy derivative in bits is a sum over
$i<j$ with coefficients
$2\sqrt{\lambda_i\lambda_j}\log_2(\lambda_i/\lambda_j)/\hbar$.
The definition of $\beta$ gives
\[
 2\sqrt{\lambda_i\lambda_j}
 |\log_2(\lambda_i/\lambda_j)|
 \le\beta(\lambda_i+\lambda_j).
\]
Weighted Cauchy--Schwarz and Hermiticity therefore yield
\begin{equation}
 |\dot S_2|\le\frac\beta\hbar
 \sqrt{\langle A^2\rangle\langle B^2\rangle}.
 \label{eq:newton-moment-rate}
\end{equation}
Here the matrix elements are in the Schmidt bases of the complete
laboratories. In particular,
$\sum_{i<j}(\lambda_i+\lambda_j)|A_{ij}|^2
 \le\operatorname{Tr}(\rho_AA^2)$, so local memories introduce no
dimension factor. Applying Cauchy--Schwarz once more to a pure-state
ensemble within a recorded history replaces the two expectations by
their values in that history's mixed state. Individual unrecorded pure
components need not obey the energy bound.

Integrating Eq.~\eqref{eq:newton-moment-rate} over the Laplace
representation now bounds the average entropy increase per interaction
time by
\[
 \frac{\beta\mathcal G}{\hbar}
 \int_0^\infty e^{-sd}Q_s\,ds
 =\beta(\kappa-\Delta\kappa_{\mathcal E}).
\]
To evaluate the difference, we use
\[
 S_s^2-L_s^2=-1+e^{s\delta_{\mathcal E}}
 +e^{-s(2\ell-\delta_{\mathcal E})}
 -e^{-s(2\ell-2\delta_{\mathcal E})}.
\]
Its Laplace integral gives Eq.~\eqref{eq:newton-energy-gap}.
The inequalities $0<Q_s<S_s^2$ prove the strict bounds on
$\Delta\kappa_{\mathcal E}$.

The kinetic-energy constraint must remain valid when passing to
continuous positions. A position grid would change that constraint, so
we regularize the entropy while leaving the physical state unchanged.
For $\varepsilon>0$, define the scalar function $g_\varepsilon$ and its
associated spectral entropy by
\[
 g_\varepsilon(x)=-(x+\varepsilon)\log_2(x+\varepsilon)
                  +\varepsilon\log_2\varepsilon,
 \qquad
 S_\varepsilon(\rho)=\operatorname{Tr}g_\varepsilon(\rho)
                         -g_\varepsilon(1).
\]
This finite, concave spectral function vanishes on pure states. Its
Schmidt-pair derivative contains
$\log_2[(\lambda_i+\varepsilon)/(\lambda_j+\varepsilon)]$, whose
magnitude is no larger than the logarithm used above. Thus the same
rate bound holds uniformly in $\varepsilon$. Local instruments cannot
increase its pure-state ensemble average, by concavity. Moreover,
$g_\varepsilon(x)-xg_\varepsilon(1)\ge0$ for $0\le x\le1$ and
converges to $-x\log_2x$. Fatou's lemma therefore passes the integrated
bound to the true final Schmidt entropies. Averaging over the recorded
histories gives
$\sum_h w_hS_2(\psi_h)\le\Gamma_{\mathcal E}\mathbb E T_{\rm int}$.
The separable-comparator and binary-data-processing argument of
Corollary~\ref{cor:newton-kl} proves
Eq.~\eqref{eq:newton-fixed-energy-confidence}. Producing $L$ Bell pairs
with error tending to zero requires average entropy at least $L-o(L)$
on their finite output registers, proving the capacity bound. The stated
protocol limits preserve the scalar probability and time inequalities.
\end{proof}

For large $\mathcal E$, the certified gap has the expansion
\[
 \Delta\kappa_{\mathcal E}
 =\frac{\mathcal G\delta_{\mathcal E}}{2\hbar}
 \left[\frac1{d^2}-\frac1{(d+2\ell)^2}\right]
 +O(\delta_{\mathcal E}^2).
\]
Thus every finite admissible budget gives a strict reduction, with this
bound approaching the ideal capacity as $\mathcal E^{-1/2}$. The exact
capacity at fixed energy and a matching confidence construction remain
open. For fixed power $0<p<1$, Eq.~\eqref{eq:newton-fixed-energy-confidence}
also gives the leading mean-time lower bound
$[p/\Gamma_{\mathcal E}+o(1)]\log_2(1/\alpha)$ as $\alpha\to0$.
A preparation-energy bound alone does not imply
Eq.~\eqref{eq:newton-active-energy}, because later controls and the
interaction can change the kinetic energy. The finite-packet
construction in Section~\ref{sec:newtonian-packets} instead has the
explicit uniform dwell-energy bound of Eq.~\eqref{eq:newton-dwell-energy}.

\subsection{A finite-barrier realization of the finite-count construction}
\label{sec:newton-finite-barriers}

We can replace the divergent walls of the smooth-packet construction
in Section~\ref{sec:newtonian-packets} by bounded local potentials and
control the resulting leakage for a fixed
interaction time. The comparison uses the residual of the compact
reference packets, so it does not require a bound on the Newtonian
potential over the entire unconfined configuration space.

Take equal masses $m$ and packet support radii $w$, with axial branch
centres $\pm a$. Recall the compact profile $f$ used in the three-dimensional
product packets. Put $\sigma=w/\sqrt3$ and
$A=\hbar^2/(2m\sigma^2)$, the energy scale of one coordinate well.
Capping its dimensionless potential $u$ at height $M$ removes a part whose
norm on the reference packet is measured by $R_M$:
\[
 u(z)=\frac{f''(z)}{f(z)}=\frac{6z^4-2}{(1-z^2)^4},\qquad
 R_M=\left[\int_{-1}^1 (u(z)-M)_+^2|f(z)|^2\,dz\right]^{1/2}.
\]
Here $M>0$ is a dimensionless barrier height and $(x)_+=\max(x,0)$.
On each occupied coordinate interval replace $Au$ by
$A\min\{u,M\}$, and set the potential to $AM$ outside those intervals.
The longitudinal coordinate has two disjoint intervals, centred at
$\pm a$; each transverse coordinate has one. The sum of these three
coordinate potentials defines the local potential on all of
$\mathbb R^3$. It is bounded above and below. On a reference packet the
local Schr\"odinger residual is the sum of the three functions
$-A(u-M)_+f$, whose norms are at most $AR_M$ each.

Let $H_M$ be the full kinetic-plus-potential Hamiltonian with these local
potentials and the three-dimensional Newtonian pair potential. Its usual
semibounded self-adjoint realization is well defined. More generally,
the argument applies to any self-adjoint pair Hamiltonian agreeing with
the Newtonian potential on the support of the reference packets, including
a finite-body continuation at overlapping configurations. Let
$\psi_{\rm ref}(t)$ be the original superposition of the four stationary
packet products, with their branch-centre phases. All these packets are
smooth and supported away from the pair singularity, so they belong to
the domain of $H_M$. Their residual obeys
\[
 \left\|\left(H_M-i\hbar\frac{d}{dt}\right)
       \psi_{\rm ref}(t)\right\|
 \le \epsilon_V+6AR_M,
 \qquad
 \epsilon_V=\frac{\mathcal G(2w)}{(d-2a)(d-2a-2w)}.
\]
The gravity term follows from the reciprocal-distance bound on the
reference support. The well term follows by the triangle inequality;
disjoint packet supports make this bound uniform in the four branch
phases. Applying variation of constants to this differentiable reference
state gives
\begin{equation}
 \|e^{-iH_Mt/\hbar}\psi_{\rm ref}(0)-\psi_{\rm ref}(t)\|
 \le\eta_M(t):=\frac{t}{\hbar}(\epsilon_V+6AR_M).
 \label{eq:newton-finite-barrier-error}
\end{equation}
This estimate uses a residual on the reference state, rather than an
operator-norm difference between the two Hamiltonians.

At $T_B=\pi/(4\kappa_a)$, use a deterministic local channel that maps
the two reference packets to the two logical basis states and extends
to a trace-preserving channel on their orthogonal complement. In
particular, amplitudes outside the reference subspace are assigned a
qubit output and are retained. The reference output is a Bell pair.
Contractivity of fidelity under this decoder and
Eq.~\eqref{eq:newton-finite-barrier-error} give
$F_{\rm Bell}\ge1-\eta_M(T_B)^2$ for $\eta_M(T_B)\le1$.
Consequently the same local Pauli verifier has
\[
 q\ge1-\frac23\eta_M(T_B)^2,\qquad b=\frac23.
\]
The finite-count guarantees above therefore apply with $\eta$ replaced
by $\eta_M(T_B)$. If an alternative implementation assigns every escape
event a nonpassing outcome, the conservative guarantee is instead
$q\ge1-\eta_M(T_B)^2$.

At the calibration in Section~\ref{sec:newtonian-packets}
($m=10^{-14}\,\mathrm{kg}$, $a=125\,\mu\mathrm m$, $w=1\,\mathrm{nm}$
and $d=450\,\mu\mathrm m$), the finite height $M=10^6$ gives
\[
 AM=1.668182576\times10^{-30}\,\mathrm J,
 \quad R_M=1.001269215\times10^{-6},
 \quad \eta_M(T_B)=8.154642576\times10^{-5}.
\]
Twelve independent trials then have size at most $0.007707346629$ and
power at least $0.999999946801$, at aggregate dwell time
$300.2147959\,\mathrm s$. The barrier modification starts
$12.6172\,\mathrm{pm}$ inside each coordinate edge. These numbers specify
a bounded-potential construction; implementing its spatial profile,
preparation and complete local decoder is an additional control task.

For a bounded finite-body interaction $V\ge-V_*$ that is nonpositive
on the initial support, the total kinetic energy also has a uniform
bound. The capped local potentials have lower bound
$-6A u_*$, where $u_*=8.5914976719\ldots$, and their initial expectation
is no larger than for the uncapped zero-energy reference state.
Conservation of energy thus gives
\[
 \langle K_A+K_B\rangle_t\le V_*+6Au_*.
\]
For equal homogeneous spheres of radius $R$, their Newtonian mutual
energy has $V_*=6\mathcal G/(5R)$, and agrees with
$-\mathcal G/L$ when their centre separation $L$ exceeds $2R$.
At density $3500\,\mathrm{kg\,m^{-3}}$ and the calibration mass,
$R=0.8802671\,\mu\mathrm m$ and the total kinetic bound is
$9.184550136\times10^{-33}\,\mathrm J$. Finite barriers permit leakage,
so this extension supplies a finite-time preparation and testing
construction; the hard-support capacity statements keep their original
confinement hypothesis.

\subsection{Near contact, three spatial levels outperform two}

The two-position optimum in Theorem~\ref{thm:newton-three-d} holds
for sufficiently separated balls. Near contact, several matched branch
pairs can each have a short separation. We show that three spatial levels
can then beat every maintained two-dimensional spatial encoding, including
encodings whose basis states are delocalized. The supports are again the
two radius-$r$ balls with centre separation $d>2r$. Let one
mass be confined throughout each interaction to a maintained spatial
subspace with isometry $W:\mathbb C^2\to\mathcal H_A$. The other mass
may use its full spatial Hilbert space, and both laboratories may have
arbitrary local ancillary memories. Put $\delta=d-2r>0$ and let
\[
 w_V=\frac{\mathcal G}{d-2r}-\frac{\mathcal G}{d+2r}
\]
be the global potential range. The compressed position POVM has the form
$W^\dagger\Pi_A(d\mathbf a)W
=d\nu(\mathbf a)[I+\mathbf n(\mathbf a)\cdot\boldsymbol\sigma]$,
where $\boldsymbol\sigma$ is the Pauli vector on the spatial qubit,
$\nu$ is a probability measure, $\int\mathbf n\,d\nu=0$ and
$|\mathbf n|\le1$. This Bloch-vector decomposition expresses every
compressed position measurement as a scalar part and a qubit part.
Writing
$V_{\mathbf a}=V(\mathbf a,X_B)$, subtract the local term
$I\otimes\int V_{\mathbf a}d\nu$ and the global range midpoint $v_0$.
The remaining interaction is
\[
 \int d\nu(\mathbf a)
 [\mathbf n(\mathbf a)\cdot\boldsymbol\sigma]
 \otimes[V_{\mathbf a}-v_0I].
\]
The two product norms are at most $1$ and $w_V/2$; their spectral
diameters are at most $2$ and $w_V$. The established ancilla-assisted
product entropy bound, integrated over $\nu$, therefore gives
\begin{equation}
 C_{E,\,2\mathrm D\,\mathrm{port}}\le
 \frac{\beta w_V}{2\hbar}.
 \label{eq:three-d-two-upper}
\end{equation}
The same proof covers a time- or classical-history-dependent maintained
subspace and ideal time sharing. The isometry acts on the spatial code;
memories are ancillary identities. It excludes leakage during interaction
and coherent memory-controlled code choices whose combined spatial range
has dimension greater than two. Such larger ranges are not a single
maintained two-dimensional spatial port. The bound is not restricted to
localized point encodings and does not require both masses to be qubits.

For the lower construction, use matching equilateral rings facing one
another. This makes the three matched-label distances equal and shorter
than all six unmatched distances. Label the vertices by $x=0,1,2$ and
choose
\[
 \mathbf a_x=(\rho\cos\theta_x,\rho\sin\theta_x,z),\qquad
 \mathbf b_x=(\rho\cos\theta_x,\rho\sin\theta_x,-z),\qquad
 \theta_x=2\pi x/3,\quad \rho^2+z^2=r^2.
\]
Matched labels have distance $s=d-2z$ and unequal labels have distance
$L=\sqrt{s^2+3\rho^2}$. On these three-dimensional spatial codes,
\[
 H=-\frac{\mathcal G}{L}I-\Delta_\triangle P_{\rm eq},\qquad
 \Delta_\triangle=\mathcal G(1/s-1/L),\quad
 P_{\rm eq}=\sum_{x=0}^2|xx\rangle\langle xx|.
\]
We choose Schmidt bases in which this equal-label projector couples
one heavily occupied Schmidt component to two weakly occupied ones.
The relative phases make both contributions to the entropy derivative
positive. Let $u_j=3^{-1/2}\sum_x e^{2\pi ijx/3}|x\rangle$,
$v_j=e^{i\phi_j}u_j^*$, $\phi=(0,\pi/2,\pi/2)$, and prepare the
Schmidt state $\sum_j\sqrt{p_j}|u_jv_j\rangle$ with
$p=(.9,.05,.05)$. Directly,
$\langle u_iv_i|P_{\rm eq}|u_jv_j\rangle=e^{i(\phi_j-\phi_i)}/3$.
Its positive entropy production is
\begin{equation}
 \Gamma_3=\frac{2\Delta_\triangle}{3\hbar}
 \sum_{j=1}^2\sqrt{p_0p_j}\log_2\frac{p_0}{p_j}
 =\frac{\sqrt2}{5}\log_2(18)\frac{\Delta_\triangle}{\hbar}.
 \label{eq:three-d-three-rate}
\end{equation}
This input is already entangled. The established entanglement-processing
framework converts its incremental entropy gain into an asymptotic net
generation rate; the finite initial seed is produced and counted, not
provided for free\cite{BennettHarrowLeungSmolin2003}.

At $d=2.001r$ and $z=.99998r$,
\[
 s=\frac{13r}{12500},\qquad L=\frac{\sqrt{302701}}{50000}r,
 \qquad
 \Gamma_3=1026.88452638\frac{\mathcal G}{\hbar r},
\]
whereas Eq.~\eqref{eq:three-d-two-upper} is
$955.89767006\mathcal G/(\hbar r)$. This is stronger than the coarse
localized two-position bound $\beta\mathcal G/(2\hbar\delta)
=956.13664448\mathcal G/(\hbar r)$. Thus the construction exceeds even
the maintained-two-dimensional-port bound by more than $7.39\%$.
Moving the ring centres
slightly inward and using sufficiently narrow smooth stationary packets
preserves the strict gap with finite energy in every approximation.
This does not give the exact near-contact capacity or an advantage at
fixed preparation energy, control bandwidth or noisy memory lifetime.
For finite bodies of radii $R_A,R_B$, the support clearance must also
satisfy $d-2r>R_A+R_B$; the near-contact example is not a feasibility estimate.

\clearpage

\bibliography{references,references_added}

\begin{thebibliography}{39}%
\makeatletter
\providecommand \@ifxundefined [1]{%
 \@ifx{#1\undefined}
}%
\providecommand \@ifnum [1]{%
 \ifnum #1\expandafter \@firstoftwo
 \else \expandafter \@secondoftwo
 \fi
}%
\providecommand \@ifx [1]{%
 \ifx #1\expandafter \@firstoftwo
 \else \expandafter \@secondoftwo
 \fi
}%
\providecommand \natexlab [1]{#1}%
\providecommand \enquote  [1]{``#1''}%
\providecommand \bibnamefont  [1]{#1}%
\providecommand \bibfnamefont [1]{#1}%
\providecommand \citenamefont [1]{#1}%
\providecommand \href@noop [0]{\@secondoftwo}%
\providecommand \href [0]{\begingroup \@sanitize@url \@href}%
\providecommand \@href[1]{\@@startlink{#1}\@@href}%
\providecommand \@@href[1]{\endgroup#1\@@endlink}%
\providecommand \@sanitize@url [0]{\catcode `\\12\catcode `\$12\catcode
  `\&12\catcode `\#12\catcode `\^12\catcode `\_12\catcode `\%12\relax}%
\providecommand \@@startlink[1]{}%
\providecommand \@@endlink[0]{}%
\providecommand \url  [0]{\begingroup\@sanitize@url \@url }%
\providecommand \@url [1]{\endgroup\@href {#1}{\urlprefix }}%
\providecommand \urlprefix  [0]{URL }%
\providecommand \Eprint [0]{\href }%
\providecommand \doibase [0]{https://doi.org/}%
\providecommand \selectlanguage [0]{\@gobble}%
\providecommand \bibinfo  [0]{\@secondoftwo}%
\providecommand \bibfield  [0]{\@secondoftwo}%
\providecommand \translation [1]{[#1]}%
\providecommand \BibitemOpen [0]{}%
\providecommand \bibitemStop [0]{}%
\providecommand \bibitemNoStop [0]{.\EOS\space}%
\providecommand \EOS [0]{\spacefactor3000\relax}%
\providecommand \BibitemShut  [1]{\csname bibitem#1\endcsname}%
\let\auto@bib@innerbib\@empty
%</preamble>
\bibitem [{\citenamefont {Bose}\ \emph {et~al.}(2017)\citenamefont {Bose},
  \citenamefont {Mazumdar}, \citenamefont {Morley}, \citenamefont {Ulbricht},
  \citenamefont {Toro{\v{s}}}, \citenamefont {Paternostro}, \citenamefont
  {Geraci}, \citenamefont {Barker}, \citenamefont {Kim},\ and\ \citenamefont
  {Milburn}}]{Bose2017}%
  \BibitemOpen
  \bibfield  {author} {\bibinfo {author} {\bibfnamefont {S.}~\bibnamefont
  {Bose}}, \bibinfo {author} {\bibfnamefont {A.}~\bibnamefont {Mazumdar}},
  \bibinfo {author} {\bibfnamefont {G.~W.}\ \bibnamefont {Morley}}, \bibinfo
  {author} {\bibfnamefont {H.}~\bibnamefont {Ulbricht}}, \bibinfo {author}
  {\bibfnamefont {M.}~\bibnamefont {Toro{\v{s}}}}, \bibinfo {author}
  {\bibfnamefont {M.}~\bibnamefont {Paternostro}}, \bibinfo {author}
  {\bibfnamefont {A.~A.}\ \bibnamefont {Geraci}}, \bibinfo {author}
  {\bibfnamefont {P.~F.}\ \bibnamefont {Barker}}, \bibinfo {author}
  {\bibfnamefont {M.~S.}\ \bibnamefont {Kim}},\ and\ \bibinfo {author}
  {\bibfnamefont {G.}~\bibnamefont {Milburn}},\ }\bibfield  {title} {\bibinfo
  {title} {Spin entanglement witness for quantum gravity},\ }\href
  {https://doi.org/10.1103/PhysRevLett.119.240401} {\bibfield  {journal}
  {\bibinfo  {journal} {Physical Review Letters}\ }\textbf {\bibinfo {volume}
  {119}},\ \bibinfo {pages} {240401} (\bibinfo {year} {2017})}\BibitemShut
  {NoStop}%
\bibitem [{\citenamefont {Marletto}\ and\ \citenamefont
  {Vedral}(2017)}]{MarlettoVedral2017}%
  \BibitemOpen
  \bibfield  {author} {\bibinfo {author} {\bibfnamefont {C.}~\bibnamefont
  {Marletto}}\ and\ \bibinfo {author} {\bibfnamefont {V.}~\bibnamefont
  {Vedral}},\ }\bibfield  {title} {\bibinfo {title} {Gravitationally induced
  entanglement between two massive particles is sufficient evidence of quantum
  effects in gravity},\ }\href {https://doi.org/10.1103/PhysRevLett.119.240402}
  {\bibfield  {journal} {\bibinfo  {journal} {Physical Review Letters}\
  }\textbf {\bibinfo {volume} {119}},\ \bibinfo {pages} {240402} (\bibinfo
  {year} {2017})}\BibitemShut {NoStop}%
\bibitem [{\citenamefont {Carney}\ \emph {et~al.}(2019)\citenamefont {Carney},
  \citenamefont {Stamp},\ and\ \citenamefont {Taylor}}]{Carney2019}%
  \BibitemOpen
  \bibfield  {author} {\bibinfo {author} {\bibfnamefont {D.}~\bibnamefont
  {Carney}}, \bibinfo {author} {\bibfnamefont {P.~C.~E.}\ \bibnamefont
  {Stamp}},\ and\ \bibinfo {author} {\bibfnamefont {J.~M.}\ \bibnamefont
  {Taylor}},\ }\bibfield  {title} {\bibinfo {title} {Tabletop experiments for
  quantum gravity: a user's manual},\ }\href
  {https://doi.org/10.1088/1361-6382/aaf9ca} {\bibfield  {journal} {\bibinfo
  {journal} {Classical and Quantum Gravity}\ }\textbf {\bibinfo {volume}
  {36}},\ \bibinfo {pages} {034001} (\bibinfo {year} {2019})}\BibitemShut
  {NoStop}%
\bibitem [{\citenamefont {Krisnanda}\ \emph {et~al.}(2020)\citenamefont
  {Krisnanda}, \citenamefont {Tham}, \citenamefont {Paternostro},\ and\
  \citenamefont {Paterek}}]{Krisnanda2020}%
  \BibitemOpen
  \bibfield  {author} {\bibinfo {author} {\bibfnamefont {T.}~\bibnamefont
  {Krisnanda}}, \bibinfo {author} {\bibfnamefont {G.~Y.}\ \bibnamefont {Tham}},
  \bibinfo {author} {\bibfnamefont {M.}~\bibnamefont {Paternostro}},\ and\
  \bibinfo {author} {\bibfnamefont {T.}~\bibnamefont {Paterek}},\ }\bibfield
  {title} {\bibinfo {title} {Observable quantum entanglement due to gravity},\
  }\href {https://doi.org/10.1038/s41534-020-0243-y} {\bibfield  {journal}
  {\bibinfo  {journal} {npj Quantum Information}\ }\textbf {\bibinfo {volume}
  {6}},\ \bibinfo {pages} {12} (\bibinfo {year} {2020})}\BibitemShut {NoStop}%
\bibitem [{\citenamefont {Lami}\ \emph {et~al.}(2024)\citenamefont {Lami},
  \citenamefont {Pedernales},\ and\ \citenamefont {Plenio}}]{Lami2024}%
  \BibitemOpen
  \bibfield  {author} {\bibinfo {author} {\bibfnamefont {L.}~\bibnamefont
  {Lami}}, \bibinfo {author} {\bibfnamefont {J.~S.}\ \bibnamefont
  {Pedernales}},\ and\ \bibinfo {author} {\bibfnamefont {M.~B.}\ \bibnamefont
  {Plenio}},\ }\bibfield  {title} {\bibinfo {title} {Testing the quantumness of
  gravity without entanglement},\ }\href
  {https://doi.org/10.1103/PhysRevX.14.021022} {\bibfield  {journal} {\bibinfo
  {journal} {Physical Review X}\ }\textbf {\bibinfo {volume} {14}},\ \bibinfo
  {pages} {021022} (\bibinfo {year} {2024})}\BibitemShut {NoStop}%
\bibitem [{\citenamefont {Ganardi}\ \emph {et~al.}(2024)\citenamefont
  {Ganardi}, \citenamefont {Panwar}, \citenamefont {Pandit}, \citenamefont
  {Wo{\l}oncewicz},\ and\ \citenamefont {Paterek}}]{Ganardi2024}%
  \BibitemOpen
  \bibfield  {author} {\bibinfo {author} {\bibfnamefont {R.}~\bibnamefont
  {Ganardi}}, \bibinfo {author} {\bibfnamefont {E.}~\bibnamefont {Panwar}},
  \bibinfo {author} {\bibfnamefont {M.}~\bibnamefont {Pandit}}, \bibinfo
  {author} {\bibfnamefont {B.}~\bibnamefont {Wo{\l}oncewicz}},\ and\ \bibinfo
  {author} {\bibfnamefont {T.}~\bibnamefont {Paterek}},\ }\bibfield  {title}
  {\bibinfo {title} {Quantitative nonclassicality of mediated interactions},\
  }\href {https://doi.org/10.1103/PRXQuantum.5.010318} {\bibfield  {journal}
  {\bibinfo  {journal} {PRX Quantum}\ }\textbf {\bibinfo {volume} {5}},\
  \bibinfo {pages} {010318} (\bibinfo {year} {2024})}\BibitemShut {NoStop}%
\bibitem [{\citenamefont {Westphal}\ \emph {et~al.}(2021)\citenamefont
  {Westphal}, \citenamefont {Hepach}, \citenamefont {Pfaff},\ and\
  \citenamefont {Aspelmeyer}}]{Westphal2021}%
  \BibitemOpen
  \bibfield  {author} {\bibinfo {author} {\bibfnamefont {T.}~\bibnamefont
  {Westphal}}, \bibinfo {author} {\bibfnamefont {H.}~\bibnamefont {Hepach}},
  \bibinfo {author} {\bibfnamefont {J.}~\bibnamefont {Pfaff}},\ and\ \bibinfo
  {author} {\bibfnamefont {M.}~\bibnamefont {Aspelmeyer}},\ }\bibfield  {title}
  {\bibinfo {title} {Measurement of gravitational coupling between
  millimetre-sized masses},\ }\href
  {https://doi.org/10.1038/s41586-021-03250-7} {\bibfield  {journal} {\bibinfo
  {journal} {Nature}\ }\textbf {\bibinfo {volume} {591}},\ \bibinfo {pages}
  {225} (\bibinfo {year} {2021})}\BibitemShut {NoStop}%
\bibitem [{\citenamefont {Fein}\ \emph {et~al.}(2019)\citenamefont {Fein},
  \citenamefont {Geyer}, \citenamefont {Zwick}, \citenamefont {Kia{\l}ka},
  \citenamefont {Pedalino}, \citenamefont {Mayor}, \citenamefont {Gerlich},\
  and\ \citenamefont {Arndt}}]{Fein2019}%
  \BibitemOpen
  \bibfield  {author} {\bibinfo {author} {\bibfnamefont {Y.~Y.}\ \bibnamefont
  {Fein}}, \bibinfo {author} {\bibfnamefont {P.}~\bibnamefont {Geyer}},
  \bibinfo {author} {\bibfnamefont {P.}~\bibnamefont {Zwick}}, \bibinfo
  {author} {\bibfnamefont {F.}~\bibnamefont {Kia{\l}ka}}, \bibinfo {author}
  {\bibfnamefont {S.}~\bibnamefont {Pedalino}}, \bibinfo {author}
  {\bibfnamefont {M.}~\bibnamefont {Mayor}}, \bibinfo {author} {\bibfnamefont
  {S.}~\bibnamefont {Gerlich}},\ and\ \bibinfo {author} {\bibfnamefont
  {M.}~\bibnamefont {Arndt}},\ }\bibfield  {title} {\bibinfo {title} {Quantum
  superposition of molecules beyond 25 {kDa}},\ }\href
  {https://doi.org/10.1038/s41567-019-0663-9} {\bibfield  {journal} {\bibinfo
  {journal} {Nature Physics}\ }\textbf {\bibinfo {volume} {15}},\ \bibinfo
  {pages} {1242} (\bibinfo {year} {2019})}\BibitemShut {NoStop}%
\bibitem [{\citenamefont {Zhou}\ \emph {et~al.}(2018)\citenamefont {Zhou},
  \citenamefont {Zhang}, \citenamefont {Preskill},\ and\ \citenamefont
  {Jiang}}]{ZhouEtAl2018}%
  \BibitemOpen
  \bibfield  {author} {\bibinfo {author} {\bibfnamefont {S.}~\bibnamefont
  {Zhou}}, \bibinfo {author} {\bibfnamefont {M.}~\bibnamefont {Zhang}},
  \bibinfo {author} {\bibfnamefont {J.}~\bibnamefont {Preskill}},\ and\
  \bibinfo {author} {\bibfnamefont {L.}~\bibnamefont {Jiang}},\ }\bibfield
  {title} {\bibinfo {title} {Achieving the {Heisenberg} limit in quantum
  metrology using quantum error correction},\ }\href
  {https://doi.org/10.1038/s41467-017-02510-3} {\bibfield  {journal} {\bibinfo
  {journal} {Nature Communications}\ }\textbf {\bibinfo {volume} {9}},\
  \bibinfo {pages} {78} (\bibinfo {year} {2018})},\ \Eprint
  {https://arxiv.org/abs/1706.02445} {arXiv:1706.02445} \BibitemShut {NoStop}%
\bibitem [{\citenamefont {Demkowicz-Dobrza{\'n}ski}\ \emph
  {et~al.}(2017)\citenamefont {Demkowicz-Dobrza{\'n}ski}, \citenamefont
  {Czajkowski},\ and\ \citenamefont {Sekatski}}]{DemkowiczDobrzanski2017}%
  \BibitemOpen
  \bibfield  {author} {\bibinfo {author} {\bibfnamefont {R.}~\bibnamefont
  {Demkowicz-Dobrza{\'n}ski}}, \bibinfo {author} {\bibfnamefont
  {J.}~\bibnamefont {Czajkowski}},\ and\ \bibinfo {author} {\bibfnamefont
  {P.}~\bibnamefont {Sekatski}},\ }\bibfield  {title} {\bibinfo {title}
  {Adaptive quantum metrology under general {Markovian} noise},\ }\href
  {https://doi.org/10.1103/PhysRevX.7.041009} {\bibfield  {journal} {\bibinfo
  {journal} {Physical Review X}\ }\textbf {\bibinfo {volume} {7}},\ \bibinfo
  {pages} {041009} (\bibinfo {year} {2017})}\BibitemShut {NoStop}%
\bibitem [{\citenamefont {Layden}\ and\ \citenamefont
  {Cappellaro}(2018)}]{LaydenCappellaro2018}%
  \BibitemOpen
  \bibfield  {author} {\bibinfo {author} {\bibfnamefont {D.}~\bibnamefont
  {Layden}}\ and\ \bibinfo {author} {\bibfnamefont {P.}~\bibnamefont
  {Cappellaro}},\ }\bibfield  {title} {\bibinfo {title} {Spatial noise
  filtering through error correction for quantum sensing},\ }\href
  {https://doi.org/10.1038/s41534-018-0082-2} {\bibfield  {journal} {\bibinfo
  {journal} {npj Quantum Information}\ }\textbf {\bibinfo {volume} {4}},\
  \bibinfo {pages} {30} (\bibinfo {year} {2018})}\BibitemShut {NoStop}%
\bibitem [{\citenamefont {Layden}\ \emph {et~al.}(2019)\citenamefont {Layden},
  \citenamefont {Zhou}, \citenamefont {Cappellaro},\ and\ \citenamefont
  {Jiang}}]{LaydenEtAl2019}%
  \BibitemOpen
  \bibfield  {author} {\bibinfo {author} {\bibfnamefont {D.}~\bibnamefont
  {Layden}}, \bibinfo {author} {\bibfnamefont {S.}~\bibnamefont {Zhou}},
  \bibinfo {author} {\bibfnamefont {P.}~\bibnamefont {Cappellaro}},\ and\
  \bibinfo {author} {\bibfnamefont {L.}~\bibnamefont {Jiang}},\ }\bibfield
  {title} {\bibinfo {title} {Ancilla-free quantum error correction codes for
  quantum metrology},\ }\href {https://doi.org/10.1103/PhysRevLett.122.040502}
  {\bibfield  {journal} {\bibinfo  {journal} {Physical Review Letters}\
  }\textbf {\bibinfo {volume} {122}},\ \bibinfo {pages} {040502} (\bibinfo
  {year} {2019})},\ \Eprint {https://arxiv.org/abs/1811.01450}
  {arXiv:1811.01450} \BibitemShut {NoStop}%
\bibitem [{\citenamefont {Wang}\ \emph {et~al.}(2026)\citenamefont {Wang},
  \citenamefont {Pedernales},\ and\ \citenamefont
  {Plenio}}]{WangPedernalesPlenio2026}%
  \BibitemOpen
  \bibfield  {author} {\bibinfo {author} {\bibfnamefont {Y.-X.}\ \bibnamefont
  {Wang}}, \bibinfo {author} {\bibfnamefont {J.~S.}\ \bibnamefont
  {Pedernales}},\ and\ \bibinfo {author} {\bibfnamefont {M.~B.}\ \bibnamefont
  {Plenio}},\ }\href {https://arxiv.org/abs/2611.05044v1} {\bibinfo {title}
  {Recoverability of gravitational entanglement under decoherence}} (\bibinfo
  {year} {2026}),\ \Eprint {https://arxiv.org/abs/2611.05044v1}
  {arXiv:2611.05044v1} \BibitemShut {NoStop}%
\bibitem [{\citenamefont {Kafri}\ \emph {et~al.}(2014)\citenamefont {Kafri},
  \citenamefont {Taylor},\ and\ \citenamefont {Milburn}}]{Kafri2014}%
  \BibitemOpen
  \bibfield  {author} {\bibinfo {author} {\bibfnamefont {D.}~\bibnamefont
  {Kafri}}, \bibinfo {author} {\bibfnamefont {J.~M.}\ \bibnamefont {Taylor}},\
  and\ \bibinfo {author} {\bibfnamefont {G.~J.}\ \bibnamefont {Milburn}},\
  }\bibfield  {title} {\bibinfo {title} {A classical channel model for
  gravitational decoherence},\ }\href
  {https://doi.org/10.1088/1367-2630/16/6/065020} {\bibfield  {journal}
  {\bibinfo  {journal} {New Journal of Physics}\ }\textbf {\bibinfo {volume}
  {16}},\ \bibinfo {pages} {065020} (\bibinfo {year} {2014})}\BibitemShut
  {NoStop}%
\bibitem [{\citenamefont {Li}\ \emph {et~al.}(2026)\citenamefont {Li},
  \citenamefont {Miki},\ and\ \citenamefont {Chen}}]{LiMikiChen2026}%
  \BibitemOpen
  \bibfield  {author} {\bibinfo {author} {\bibfnamefont {A.}~\bibnamefont
  {Li}}, \bibinfo {author} {\bibfnamefont {D.}~\bibnamefont {Miki}},\ and\
  \bibinfo {author} {\bibfnamefont {Y.}~\bibnamefont {Chen}},\ }\href
  {https://arxiv.org/abs/2605.26215} {\bibinfo {title} {Universal bound for
  entanglement generation}} (\bibinfo {year} {2026}),\ \Eprint
  {https://arxiv.org/abs/2605.26215} {arXiv:2605.26215} \BibitemShut {NoStop}%
\bibitem [{\citenamefont {Bennett}\ \emph {et~al.}(2003)\citenamefont
  {Bennett}, \citenamefont {Harrow}, \citenamefont {Leung},\ and\ \citenamefont
  {Smolin}}]{BennettHarrowLeungSmolin2003}%
  \BibitemOpen
  \bibfield  {author} {\bibinfo {author} {\bibfnamefont {C.~H.}\ \bibnamefont
  {Bennett}}, \bibinfo {author} {\bibfnamefont {A.~W.}\ \bibnamefont {Harrow}},
  \bibinfo {author} {\bibfnamefont {D.~W.}\ \bibnamefont {Leung}},\ and\
  \bibinfo {author} {\bibfnamefont {J.~A.}\ \bibnamefont {Smolin}},\ }\bibfield
   {title} {\bibinfo {title} {On the capacities of bipartite hamiltonians and
  unitary gates},\ }\href {https://doi.org/10.1109/TIT.2003.814935} {\bibfield
  {journal} {\bibinfo  {journal} {IEEE Transactions on Information Theory}\
  }\textbf {\bibinfo {volume} {49}},\ \bibinfo {pages} {1895} (\bibinfo {year}
  {2003})},\ \Eprint {https://arxiv.org/abs/quant-ph/0205057v4}
  {arXiv:quant-ph/0205057v4} \BibitemShut {NoStop}%
\bibitem [{\citenamefont {D{\"u}r}\ \emph {et~al.}(2001)\citenamefont
  {D{\"u}r}, \citenamefont {Vidal}, \citenamefont {Cirac}, \citenamefont
  {Linden},\ and\ \citenamefont {Popescu}}]{DuerVidalCiracLindenPopescu2001}%
  \BibitemOpen
  \bibfield  {author} {\bibinfo {author} {\bibfnamefont {W.}~\bibnamefont
  {D{\"u}r}}, \bibinfo {author} {\bibfnamefont {G.}~\bibnamefont {Vidal}},
  \bibinfo {author} {\bibfnamefont {J.~I.}\ \bibnamefont {Cirac}}, \bibinfo
  {author} {\bibfnamefont {N.}~\bibnamefont {Linden}},\ and\ \bibinfo {author}
  {\bibfnamefont {S.}~\bibnamefont {Popescu}},\ }\bibfield  {title} {\bibinfo
  {title} {Entanglement capabilities of nonlocal hamiltonians},\ }\href
  {https://doi.org/10.1103/PhysRevLett.87.137901} {\bibfield  {journal}
  {\bibinfo  {journal} {Physical Review Letters}\ }\textbf {\bibinfo {volume}
  {87}},\ \bibinfo {pages} {137901} (\bibinfo {year} {2001})},\ \Eprint
  {https://arxiv.org/abs/quant-ph/0006034} {arXiv:quant-ph/0006034}
  \BibitemShut {NoStop}%
\bibitem [{\citenamefont {Childs}\ \emph {et~al.}(2003)\citenamefont {Childs},
  \citenamefont {Leung}, \citenamefont {Verstraete},\ and\ \citenamefont
  {Vidal}}]{ChildsLeungVerstraeteVidal2003}%
  \BibitemOpen
  \bibfield  {author} {\bibinfo {author} {\bibfnamefont {A.~M.}\ \bibnamefont
  {Childs}}, \bibinfo {author} {\bibfnamefont {D.~W.}\ \bibnamefont {Leung}},
  \bibinfo {author} {\bibfnamefont {F.}~\bibnamefont {Verstraete}},\ and\
  \bibinfo {author} {\bibfnamefont {G.}~\bibnamefont {Vidal}},\ }\bibfield
  {title} {\bibinfo {title} {Asymptotic entanglement capacity of the {Ising}
  and anisotropic {Heisenberg} interactions},\ }\href
  {https://doi.org/10.26421/QIC3.2-1} {\bibfield  {journal} {\bibinfo
  {journal} {Quantum Information and Computation}\ }\textbf {\bibinfo {volume}
  {3}},\ \bibinfo {pages} {97} (\bibinfo {year} {2003})},\ \Eprint
  {https://arxiv.org/abs/quant-ph/0207052v2} {arXiv:quant-ph/0207052v2}
  \BibitemShut {NoStop}%
\bibitem [{\citenamefont {Childs}\ \emph {et~al.}(2004)\citenamefont {Childs},
  \citenamefont {Leung},\ and\ \citenamefont {Vidal}}]{ChildsLeungVidal2004}%
  \BibitemOpen
  \bibfield  {author} {\bibinfo {author} {\bibfnamefont {A.~M.}\ \bibnamefont
  {Childs}}, \bibinfo {author} {\bibfnamefont {D.~W.}\ \bibnamefont {Leung}},\
  and\ \bibinfo {author} {\bibfnamefont {G.}~\bibnamefont {Vidal}},\ }\bibfield
   {title} {\bibinfo {title} {Reversible simulation of bipartite product
  hamiltonians},\ }\href {https://doi.org/10.1109/TIT.2004.828069} {\bibfield
  {journal} {\bibinfo  {journal} {IEEE Transactions on Information Theory}\
  }\textbf {\bibinfo {volume} {50}},\ \bibinfo {pages} {1189} (\bibinfo {year}
  {2004})},\ \Eprint {https://arxiv.org/abs/quant-ph/0303097v1}
  {arXiv:quant-ph/0303097v1} \BibitemShut {NoStop}%
\bibitem [{\citenamefont {Frank}\ and\ \citenamefont
  {Lieb}(2010)}]{FrankLieb2010}%
  \BibitemOpen
  \bibfield  {author} {\bibinfo {author} {\bibfnamefont {R.~L.}\ \bibnamefont
  {Frank}}\ and\ \bibinfo {author} {\bibfnamefont {E.~H.}\ \bibnamefont
  {Lieb}},\ }\bibfield  {title} {\bibinfo {title} {Inversion positivity and the
  sharp {Hardy--Littlewood--Sobolev} inequality},\ }\href
  {https://doi.org/10.1007/s00526-009-0302-x} {\bibfield  {journal} {\bibinfo
  {journal} {Calculus of Variations and Partial Differential Equations}\
  }\textbf {\bibinfo {volume} {39}},\ \bibinfo {pages} {85} (\bibinfo {year}
  {2010})},\ \Eprint {https://arxiv.org/abs/0904.4275} {arXiv:0904.4275
  [math.FA]} \BibitemShut {NoStop}%
\bibitem [{\citenamefont {Bravyi}(2007)}]{Bravyi2007}%
  \BibitemOpen
  \bibfield  {author} {\bibinfo {author} {\bibfnamefont {S.}~\bibnamefont
  {Bravyi}},\ }\bibfield  {title} {\bibinfo {title} {Upper bounds on entangling
  rates of bipartite {Hamiltonians}},\ }\href
  {https://doi.org/10.1103/PhysRevA.76.052319} {\bibfield  {journal} {\bibinfo
  {journal} {Physical Review A}\ }\textbf {\bibinfo {volume} {76}},\ \bibinfo
  {pages} {052319} (\bibinfo {year} {2007})}\BibitemShut {NoStop}%
\bibitem [{\citenamefont {Van~Acoleyen}\ \emph {et~al.}(2013)\citenamefont
  {Van~Acoleyen}, \citenamefont {Mari{\"e}n},\ and\ \citenamefont
  {Verstraete}}]{VanAcoleyen2013}%
  \BibitemOpen
  \bibfield  {author} {\bibinfo {author} {\bibfnamefont {K.}~\bibnamefont
  {Van~Acoleyen}}, \bibinfo {author} {\bibfnamefont {M.}~\bibnamefont
  {Mari{\"e}n}},\ and\ \bibinfo {author} {\bibfnamefont {F.}~\bibnamefont
  {Verstraete}},\ }\bibfield  {title} {\bibinfo {title} {Entanglement rates and
  area laws},\ }\href {https://doi.org/10.1103/PhysRevLett.111.170501}
  {\bibfield  {journal} {\bibinfo  {journal} {Physical Review Letters}\
  }\textbf {\bibinfo {volume} {111}},\ \bibinfo {pages} {170501} (\bibinfo
  {year} {2013})}\BibitemShut {NoStop}%
\bibitem [{\citenamefont {Knill}\ and\ \citenamefont
  {Laflamme}(1997)}]{KnillLaflamme1997}%
  \BibitemOpen
  \bibfield  {author} {\bibinfo {author} {\bibfnamefont {E.}~\bibnamefont
  {Knill}}\ and\ \bibinfo {author} {\bibfnamefont {R.}~\bibnamefont
  {Laflamme}},\ }\bibfield  {title} {\bibinfo {title} {Theory of quantum
  error-correcting codes},\ }\href {https://doi.org/10.1103/PhysRevA.55.900}
  {\bibfield  {journal} {\bibinfo  {journal} {Physical Review A}\ }\textbf
  {\bibinfo {volume} {55}},\ \bibinfo {pages} {900} (\bibinfo {year}
  {1997})}\BibitemShut {NoStop}%
\bibitem [{\citenamefont {Jain}\ \emph {et~al.}(2024)\citenamefont {Jain},
  \citenamefont {Iosue}, \citenamefont {Barg},\ and\ \citenamefont
  {Albert}}]{JainEtAl2024}%
  \BibitemOpen
  \bibfield  {author} {\bibinfo {author} {\bibfnamefont {S.~P.}\ \bibnamefont
  {Jain}}, \bibinfo {author} {\bibfnamefont {J.~T.}\ \bibnamefont {Iosue}},
  \bibinfo {author} {\bibfnamefont {A.}~\bibnamefont {Barg}},\ and\ \bibinfo
  {author} {\bibfnamefont {V.~V.}\ \bibnamefont {Albert}},\ }\bibfield  {title}
  {\bibinfo {title} {Quantum spherical codes},\ }\href
  {https://doi.org/10.1038/s41567-024-02496-y} {\bibfield  {journal} {\bibinfo
  {journal} {Nature Physics}\ }\textbf {\bibinfo {volume} {20}},\ \bibinfo
  {pages} {1300} (\bibinfo {year} {2024})},\ \Eprint
  {https://arxiv.org/abs/2302.11593} {arXiv:2302.11593} \BibitemShut {NoStop}%
\bibitem [{\citenamefont {Tilly}\ \emph {et~al.}(2021)\citenamefont {Tilly},
  \citenamefont {Marshman}, \citenamefont {Mazumdar},\ and\ \citenamefont
  {Bose}}]{TillyEtAl2021}%
  \BibitemOpen
  \bibfield  {author} {\bibinfo {author} {\bibfnamefont {J.}~\bibnamefont
  {Tilly}}, \bibinfo {author} {\bibfnamefont {R.~J.}\ \bibnamefont {Marshman}},
  \bibinfo {author} {\bibfnamefont {A.}~\bibnamefont {Mazumdar}},\ and\
  \bibinfo {author} {\bibfnamefont {S.}~\bibnamefont {Bose}},\ }\bibfield
  {title} {\bibinfo {title} {Qudits for witnessing quantum-gravity-induced
  entanglement of masses under decoherence},\ }\href
  {https://doi.org/10.1103/PhysRevA.104.052416} {\bibfield  {journal} {\bibinfo
   {journal} {Physical Review A}\ }\textbf {\bibinfo {volume} {104}},\ \bibinfo
  {pages} {052416} (\bibinfo {year} {2021})},\ \Eprint
  {https://arxiv.org/abs/2101.08086} {arXiv:2101.08086 [quant-ph]} \BibitemShut
  {NoStop}%
\bibitem [{\citenamefont {Nguyen}\ and\ \citenamefont
  {Bernards}(2020)}]{NguyenBernards2020}%
  \BibitemOpen
  \bibfield  {author} {\bibinfo {author} {\bibfnamefont {H.~C.}\ \bibnamefont
  {Nguyen}}\ and\ \bibinfo {author} {\bibfnamefont {F.}~\bibnamefont
  {Bernards}},\ }\bibfield  {title} {\bibinfo {title} {Entanglement dynamics of
  two mesoscopic objects with gravitational interaction},\ }\href
  {https://doi.org/10.1140/epjd/e2020-10077-8} {\bibfield  {journal} {\bibinfo
  {journal} {The European Physical Journal D}\ }\textbf {\bibinfo {volume}
  {74}},\ \bibinfo {pages} {69} (\bibinfo {year} {2020})},\ \Eprint
  {https://arxiv.org/abs/1906.11184} {arXiv:1906.11184 [quant-ph]} \BibitemShut
  {NoStop}%
\bibitem [{\citenamefont {Paulson}\ \emph {et~al.}(2021)\citenamefont
  {Paulson}, \citenamefont {Panwar}, \citenamefont {Banerjee},\ and\
  \citenamefont {Srikanth}}]{PaulsonEtAl2021}%
  \BibitemOpen
  \bibfield  {author} {\bibinfo {author} {\bibfnamefont {K.~G.}\ \bibnamefont
  {Paulson}}, \bibinfo {author} {\bibfnamefont {E.}~\bibnamefont {Panwar}},
  \bibinfo {author} {\bibfnamefont {S.}~\bibnamefont {Banerjee}},\ and\
  \bibinfo {author} {\bibfnamefont {R.}~\bibnamefont {Srikanth}},\ }\bibfield
  {title} {\bibinfo {title} {Hierarchy of quantum correlations under
  non-markovian dynamics},\ }\href {https://doi.org/10.1007/s11128-021-03061-9}
  {\bibfield  {journal} {\bibinfo  {journal} {Quantum Information Processing}\
  }\textbf {\bibinfo {volume} {20}},\ \bibinfo {pages} {141} (\bibinfo {year}
  {2021})},\ \Eprint {https://arxiv.org/abs/2004.11208} {arXiv:2004.11208
  [quant-ph]} \BibitemShut {NoStop}%
\bibitem [{\citenamefont {Wiseman}(1994)}]{WisemanMilburn1994}%
  \BibitemOpen
  \bibfield  {author} {\bibinfo {author} {\bibfnamefont {H.~M.}\ \bibnamefont
  {Wiseman}},\ }\bibfield  {title} {\bibinfo {title} {Quantum theory of
  continuous feedback},\ }\href {https://doi.org/10.1103/PhysRevA.49.2133}
  {\bibfield  {journal} {\bibinfo  {journal} {Physical Review A}\ }\textbf
  {\bibinfo {volume} {49}},\ \bibinfo {pages} {2133} (\bibinfo {year}
  {1994})}\BibitemShut {NoStop}%
\bibitem [{\citenamefont {Horodecki}\ \emph {et~al.}(1997)\citenamefont
  {Horodecki}, \citenamefont {Horodecki},\ and\ \citenamefont
  {Horodecki}}]{HorodeckiDistill1997}%
  \BibitemOpen
  \bibfield  {author} {\bibinfo {author} {\bibfnamefont {M.}~\bibnamefont
  {Horodecki}}, \bibinfo {author} {\bibfnamefont {P.}~\bibnamefont
  {Horodecki}},\ and\ \bibinfo {author} {\bibfnamefont {R.}~\bibnamefont
  {Horodecki}},\ }\bibfield  {title} {\bibinfo {title} {Inseparable two
  spin-{$1/2$} density matrices can be distilled to a singlet form},\ }\href
  {https://doi.org/10.1103/PhysRevLett.78.574} {\bibfield  {journal} {\bibinfo
  {journal} {Physical Review Letters}\ }\textbf {\bibinfo {volume} {78}},\
  \bibinfo {pages} {574} (\bibinfo {year} {1997})},\ \Eprint
  {https://arxiv.org/abs/quant-ph/9607009} {arXiv:quant-ph/9607009}
  \BibitemShut {NoStop}%
\bibitem [{\citenamefont {Devetak}\ and\ \citenamefont
  {Winter}(2005)}]{DevetakWinter2005}%
  \BibitemOpen
  \bibfield  {author} {\bibinfo {author} {\bibfnamefont {I.}~\bibnamefont
  {Devetak}}\ and\ \bibinfo {author} {\bibfnamefont {A.}~\bibnamefont
  {Winter}},\ }\bibfield  {title} {\bibinfo {title} {Distillation of secret key
  and entanglement from quantum states},\ }\href
  {https://doi.org/10.1098/rspa.2004.1372} {\bibfield  {journal} {\bibinfo
  {journal} {Proceedings of the Royal Society A}\ }\textbf {\bibinfo {volume}
  {461}},\ \bibinfo {pages} {207} (\bibinfo {year} {2005})}\BibitemShut
  {NoStop}%
\bibitem [{\citenamefont {Zhu}\ and\ \citenamefont
  {Hayashi}(2019)}]{ZhuHayashi2019}%
  \BibitemOpen
  \bibfield  {author} {\bibinfo {author} {\bibfnamefont {H.}~\bibnamefont
  {Zhu}}\ and\ \bibinfo {author} {\bibfnamefont {M.}~\bibnamefont {Hayashi}},\
  }\bibfield  {title} {\bibinfo {title} {Optimal verification and fidelity
  estimation of maximally entangled states},\ }\href
  {https://doi.org/10.1103/PhysRevA.99.052346} {\bibfield  {journal} {\bibinfo
  {journal} {Physical Review A}\ }\textbf {\bibinfo {volume} {99}},\ \bibinfo
  {pages} {052346} (\bibinfo {year} {2019})},\ \Eprint
  {https://arxiv.org/abs/1901.09772} {arXiv:1901.09772} \BibitemShut {NoStop}%
\bibitem [{\citenamefont {Pallister}\ \emph {et~al.}(2018)\citenamefont
  {Pallister}, \citenamefont {Linden},\ and\ \citenamefont
  {Montanaro}}]{PallisterLindenMontanaro2018}%
  \BibitemOpen
  \bibfield  {author} {\bibinfo {author} {\bibfnamefont {S.}~\bibnamefont
  {Pallister}}, \bibinfo {author} {\bibfnamefont {N.}~\bibnamefont {Linden}},\
  and\ \bibinfo {author} {\bibfnamefont {A.}~\bibnamefont {Montanaro}},\
  }\bibfield  {title} {\bibinfo {title} {Optimal verification of entangled
  states with local measurements},\ }\href
  {https://doi.org/10.1103/PhysRevLett.120.170502} {\bibfield  {journal}
  {\bibinfo  {journal} {Physical Review Letters}\ }\textbf {\bibinfo {volume}
  {120}},\ \bibinfo {pages} {170502} (\bibinfo {year} {2018})},\ \Eprint
  {https://arxiv.org/abs/1709.03353v3} {arXiv:1709.03353v3} \BibitemShut
  {NoStop}%
\bibitem [{\citenamefont {Holevo}\ \emph {et~al.}(2005)\citenamefont {Holevo},
  \citenamefont {Shirokov},\ and\ \citenamefont
  {Werner}}]{HolevoShirokovWerner2005}%
  \BibitemOpen
  \bibfield  {author} {\bibinfo {author} {\bibfnamefont {A.~S.}\ \bibnamefont
  {Holevo}}, \bibinfo {author} {\bibfnamefont {M.~E.}\ \bibnamefont
  {Shirokov}},\ and\ \bibinfo {author} {\bibfnamefont {R.~F.}\ \bibnamefont
  {Werner}},\ }\href {https://arxiv.org/abs/quant-ph/0504204} {\bibinfo {title}
  {Separability and entanglement-breaking in infinite dimensions}} (\bibinfo
  {year} {2005}),\ \Eprint {https://arxiv.org/abs/quant-ph/0504204}
  {arXiv:quant-ph/0504204} \BibitemShut {NoStop}%
\bibitem [{\citenamefont {Ozawa}(1984)}]{Ozawa1984}%
  \BibitemOpen
  \bibfield  {author} {\bibinfo {author} {\bibfnamefont {M.}~\bibnamefont
  {Ozawa}},\ }\bibfield  {title} {\bibinfo {title} {Quantum measuring processes
  of continuous observables},\ }\href {https://doi.org/10.1063/1.526000}
  {\bibfield  {journal} {\bibinfo  {journal} {Journal of Mathematical Physics}\
  }\textbf {\bibinfo {volume} {25}},\ \bibinfo {pages} {79} (\bibinfo {year}
  {1984})}\BibitemShut {NoStop}%
\bibitem [{\citenamefont {Friedland}\ and\ \citenamefont
  {Lim}(2018)}]{FriedlandLim2018}%
  \BibitemOpen
  \bibfield  {author} {\bibinfo {author} {\bibfnamefont {S.}~\bibnamefont
  {Friedland}}\ and\ \bibinfo {author} {\bibfnamefont {L.-H.}\ \bibnamefont
  {Lim}},\ }\bibfield  {title} {\bibinfo {title} {Nuclear norm of higher-order
  tensors},\ }\href {https://doi.org/10.1090/mcom/3239} {\bibfield  {journal}
  {\bibinfo  {journal} {Mathematics of Computation}\ }\textbf {\bibinfo
  {volume} {87}},\ \bibinfo {pages} {1255} (\bibinfo {year}
  {2018})}\BibitemShut {NoStop}%
\bibitem [{\citenamefont {Bennett}\ \emph {et~al.}(1996)\citenamefont
  {Bennett}, \citenamefont {DiVincenzo}, \citenamefont {Smolin},\ and\
  \citenamefont {Wootters}}]{BennettDiVincenzoSmolinWootters1996}%
  \BibitemOpen
  \bibfield  {author} {\bibinfo {author} {\bibfnamefont {C.~H.}\ \bibnamefont
  {Bennett}}, \bibinfo {author} {\bibfnamefont {D.~P.}\ \bibnamefont
  {DiVincenzo}}, \bibinfo {author} {\bibfnamefont {J.~A.}\ \bibnamefont
  {Smolin}},\ and\ \bibinfo {author} {\bibfnamefont {W.~K.}\ \bibnamefont
  {Wootters}},\ }\bibfield  {title} {\bibinfo {title} {Mixed-state entanglement
  and quantum error correction},\ }\href
  {https://doi.org/10.1103/PhysRevA.54.3824} {\bibfield  {journal} {\bibinfo
  {journal} {Physical Review A}\ }\textbf {\bibinfo {volume} {54}},\ \bibinfo
  {pages} {3824} (\bibinfo {year} {1996})},\ \Eprint
  {https://arxiv.org/abs/quant-ph/9604024} {arXiv:quant-ph/9604024}
  \BibitemShut {NoStop}%
\bibitem [{\citenamefont {B{\'e}ny}\ and\ \citenamefont
  {Oreshkov}(2010)}]{BenyOreshkov2010}%
  \BibitemOpen
  \bibfield  {author} {\bibinfo {author} {\bibfnamefont {C.}~\bibnamefont
  {B{\'e}ny}}\ and\ \bibinfo {author} {\bibfnamefont {O.}~\bibnamefont
  {Oreshkov}},\ }\bibfield  {title} {\bibinfo {title} {General conditions for
  approximate quantum error correction and near-optimal recovery channels},\
  }\href {https://doi.org/10.1103/PhysRevLett.104.120501} {\bibfield  {journal}
  {\bibinfo  {journal} {Physical Review Letters}\ }\textbf {\bibinfo {volume}
  {104}},\ \bibinfo {pages} {120501} (\bibinfo {year} {2010})},\ \Eprint
  {https://arxiv.org/abs/0907.5391} {arXiv:0907.5391 [quant-ph]} \BibitemShut
  {NoStop}%
\bibitem [{\citenamefont {Das}\ \emph {et~al.}(2020)\citenamefont {Das},
  \citenamefont {B{\"a}uml},\ and\ \citenamefont {Wilde}}]{DasBaumlWilde2020}%
  \BibitemOpen
  \bibfield  {author} {\bibinfo {author} {\bibfnamefont {S.}~\bibnamefont
  {Das}}, \bibinfo {author} {\bibfnamefont {S.}~\bibnamefont {B{\"a}uml}},\
  and\ \bibinfo {author} {\bibfnamefont {M.~M.}\ \bibnamefont {Wilde}},\
  }\bibfield  {title} {\bibinfo {title} {Entanglement and secret-key-agreement
  capacities of bipartite quantum interactions and read-only memory devices},\
  }\href {https://doi.org/10.1103/PhysRevA.101.012344} {\bibfield  {journal}
  {\bibinfo  {journal} {Physical Review A}\ }\textbf {\bibinfo {volume}
  {101}},\ \bibinfo {pages} {012344} (\bibinfo {year} {2020})},\ \Eprint
  {https://arxiv.org/abs/1712.00827v3} {arXiv:1712.00827v3} \BibitemShut
  {NoStop}%
\bibitem [{\citenamefont {Harrow}\ and\ \citenamefont
  {Lowe}(2025)}]{HarrowLowe2025}%
  \BibitemOpen
  \bibfield  {author} {\bibinfo {author} {\bibfnamefont {A.~W.}\ \bibnamefont
  {Harrow}}\ and\ \bibinfo {author} {\bibfnamefont {A.}~\bibnamefont {Lowe}},\
  }\bibfield  {title} {\bibinfo {title} {Optimal quantum circuit cuts with
  application to clustered hamiltonian simulation},\ }\href
  {https://doi.org/10.1103/PRXQuantum.6.010316} {\bibfield  {journal} {\bibinfo
   {journal} {PRX Quantum}\ }\textbf {\bibinfo {volume} {6}},\ \bibinfo {pages}
  {010316} (\bibinfo {year} {2025})},\ \Eprint
  {https://arxiv.org/abs/2403.01018v2} {arXiv:2403.01018v2} \BibitemShut
  {NoStop}%
\end{thebibliography}%
\end{document}